\documentclass[amsmath,amssymb,10pt,aps,twocolumn,nofootinbib]{revtex4-2} 
\usepackage[colorlinks=true]{hyperref}
 \usepackage{lipsum}
 \usepackage{graphicx}
 \usepackage{textcomp}
\usepackage[dvipsnames]{xcolor}
\usepackage{latexsym}
\usepackage{amsmath}
\usepackage{amsthm}
\usepackage{amssymb}
\usepackage{epstopdf}
\usepackage{enumitem}
\usepackage{orcidlink}
\usepackage{setspace} 
\usepackage{dcolumn} 
\usepackage{bm} 
\usepackage{setspace} 
\usepackage{slashed}
\usepackage{color}
\usepackage{xcolor}
\usepackage{youngtab}
\usepackage{tikz}
\usepackage{braket}
\usepackage[version=4]{mhchem}

\usepackage[tabbotcap]{subfigure}

\allowdisplaybreaks
 
\begin{document}
\title{Quantum impurity model for two-stage multipolar ordering and Fermi surface reconstruction
}

\author{Daniel J. Schultz\,\orcidlink{0000-0003-0567-850X}}
\thanks{These authors contributed equally to this work.}
\affiliation{Department of Physics, University of Toronto, Toronto, Ontario M5S 1A7, Canada}

\author{SangEun Han\,\orcidlink{0000-0003-3141-1964}}
\thanks{These authors contributed equally to this work.}
\affiliation{Department of Physics, University of Toronto, Toronto, Ontario M5S 1A7, Canada}

\author{Yong Baek Kim}
\affiliation{Department of Physics, University of Toronto, Toronto, Ontario M5S 1A7, Canada}

\date{\today}   

\begin{abstract}

Classification and understanding of quantum phase transitions and critical phenomena in itinerant electron systems are outstanding questions in quantum materials research. Recent experiments on heavy fermion systems with higher-rank multipolar local moments provide a new platform to study such questions. In particular, experiments on \ce{Ce3Pd20(Si,Ge)6} show novel quantum critical behaviors via two consecutive magnetic field-driven quantum phase transitions. At each transition, the derivative of the Hall resistivity jumps discontinuously, which was attributed to sequential Fermi surface reconstructions. Motivated by this discovery, we consider an effective quantum impurity model of itinerant electrons coupled to local dipolar, quadrupolar, and octupolar moments arising from \ce{Ce^{3+}} ions. Using renormalization group analyses, we demonstrate that two-stage multipolar ordering and Fermi surface reconstruction arise depending on which multipolar moments participate in the Fermi surface and which other moments are decoupled via Kondo destruction.
\end{abstract}

\maketitle

\paragraph*{Introduction.}
Experimental work on rare-earth metallic systems has shown a wide variety of quantum phases of matter and novel quantum phase transitions (QPTs). In some systems, $f$-electrons give rise to multipolar moments, and these moments couple to itinerant conduction electrons; this situation is described by a multipolar Kondo lattice model. Since we have a large number of degrees of freedom and constraining crystal field symmetries, Kondo couplings become highly anisotropic in contrast to the conventional dipolar Kondo lattice model. Hence we can expect to find more diverse novel quantum phenomena such as the emergence of the new types of Kondo phases, RKKY mediated multipolar ordered phases \cite{Doniach1977a, Coleman1983}, and novel quantum criticality from the competition between them \cite{Schroder2000, Aynajian2012, Yang2017, Kumar2021, Si2010a, Maksimovic2022, Prokleska2015, Inui2020, Knebel2011, Seiro2018, Oh2019, Fuhrman2021}. Many of these phenomena are not yet well understood, and researchers face ongoing challenges to theoretically describe and experimentally detect these multipolar quantum phases \cite{Sorensen2021, Thalmeier2008, Koitzsch2016, Barman2019a, Thalmeier2014, Patri2020d, Patri2020e, Schultz2021b, Sakai2011b, Thalmeier2021, Iwasa2018, Patri2022, Sato2012a, Fu2020a, Shimura2019, Matsubayashi2012c, Tsujimoto2014a, Sakai2012a}. 

One class of metallic systems which contains multipolar moments is \ce{Ce3Pd20(Si,Ge)6}. Here, the magnetically active Ce$^{3+}$ ions ($4f^1$ configuration) are surrounded by a tetrahedral crystal field which constrains the \ce{Ce^{3+}} ground state to be a fourfold degenerate quartet \cite{Kitagawa1997, Goto2009, Paschen2008}. The four states consist of two degenerate Kramers doublets, and support a large number of the multipolar moments (see Table~\ref{tab:multipolar_moments}) \cite{Shiina1997}. 

\begin{figure}[t]
    \centering
    \includegraphics[height=0.55\linewidth]{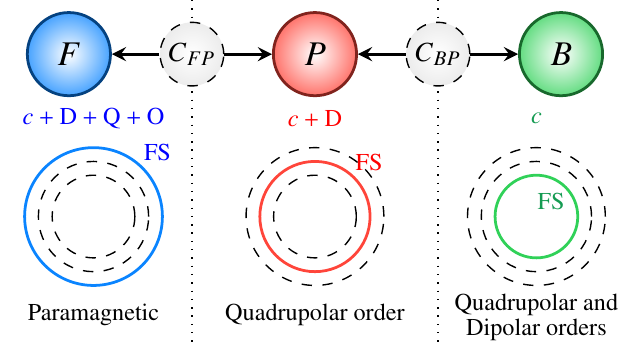} 
    \caption{A schematic diagram for QPT with two-stage Kondo destruction. 
    $F$, $P$, and $B$ stand for the fermionic Kondo, partially Kondo destroyed, and magnetically ordered phases, respectively.
    The second row stands for which degrees of freedom participate in the formation of the Fermi surface. $c$, D, Q, and O mean conduction electrons, dipolar, quadrupolar and octupolar moments, respectively. The circles in the third row show the schematic size difference of the Fermi surface between each phase depending on how many degrees of freedom participated in the formation of the Fermi surface. The last row means the multipolar ordering at each phase.
    }
    \label{fig:2kd}
\end{figure}

Experimental studies on \ce{Ce3Pd20Si6} in particular show novel quantum critical behaviors corresponding to two consecutive field-induced QPTs \cite{Strydom2006, Goto2009, Mitamura2010, Ono2013, Martelli2019, Mazza2022}. At zero magnetic field, the system exhibits coexisting antiferromagnetic and antiferroquadrupolar order, and by increasing the external magnetic field along $[0\,0\,1]$, the antiferromagnetic order disappears but the antiferroquadrupolar order remains. Upon increasing the field further, the system arrives at another phase that has not been clearly identified yet. 
Interestingly, the experiment observed that the derivative of the Hall resistivity with respect to magnetic field, when extrapolated to zero temperature, jumps at both phase transitions. These jumps indicate sequential Fermi surface reconstruction \cite{Coleman2001,Si2001, Paschen2004, Steglich2014, Custers2012, Friedemann2010,Friedemann2010a,Liu2021}. 

In this work, we study a theoretical model for the existence of such a two-stage QPT and extract the types of magnetic order in the context of experiments on \ce{Ce3Pd20Si6}, as well as suggest experimental signatures of the quantum critical points. 
For simplicity, we construct a multipolar Bose-Fermi Kondo model of 15 multipolar moments of the \ce{Ce^{3+}} quartet coupled to both $p$-wave conduction electrons and a dynamical bosonic bath representing RKKY interactions \cite{Si1996,Smith1999,Sengupta2000, Zhu2002, Zarand2002, Kirchner2005}. Despite now being a local approximation of the Kondo lattice in the form of an impurity model in the spirit of dynamical mean field theory, the fact that the impurity is coupled to a bosonic bath via Bose-Kondo couplings means that magnetic fluctuations due to other sites are included in addition to the usual Fermi-Kondo effect with the fermionic (conduction electrons) bath. 
This model therefore facilitates a study of the competition between the Kondo effect and magnetic ordering. We use the renormalization group approach to determine the permissible types of magnetic order and Kondo destruction pathways purely on the basis of local symmetry. Specifically, we examine which local moments participate in the Fermi surface and which local moments are ordered and decouple from the conduction electrons in each part of the zero temperature phase diagram. We then discuss experimental consequences of our findings.

\paragraph*{Models.}
\label{sec:models}
As mentioned in the introduction, our effective model for this system consists of a single local multipolar moment coupled to conduction electrons and a dynamical bosonic bath representing RKKY magnetic fluctuations. To construct the model, we consider the local symmetry at the \ce{Ce^{3+}} ion site in \ce{Ce3Pd20(Si,Ge)6}. For this family of materials, there are two crystallographically distinct sites for Ce ions: the 4a and 8c sites \cite{Martelli2019, Donni2000}. The magnetically active Ce ions occupy the 8c sites and are surrounded by a \ce{Pd16} cage, which has tetrahedral $T_d$ symmetry. We construct the Bose-Fermi Kondo model for the system to include all symmetry-allowed interactions. To find such symmetry-allowed interactions, we list the transformations of constituent elements in the model under the tetrahedral group $T_d$ and time-reversal in Supplementary Materials~\cite{SM}.

The degenerate ground states of an ion in a vacuum can be described by an effective higher-spin system through Hund's rules. For the case of a \ce{Ce^{3+}} ion, it has a $4f^1$ configuration and resulting $J=5/2$ moment. In the presence of the $T_d$ CEF, the 6 degenerate states split and a $\Gamma_8$ quartet ground state is formed. The states of this quartet are listed in Supplementary Materials~\cite{SM}. Since there are 4 degenerate ground states, numerous multipolar moments can be formed; in particular we have 3 dipolar, 5 quadrupolar, and 7 octupolar moments \cite{Shiina1997}, which are tabulated in Table \ref{tab:multipolar_moments}. In the table they are classified by time-reversal and by irreps of $T_{d}$. The details of how the multipolar operators are constructed from the quartet states, as well as how to represent the multipolar operators by Abrikosov pseudofermions (which is required for the renormalization group analysis), are in the Supplemental Material \cite{SM}. 

\begin{table}[!t]
\centering
\begin{tabular}{|c|c|c|c|}
\hline
Irrep. & Stevens & In terms of $J_x,J_y,J_z$ & Moment \\ \hline \hline
$T_{1}$ & $J_x$ & $J_x$ & D \\ [5pt]
$T_{1}$ & $J_y$ & $J_y$ & D \\ [5pt]
$T_{1}$ & $J_z$ & $J_z$ & D\\ [5pt]
$E$ & $\mathcal{O}_{22}$ & $\frac{\sqrt{3}}{2}(J_x^2-J_y^2)$ & Q\\ [5pt]
$E$ & $\mathcal{O}_{20}$ & $\frac{1}{2}(3J_z^2 - \bm{J}^2)$ & Q \\ [5pt]
$T_{2+}$ & $\mathcal{O}_{yz}$ & $\frac{\sqrt{3}}{2}\overline{J_yJ_z}$ & Q \\ [5pt]
$T_{2+}$ & $\mathcal{O}_{zx}$ & $\frac{\sqrt{3}}{2}\overline{J_zJ_x}$ & Q \\ [5pt]
$T_{2+}$ & $\mathcal{O}_{xy}$ & $\frac{\sqrt{3}}{2}\overline{J_xJ_y}$ & Q\\ [5pt]
$A_{2}$ & $\mathcal{T}_{xyz}$ & $\frac{\sqrt{15}}{6}\overline{J_xJ_yJ_z}$ & O \\ [5pt]
$T_{1} $ & $\mathcal{T}^\alpha_x$ & $\frac{1}{2}(2J_x^3 - \overline{J_xJ_y^2} - \overline{J_z^2J_x})$ & O\\ [5pt]
$T_{1} $ & $\mathcal{T}^\alpha_y$ & $\frac{1}{2}(2J_y^3 - \overline{J_yJ_z^2} - \overline{J_x^2J_y})$ & O\\ [5pt]
$T_{1} $ & $\mathcal{T}^\alpha_z$ & $\frac{1}{2}(2J_z^3 - \overline{J_zJ_x^2} - \overline{J_y^2J_z})$ & O\\ [5pt]
$T_{2-}$ & $\mathcal{T}^\beta_x$ & $\frac{\sqrt{15}}{6}(\overline{J_xJ_y^2} - \overline{J_z^2J_x})$ & O\\ [5pt]
$T_{2-}$ & $\mathcal{T}^\beta_y$ & $\frac{\sqrt{15}}{6}(\overline{J_yJ_z^2} - \overline{J_x^2J_y})$ & O\\ [5pt]
$T_{2-}$ & $\mathcal{T}^\beta_z$ & $\frac{\sqrt{15}}{6}(\overline{J_zJ_x^2} - \overline{J_y^2J_z})$ & O\\\hline
\end{tabular}
\caption{Multipolar Moments, $J_x,J_y,J_z$ are $J=5/2$ operators. The overline notation means full symmetrization. For example $\overline{AB} = AB + BA$, $\overline{A^2B} = A^2B + ABA + BA^2$, and $\overline{ABC} = ABC + ACB + BAC + BCA + CAB + CBA$. The irrep.~column denotes irreducible representations of $T_d$, and the Stevens column contains Stevens operators. The $+\-$ subscripts on the $T_2$ moments denote the time reversal even/odd nature of the moments for quadrupole/octupoles. We do not include a $+\-$ label if there is no ambiguity. In the moment column, we indicate if the moment is dipolar (D), quadrupolar (Q), or octupolar (O). \label{tab:multipolar_moments}}
\end{table}

We now turn to the Fermi-Kondo model, where we couple the local moments to conduction electrons. The conduction electron wave functions are considered to be molecular orbitals centred on the \ce{Ce} ion and constructed from electrons hopping on the \ce{Pd16} cage. The resulting wave functions are classifiable according to irreducible representations of $T_d$. We construct a model of 3 degenerate bands of conduction electrons, made up of Wannier functions which lie in the $T_2$ irrep.~of $T_d$. We may use $p$-wave $\{x,y,z\}$ orbitals, or $d$-wave $T_2$ $\{yz,zx,xy\}$ orbitals; the results are identical with either choice and we use $p$-wave in this work. We assume a constant density of states for the conduction electrons, and couple them to the local moment in the maximal way allowed by symmetry; this leads to 15 coupling constants \cite{SM} (this is unrelated to the fact that there are 15 multipolar moments).

Lastly, we construct the Bose-Kondo part of the model, where the local moments are coupled to the fluctuating bosonic bath. In the case of a lattice of multipolar moments, the spin bilinear RKKY interaction is induced by the conduction electrons. As mentioned before, we capture such an interaction in the Bose-Fermi Kondo model by replacing it with a bosonic bath, which can be thought of as a dynamical Weiss mean field. 
The Bose-Kondo couplings are determined by the number of independent irreps., so we have 6 bosonic couplings because 2 of the 4 irreps are counted twice, namely $T_1$ and $T_2$ (see Table~\ref{tab:multipolar_moments}). 
The details of deriving the symmetry-allowed bosonic couplings for the model are given in Supplementary Materials \cite{SM}. 
The Hamiltonian for the kinetic part of the bosonic bath is shown as $H^B_0 = \sum_{i,\mathbf{k}}\Omega_\mathbf{k}\phi^{i\dagger}_\mathbf{k} \phi^i_\mathbf{k}$, where we assume that all flavors are degenerate for simplicity. The index $i=1,\dots,15$ runs over all bosonic baths, and $\Omega_\mathbf{k}$ is the dispersion of the bosonic fields. In order to perform the RG analysis, we set up an $\epsilon$ expansion, where $\epsilon$ controls the sub-linearity of the spectral function of the bosonic bath: 
\begin{equation}
\sum_\mathbf{k}[\delta(\omega-\Omega_\mathbf{k}) - \delta(\omega+\Omega_\mathbf{k})] = \frac{N_1^2}{2}|\omega|^{1-\epsilon}\text{sgn}(\omega). \label{eq:bosonic_spectral}
\end{equation}
Because we assumed that all flavors of the bosonic bath were degenerate, they also all have the same $\epsilon$ controlling their densities of states.

As a result, by adding the Fermi-Kondo and Bose-Kondo interactions, we construct the full Bose-Fermi Kondo model, which yields a model with a grand total of $15+6=21$ coupling constants, and we compute the full beta functions \cite{Ellis2017,SM}.

\paragraph*{Fixed points of the Bose-Fermi Kondo model and Two-Stage Kondo Destruction QPT.} \label{sec:fermi_betas}

From the beta functions we calculate, we can find a number of stable fixed points, but not all are physically important. In the following, we will discuss three stable fixed points, $F$, $B$, $P$, and two critical points, $C_{FP}$ and $C_{PB}$, between them, and how they are connected by the two-stage Kondo destruction QPT.

The first type of stable fixed point is a Fermi-Kondo fixed point, whose representative is $F$. 
This type of phase has local moments hybridized into the Fermi surface and hence has the largest Fermi surface. It is paramagnetic, may be Fermi or non-Fermi liquid \cite{Affleck1991a,Affleck1993b,Ludwig1991,Mora2009,Kimura2017b}, and can be found within the Fermi-Kondo models. 
These points have nonzero Fermi-Kondo couplings, while their Bose-Kondo couplings are all zero.
For the fixed point $F$ in particular, its fixed point Hamiltonian $H_F$ corresponds to a 6 generator truncated SU(4) fixed point \cite{Schultz2021c}, and thus $H_F$ can be written as follows:

    \begin{align}
        H_{F}={}&\frac{1}{2}\sum_{\rho,\tau=1}^4\tilde{\psi}_{\rho}^{\dagger}[(\sigma^{0}\otimes\vec{\sigma})_{\rho\tau}\cdot(Q^{x},O^{y},Q^{z})]\tilde{\psi}_{\tau}\notag\\
        &+\sum_{\rho,\tau=1}^2\psi_{\rho}^{\dagger}[\vec{\sigma}_{\rho\tau}\cdot(D^{x},D^{y},D^{z})]\psi_{\tau},
    \end{align}
    where $\vec{\sigma}=(\sigma^{x},\sigma^{y},\sigma^{z})$, $\tilde{\psi}$ and $\psi$ are 4 and 2 component spinors, respectively (see Supplementary Materials for the relationship between these new electrons and the original $p$-wave electrons \cite{SM}), which are related to the original $p$-wave conduction electrons via a change of basis \cite{Patri2020e,Schultz2021b,SM}. Here, $\{Q^{x},O^{y},Q^{z}\}\sim\{\mathcal{O}_{22},\mathcal{T}_{xyz},\mathcal{O}_{20}\}$, and $D^{x,y,z}\sim\{J_{x,y,z}\}$.
    Each set of three multipolar moments satisfies an SU(2) algebra, and the two SU(2) algebras are mutually commuting. Although the two components commute and thus appear decoupled, the leading irrelevant operator does not commute with either the two-channel or one-channel component, which means that they are coupled at any nonzero distance from the fixed point. This fixed point Hamiltonian has two-fold degenerate ground states in the strong coupling limit \cite{Lavagna2003, Bensimon2006a}, so the IR fixed point is valid and likely shows non-Fermi liquid behavior \cite{Lavagna2003, Bensimon2006a,Affleck1991a,Affleck1993b,Ludwig1991,Kimura2017b}. 

Secondly, there are the Bose-Kondo fixed points, whose representative is $B$. These phases have all local moments decoupled from the Fermi surface, and hence have the smallest Fermi surface. 
These points have nonzero Bose-Kondo couplings, while their Fermi-Kondo couplings are all zero. It means that they are magnetically ordered phases, and can be found within the Bose-Kondo model. 
In particular, the fixed point $B$ is a multipolar ordered fixed point that has quadrupolar ordering and  dipolar ordering with $E$ and $T_{1}$ irreps respectively. 

Thirdly, we find a (stable) partially Kondo-destroyed fixed point $P$. 
For a general partially Kondo-destroyed point, the multipolar moments are coupled to bosonic baths as well as conduction electrons, so some local moments are absorbed into the Fermi surface (and behave paramagnetically), while others decouple from the Fermi surface and magnetically order. This is only possible due to the large number of local states. 
In the case of $P$, the dipolar local moments are absorbed into the Fermi surface whereas the quadrupolar moments undergo magnetic ordering, so this point has a partially shrunk Fermi surface. 
Note that more details of the fixed points are given in Supplementary Materials \cite{SM}.
    
Furthermore, we find critical points $C_{FP}$ between $F$ and $P$, and $C_{BP}$ between $B$ and $P$. These critical points and stable fixed points are connected by the path of QPTs, $F\leftarrow C_{FP}\rightarrow P\leftarrow C_{BP}\rightarrow B$, represented pictorially in Fig.~\ref{fig:2kd}. The physical interpretation is as follows.

At $F$, as mentioned before, the system is paramagnetic with a large Fermi surface due to Kondo hybridization by one- and two-channel Kondo interactions and absence of Bose-Kondo coupling. 
When we pass through $C_{FP}$ from $F$ to $P$, the two-channel Kondo interaction vanishes, so the quadrupolar and octupolar moments are decoupled from the conduction electrons. During this, the quadrupolar Kondo coupling induces non-zero Bose-Kondo coupling for the quadrupolar moments, so we can reach $P$.
At $P$, since the conduction electrons decouple from the quadrupolar and octupolar moments, the Kondo effect is partially destroyed, but we still have nonzero Kondo couplings with dipolar moments. This corresponds to the Fermi surface shrinking one time, so it has a medium size of the Fermi surface. Furthermore, the decoupled quadrupolar moments order, so it has the quadrupolar order parameters $\sim\{J_x^2-J_y^2,3J_z^2-\bm{J}^2\}$.
Next, when we pass through $C_{BP}$ between $P$ and $B$, the remaining (dipolar) Kondo hybridization vanishes and thereby induces a non-zero Bose-Kondo coupling for the dipolar moments. This corresponds to a second shrinking of the Fermi surface, as well as magnetic ordering of the dipolar moments. 
At $B$, since all the Kondo couplings are zero, the Kondo effect is completely destroyed, and it has a small Fermi surface; the full picture of going from $F$ to $B$ is therefore two-stage Kondo destruction (Fig.~\ref{fig:2kd}). Moreover, at $B$, we have coexistence between quadrupolar ordering and dipolar ordering. 
The two-stage Kondo destruction QPT is consistent with experiment in the sense that the type of ordering we find matches the observed order parameters. Furthermore, the jump in the derivative of the Hall resistivity is consistent with the Kondo destruction phase transitions we have found. We note that, strictly speaking, the symmetry breaking character of the external magnetic field would modify the analysis, but we have neglected it for simplicity.
Note that the RG flow diagrams between the stable fixed points and critical points are presented in the Supplementary Materials~\cite{SM}.

\paragraph*{Ultrasound Measurement of Multipolar Susceptibility.}\label{sec:mul_suscep}

In addition to the qualitative signature of Fermi surface reconstruction, the multipolar susceptibility exponent can be used to quantitatively identify the fixed points. The multipolar susceptibility is defined by $\chi_{i}(\tau)=\braket{T_{\tau}S^{i}(\tau)S^{i}(0)}\sim \left(\tau_{0}/|\tau|\right)^{\gamma_{i}}$, where $S^{i}$ is the multipolar moment, $\gamma_{i}$ is the multipolar susceptibility exponent, $i$ is an index for the irrep., $\tau$ is imaginary time, and $\tau\gg\tau_{0}$ with the cutoff $\tau_{0}=1/\Lambda\sim 1/\mu$. We label the spin operators by irrep.~because any representative from the irrep.~yields the same result.
From the beta functions, we can compute the multipolar susceptibility exponent \cite{SM}. 
The resulting susceptibility exponents $\gamma_{i}$ are presented in Supplementary Materials~\cite{SM}. 
By assuming that the multipolar moments are primary fields with conformal dimension $\gamma_{i}/2$, the finite temperature scaling of the multipolar susceptibility is given by \cite{Zarand2002,Han2022}
\begin{align} 
    \chi_{i}'(\omega,T)\sim&\begin{cases}
    T^{\gamma_{i}-1}\left(1+C_{\text{Re}1}\left(\frac{\omega}{T}\right)^{2}\right),&|\omega/T|\ll1,\\
    \omega^{\gamma_{i}-1},&|\omega/T|\gg1,
    \end{cases}\\
    \chi_{i}''(\omega,T)\sim&\begin{cases}
    T^{\gamma_{i}-1}\left(\frac{\omega}{T}\right),&|\omega/T|\ll1,\\
    \omega^{\gamma_{i}-1},&|\omega/T|\gg1,
    \end{cases},
\end{align}
where $\chi_{i}(\omega,T)=\chi_{i}'(\omega,T)+i\chi_{i}''(\omega,T)$ and $C_{\text{Re}1}$ is a real constant. 
Although the dipolar susceptibilities can be detected by conventional techniques, the purely multipolar susceptibilities require elastic measurements. One way to achieve this is through ultrasound experiments. The symmetry-allowed free energy produces a linear coupling between strain and quadrupolar moments, which facilitates a relationship between elastic constants and quadrupolar susceptibilities. Furthermore, in the presence of an external magnetic field, a product of magnetic field and strain couples linearly to octupolar moments, adding octupolar susceptibility corrections to the elastic constants (see the Supplementary Materials for details on how this coupling arises through symmetry considerations \cite{SM}). 
Then, the resulting renormalized elastic constants in the presence of a small magnetic field $\mathbf{h} = (0,0,h_z)$ are then given by second-order perturbation theory as \cite{Patri2019d,Sorensen2021,Han2022}:
\begin{align}
    C_{11}-C_{12}={}&C_{11}^{0}-C_{12}^{0}-s_{E}^{2}\chi'_{Q_E}-2s_{-}^{2}h_{z}^{2}\chi_{O_{2}}',\\
    C_{44}={}&C_{44}^{0}-s_{+}^{2}\chi'_{Q_{2}}-s_{A}^{2}h_{z}^{2}\chi'_{O_A},
\end{align}
where $C_{mn}^{0}$ and $C_{mn}$ are the bare and renormalized elastic constants, and $s_{E,A,\pm}$ are the coupling strengths between multipolar moments and the elastic tensors/external magnetic field. $\chi_{Q_{E}}',\chi_{Q_{2}}',\chi_{O_{2}}',\chi_{O_{A}}'$ are the multipolar susceptibilities for quadrupolar moments in $E$ and $T_{2+}$ irreps., and octupolar moments in $T_{2-}$ and $A$ irreps., respectively.
We see that the multipolar susceptibilities $\chi_{Q_{E}}'$ and $\chi_{Q_{2}}'$ can both be measured without an external magnetic field. Once these are determined, the susceptibilities $\chi_{O_{A}}'$ and $\chi_{O_{-}}'$ can then be found. In the case of the dipolar susceptibilities, they couple to the magnetic field linearly, and can be measured by conventional magnetic susceptibility probes such as neutron scattering.

\paragraph*{Discussions.}
\label{sec:discussions}
In this work, we provided a detailed perturbative renormalization group analysis of the Bose-Fermi Kondo model describing a quartet of local states from \ce{Ce3Pd20(Si,Ge)6} \cite{Shiina1997} coupled to 3 bands of $p$-wave conduction electrons. The primary result we find is a two-stage Kondo destruction pair of QPT wherein the Fermi surface shrinks twice as local moments decouple from the Fermi surface and undergo magnetic ordering. The phase with smallest Fermi surface is particularly relevant to recent experiments, and exhibits the coexistence of quadrupolar $\{J_x^2-J_y^2, 3J_z^2-\bm{J}^2\}$ and dipolar $\{J_{x,y,z}\}$ order. This is similar to the low temperature and zero magnetic field phase of \ce{Ce3Pd20Si6}, which has coexistence of antiferromagnetic $\{J_z\}$ and antiferroquadrupolar $\{3J_z^2-\bm{J}^2\}$ orders \cite{Mazza2022, Custers2012, Ono2013}.

Connected to this magnetically ordered phase is a partially Kondo destroyed phase wherein the dipolar moments hybridize with and enlarge the Fermi surface, whereas the quadrupolar moments remain ordered and decoupled from the conduction electrons. This partially Kondo destroyed phase is potentially related to the quadrupolar ordered phase observed in the experiment at low temperatures for magnetic fields between 1T and 2T \cite{Martelli2019}. Our results show a further phase transition to a paramagnetic phase, where the quadrupolar moments also get hybridized with the Fermi surface and enlarge it a second time. Interestingly, all three of these phases are also observed experimentally at zero magnetic field as a function of temperature; indeed the paramagnetic $F$ phase we calculate could be the experimentally observed paramagnetic phase at zero magnetic field and temperatures above $T_{Q}$.
Experimentally, (at zero temperature) additional reconstruction of the Fermi surface is observed above 2T. However, this unidentified phase above 2T is not connected to the phases observed at zero magnetic field and its explanation may require explicit inclusion of the magnetic field, which is beyond the scope of the current work. Our results expand on the previous toy model study demonstrating the possibility of two consecutive Kondo destruction phase transitions \cite{Liu2021}. Notice, however, that the previous toy model study did not identify the types of multipolar order, and also did not suggest the coexistence of quadrupolar with dipolar order in the fully Kondo destroyed phase. We also discuss ultrasound measurements as an experimental probe of the multipolar susceptibilities at the different quantum critical points.

Here, we have solved a local version of the multipolar Kondo lattice model, but in future work it would be interesting to understand the full lattice problem with all the allowed multipolar moments, and determine whether quantum fluctuations beyond the local approximation are important or not. A further extension for our work is motivated by the fact that the phase transitions observed in experiments on \ce{Ce3Pd20Si6} are tuned by the magnetic field \cite{Martelli2019,Mazza2022}. Our model does not include the magnetic field explicitly, when in fact its effect on the Fermi surface, splitting of local moment states, and tetrahedral symmetry breaking nature may be important for connections with experiment. 
Another possible direction of theoretical inquiry is provided by the fact that, when the Kondo effect is destroyed, not every moment that was initially hybridizing with the conduction electron becomes ordered. The remaining moments may enter a (potentially multipolar) spin liquid phase \cite{Senthil2003,Senthil2004}, with interactions mediated by the RKKY coupling providing a mechanism for frustration.

Although the construction of our model in the tetrahedral $T_d$ environment was inspired by work on \ce{Ce3Pd20(Si,Ge)6}, the results apply equally as well to other materials with a $\Gamma_8$ quartet in a cubic environment. This quartet can also arise in the presence of an octahedral $O_h$ crystal field, as is the case for the ground state of \ce{Ce^{3+}} in \ce{CeB6} \cite{Shiina1997}. In fact, it is likely that such a rich phase diagram with possibility of both single and two-stage Kondo destruction is the case in any rare earth metallic system with a quartet of local moment ground states; this even applies to compounds with accidental fourfold degeneracy like \ce{YbRu2Ge2} \cite{Rosenberg2019c, Jeevan2006}. This work therefore demonstrates the striking details one can uncover about exotic Kondo physics, the symmetries of multipolar ordering, and Fermi surface reconstruction based purely on local symmetry, and opens a new route to study novel quantum criticality in multipolar heavy fermion systems.

\begin{acknowledgements}
We thank Silke Paschen and Qimiao Si for letting us know about their experimental and theoretical works on Ce-based heavy fermion systems and for useful discussion.
This work was supported by NSERC of Canada and the
Center for Quantum Materials at the University of Toronto. This work was initiated in a winter conference in January 2022 at the Aspen Center for Physics, which is supported by National Science Foundation grant PHY-1607611. Y.B.K. is also supported by the Simons Fellowship from the Simons Foundation and the Guggenheim Fellowship from the John Simon Guggenheim Memorial Foundation. D.S. is supported by the Ontario Graduate Scholarship.
\end{acknowledgements}

%

\newpage
\onecolumngrid
\clearpage
\begin{center}
\textbf{\large Supplemental Materials for ``Theory of a quantum impurity model \\for two-stage multipolar ordering and Fermi surface reconstruction"}
\end{center}
\begin{center}
\renewcommand{\thefootnote}{\fnsymbol{footnote}}
Daniel J. Schultz
\footnote{\label{coauthor}These authors contributed equally to this work.}, SangEun Han\textsuperscript{\ref{coauthor}}, and Yong Baek Kim
\vspace{0.2cm}

{\it Department of Physics, University of Toronto, Toronto, Ontario M5S 1A7, Canada}
\end{center}

\setcounter{equation}{0}
\setcounter{figure}{0}
\setcounter{table}{0}
\setcounter{page}{1}
\setcounter{section}{0}
\setcounter{subsection}{0}

\makeatletter
\renewcommand*{\thesection}{S\arabic{section}}
\renewcommand*{\thesubsection}{\thesection.\Alph{subsection}}
\renewcommand*{\p@subsection}{}
\renewcommand*{\thesubsubsection}{\thesubsection.\arabic{subsubsection}}
\renewcommand*{\p@subsubsection}{}
\renewcommand{\theequation}{S\arabic{equation}}
\renewcommand{\thefigure}{S\arabic{figure}}
\renewcommand{\thetable}{S\arabic{table}}
\renewcommand{\@seccntformat}[1]{%
  \ifcsname prefix@#1\endcsname
    \csname prefix@#1\endcsname
  \else
    \csname the#1\endcsname\quad
  \fi}
\makeatother

\section{\texorpdfstring{$\Gamma_8$}{Gamma8} quartet wave functions}\label{app:quartet}
In a vacuum, a \ce{Ce^{3+}} ion forms an effective spin-$J=5/2$ system by Hund's rules. In the presence of a tetrahedral ($T_d$) crystal field, these 6 degenerate states are split, and in \ce{Ce3Pd20(Si,Ge)6}, the resulting ground state of the \ce{Ce^{3+}} at the 8c site is a quartet spanned by the following four states \cite{Shiina1997_SM}:

\begin{align}
\ket{\Gamma^{(1)}_8} ={}& \sqrt{\frac{5}{6}} \ket{5/2} + \sqrt{\frac{1}{6}} \ket{-3/2}, \\
\ket{\Gamma^{(2)}_8} ={}& \sqrt{\frac{1}{6}} \ket{3/2} + \sqrt{\frac{5}{6}} \ket{-5/2}, \\
\ket{\Gamma^{(3)}_8} ={}& \ket{1/2}, \\
\ket{\Gamma^{(4)}_8} ={}& \ket{-1/2}.
\end{align}

To determine which multipolar moments are supported by these wave functions, we can compute the matrix elements of Stevens operators in the quartet $\{\ket{\Gamma^{(1)}_8}, \ket{\Gamma^{(2)}_8}, \ket{\Gamma^{(3)}_8}, \ket{\Gamma^{(4)}_8}\}$. More specifically, we compute 
\begin{equation}
\Lambda^i_{\rho\tau} = \bra{\Gamma^{(\rho)}_8} \mathcal{O}^i \ket{\Gamma^{(\tau)}_8},
\end{equation}
where $\mathcal{O}^i$ is any (rescaled or linear combination of) Stevens operator (from ``As Stevens" column in Table \ref{tab:spin_operators}), and $\Lambda^i$ is some $\text{SU}(4)$ generator. Note that, in the Abrikosov pseudofermion representation, the spin operators are then constructed as $S^i = \sum_{\rho,\tau=1}^4 f^\dagger_\rho \Lambda^i_{\rho\tau} f_\tau$. 

\section{Multipolar moments and Pseudofermion representation}
\subsection{Redefining Multipolar moments}
The multipolar moments presented in Table 1 in the main text, after being projected into the quartet subspace, are not orthogonal under the trace, by which we mean they do not satisfy $\text{Tr}[M^{i}M^{j}]\propto\delta^{ij}$ (where $M^i_{\rho\tau} = \bra{\Gamma^{(\rho)}_8} \mathcal{O}^i \ket{\Gamma^{(\tau)}_8}$ where in this case $\mathcal{O}^i$ is one of the original Stevens operators in Table 1, not the rescaled ones in Table~\ref{tab:spin_operators}). Thus, in the actual computation, we used the linear combinations of the rescaled multipolar moments in Table~\ref{tab:spin_operators}.
Note that $S^{1,2,3}$ and $S^{10,11,12}$ in Table~\ref{tab:spin_operators} are mixtures between dipolar and octupolar moments, but we will treat them as dipolar moments in the main text and Supplementary Materials because they are not practically distinguishable since they share precisely the same symmetries.

\begin{table}[ht!]
\centering
\begin{tabular}{|c|c|c|c|c|c|}
\hline
Irrep. & $S^i$ & As Stevens & Moment & $K_i$ & $g_i$ \\ \hline \hline
$T_{1a}$ & $S^1$ & $-\frac{1}{15}J_x + \frac{7}{90}\mathcal{T}^\alpha_x$ & D & & \\ [5pt]
$T_{1a}$ & $S^2$ & $-\frac{1}{15}J_y + \frac{7}{90}\mathcal{T}^\alpha_y$ & D & $K_{a(1,2,3,4)}$ & $g_{a}$ \\ [5pt]
$T_{1a}$ & $S^3$ & $-\frac{1}{15}J_z + \frac{7}{90}\mathcal{T}^\alpha_z$ & D & & \\ [5pt] \hline
& & & & & \\ [-8 pt]
$E$ & $S^4$ & $\frac{1}{8}\mathcal{O}_{22}$ & Q & $K_{E(a,b)}$ & $g_{E}$ \\ [5pt]
$E$ & $S^5$ & $\frac{1}{8}\mathcal{O}_{20}$ & Q & & \\ [5pt] \hline
& & & & & \\ [-8 pt]
$T_{2+}$ & $S^6$ & $\frac{1}{2}\mathcal{O}_{yz}$ & Q & & \\ [5pt]
$T_{2+}$ & $S^7$ & $\frac{1}{2}\mathcal{O}_{zx}$ & Q & $K_{2(\alpha,\beta)+}$ & $g_{2+}$ \\ [5pt]
$T_{2+}$ & $S^8$ & $\frac{1}{2}\mathcal{O}_{xy}$ & Q & & \\ [5pt] \hline
& & & & & \\ [-8 pt]
$A_{2}$ & $S^9$ & $\frac{1}{9\sqrt{5}}\mathcal{T}_{xyz}$ & O & $K_{A}$ & $g_{A}$ \\ [5pt] \hline
& & & & & \\ [-8 pt]
$T_{1b} $ & $S^{10}$ & $-\frac{7}{15}J_x + \frac{2}{45}\mathcal{T}^\alpha_x$ & D & & \\ [5pt]
$T_{1b} $ & $S^{11}$ & $-\frac{7}{15}J_y + \frac{2}{45}\mathcal{T}^\alpha_y$ & D & $K_{b(1,2,3,4)}$ & $g_{b}$ \\ [5pt]
$T_{1b} $ & $S^{12}$ & $-\frac{7}{15}J_z + \frac{2}{45}\mathcal{T}^\alpha_z$ & D & & \\ [5pt] \hline
& & & & & \\ [-8 pt]
$T_{2-}$ & $S^{13}$ & $\frac{1}{6\sqrt{5}}\mathcal{T}^\beta_x$ & O & & \\ [5pt]
$T_{2-}$ & $S^{14}$ & $\frac{1}{6\sqrt{5}}\mathcal{T}^\beta_y$ & O & $K_{2(\alpha,\beta)-}$ & $g_{2-}$ \\ [5pt]
$T_{2-}$ & $S^{15}$ & $\frac{1}{6\sqrt{5}}\mathcal{T}^\beta_z$ & O & & \\ \hline
\end{tabular}
\caption{The multipolar moments used in the Hamiltonians, expressed in terms of $J=5/2$ Stevens operators. The operators $S^i$, $i=1,\dots,15$ are the special linear combinations of Stevens operators chosen such that $\text{tr}(\Lambda^{i}\Lambda^{j})=\delta_{ij}$ where $\Lambda^{i}$ is the traceless $4\times4$ matrix defined by $S^{i}=\sum_{\rho,\tau=1}^{4}(f_{\rho}^{\dagger}\Lambda^{i}_{\rho\tau}f_{\tau})$. We redefine and classify the six $T_1$ moments as $T_{1a}$ and $T_{1b}$ \cite{Shiina1997_SM}. In the moment column, we indicate if the moment is a dipolar (D), quadrupolar (Q), or pure octupolar (O). Note that the dipolar moments actually correspond to a mixed dipolar and octupolar moments because they are linear combinations between dipolar and octupolar moments. However, we can regard them as dipolar moments because the $T_{1a}$ and $T_{1b}$ octupoles cannot practically be distinguished from the dipoles.
In the $K_{i}$ and $g_{i}$ columns, we list the Fermi-Kondo $K_{i}$ and Bose-Kondo $g_{i}$ (bosonic bath) couplings which couple to moments in the corresponding irrep. In total, we have 15 Fermi-Kondo couplings and 6 Bose-Kondo couplings.
\label{tab:spin_operators}}
\end{table}

\subsection{Pseudofermion representation for multipolar moments}
In order to later perform the RG analysis, we write the multipolar spin operators in terms of pseudofermions. This is done by writing the operators as $S^i=\sum_{\rho,\tau=1}^4(f_{\rho}^{\dagger}\Lambda^i_{\rho\tau}f_{\tau})$, with $\Lambda^i$ being a $4\times4$ traceless Hermitian matrix which corresponds to the specific form of the multipolar operator. Note that the rescalings and linear combinations ensure that $\text{tr}(\Lambda^i\Lambda^j) = \delta_{ij}$. In order to restrict the Hilbert space of the impurity states to the original 4 states, we impose the single occupation condition $\sum_{\rho=1}^4f_{\rho}^{\dagger}f_{\rho}=1$. This means that we introduce a term $\lambda \sum_{\rho=1}^4f_{\rho}^{\dagger}f_{\rho}$ to the Hamiltonian, and take $\lambda\to\infty$ at the end of the calculation.

\section{Hamiltonian Derivation}

\subsection{Action of the Tetrahedral group} \label{app:symmetries}
To derive the symmetry-allowed Hamiltonian, we need to test all possible candidate terms and determine which remain invariant under action of the tetrahedral group $T_d$, and under time-reversal $\mathcal{T}$. The most efficient way to verify that a term is invariant under all elements of $T_d$ is to pick two of its generators,  which we select to be $\mathcal{C}_{31}$ and $\mathcal{S}_{4z}$. $\mathcal{C}_{31}$ is a rotation by $2\pi/3$ about the $(1,1,1)$ axis, and $\mathcal{S}_{4z}$ is a rotation by $\pi/2$ about the $z$-axis followed by a mirror reflection across the $xy$ plane. Both of these transformations map a tetrahedron to itself. Adding in time-reversal symmetry yields Table~\ref{tab:symmetries}. Checking all possible Hermitian Kondo terms respecting these symmetries yields Eqs.~\eqref{eq:HKT1a}-\eqref{eq:HKT2mb}.

\begin{table}[t]
\centering
\begin{tabular}{|c|c|c|c|}
\hline
Object & $\mathcal{S}_{4z}$ & $\mathcal{C}_{31}$ & $\mathcal{T}$ \\ \hline \hline
$x$ & $-y$ & $y$ & $x$ \\
$y$ & $x$ & $z$ & $y$ \\
$z$ & $-z$ & $x$ & $z$ \\ \hline
$\sigma^0$ & $\sigma^0$ & $\sigma^0$ & $\sigma^0$ \\
$\sigma^x$ & $\sigma^y$ & $\sigma^y$ & $-\sigma^x$ \\
$\sigma^y$ & $-\sigma^x$ & $\sigma^z$ & $-\sigma^y$ \\
$\sigma^z$ & $\sigma^z$ & $\sigma^x$ & $-\sigma^z$ \\ \hline
$J_x$ & $J_y$ & $J_y$ & $-J_x$ \\
$J_y$ & $-J_x$ & $J_z$ & $-J_y$ \\
$J_z$ & $J_z$ & $J_x$ & $-J_z$ \\ \hline
$S^1$ & $S^2$ & $S^2$ & $-S^1$ \\ 
$S^2$ & $-S^1$ & $S^3$ & $-S^2$ \\ 
$S^3$ & $S^3$ & $S^1$ & $-S^3$ \\ 
$S^4$ & $-S^4$ & $-\frac{1}{2}S^4 - \frac{\sqrt{3}}{2}S^5$ & $S^4$ \\ 
$S^5$ & $S^5$ & $\frac{\sqrt{3}}{2}S^4 - \frac{1}{2}S^5$ & $S^5$ \\ 
$S^6$ & $-S^7$ & $S^7$ & $S^6$ \\ 
$S^7$ & $S^6$ & $S^8$ & $S^7$ \\ 
$S^8$ & $-S^8$ & $S^6$ & $S^8$ \\ 
$S^9$ & $-S^9$ & $S^9$ & $-S^9$ \\ 
$S^{10}$ & $S^{11}$ & $S^{11}$ & $-S^{10}$ \\ 
$S^{11}$ & $-S^{10}$ & $S^{12}$ & $-S^{11}$ \\ 
$S^{12}$ & $S^{12}$ & $S^{10}$ & $-S^{12}$ \\ 
$S^{13}$ & $-S^{14}$ & $S^{14}$ & $-S^{13}$ \\ 
$S^{14}$ & $S^{13}$ & $S^{15}$ & $-S^{14}$ \\ 
$S^{15}$ & $-S^{15}$ & $S^{13}$ & $-S^{15}$ \\\hline
\end{tabular}
\caption{Symmetry transformations of orbitals, Pauli matrices, and multipolar moments, under two generators of the tetrahedral group as well as time-reversal.}\label{tab:symmetries}
\end{table}

\subsection{SU(3) Gell-Mann matrices}\label{app:gell_mann}
When constructing the multipolar Kondo models for $p$-wave electrons coupled to the local moment, we have to consider all possible fermion bilinears of these 3 orbitals. For this, we use the generators of $\text{SU}(3)$, normalized by $\text{tr}(\lambda^i\lambda^j) = 2\delta_{ij}$. The $3\times3$ Gell-Mann matrices are enumerated in Eqs.~\eqref{eq:gell_mann0}-\eqref{eq:gell_mann8}

\begin{align}
\lambda^0 ={}& \sqrt{\frac{2}{3}}\begin{pmatrix} 1 & 0 & 0 \\ 0 & 1 & 0 \\ 0 & 0 & 1 \end{pmatrix}, & \lambda^1 ={}& \begin{pmatrix} 0 & 1 & 0 \\ 1 & 0 & 0 \\ 0 & 0 & 0 \end{pmatrix}, \label{eq:gell_mann0} \\
\lambda^2 ={}& \begin{pmatrix} 0 & -i & 0 \\ i & 0 & 0 \\ 0 & 0 & 0 \end{pmatrix}, &
\lambda^3 ={}& \begin{pmatrix} 1 & 0 & 0 \\ 0 & -1 & 0 \\ 0 & 0 & 0 \end{pmatrix}, \\
\lambda^4 ={}& \begin{pmatrix} 0 & 0 & 1 \\ 0 & 0 & 0 \\ 1 & 0 & 0 \end{pmatrix}, &
\lambda^5 ={}& \begin{pmatrix} 0 & 0 & -i \\ 0 & 0 & 0 \\ i & 0 & 0 \end{pmatrix}, \\
\lambda^6 ={}& \begin{pmatrix} 0 & 0 & 0 \\ 0 & 0 & 1 \\ 0 & 1 & 0 \end{pmatrix}, &
\lambda^7 ={}& \begin{pmatrix} 0 & 0 & 0 \\ 0 & 0 & -i \\ 0 & i & 0 \end{pmatrix}, \\
\lambda^8 ={}& \frac{1}{\sqrt{3}}\begin{pmatrix} 1 & 0 & 0 \\ 0 & 1 & 0 \\ 0 & 0 & -2 \end{pmatrix} \label{eq:gell_mann8}
\end{align} 

\subsection{Fermi-Kondo Coupling \label{app:fermi_hamiltonians}}
The Kondo Hamiltonians coupling the conduction electrons to local moments described in the main text are given by Eqs.~\eqref{eq:HKT1a}-\eqref{eq:HKT2mb}. In these Hamiltonians, $c^\dagger$ and $c$ are conduction electron creation and annihilation operators, the subscript 0 on the conduction electron operators denotes the interaction taking place at the impurity site, which we choose to be the origin. The Latin indices sum over orbitals, $m,n = x,y,z$, and the Greek indices sum over spins, $\rho,\tau=\uparrow,\downarrow$. The $\sigma$ are the standard Pauli matrices satisfying $[\sigma^i,\sigma^j] = 2i\epsilon_{ijk}\sigma^k$, and the $\lambda^i$ are the $3\times 3$ Gell-Mann matrices, listed in Supplementary Materials \ref{app:gell_mann}. 

\begin{align}
H_{a1} ={}& K_{a1}\sum_{mn\rho\tau} c^\dagger_{0m\rho} c_{0n\tau}\lambda^0_{mn}\left(\sigma^x_{\rho\tau}S^1 + \sigma^y_{\rho\tau}S^2 + \sigma^z_{\rho\tau}S^3\right) \label{eq:HKT1a} ,\\
H_{b1} ={}& K_{b1}\sum_{mn\rho\tau}c^\dagger_{0m\rho} c_{0n\tau}\lambda^0_{mn}\left(\sigma^x_{\rho\tau}S^{10} + \sigma^y_{\rho\tau}S^{11} + \sigma^z_{\rho\tau}S^{12}\right) \label{eq:HKT1b} ,\\
H_{a2} ={}& K_{a2}\sum_{mn\rho\tau}c^\dagger_{0m\rho} c_{0n\tau}\sigma^0_{\rho\tau}\left(\lambda^7_{mn}S^{1} - \lambda^5_{mn}S^{2} + \lambda^2_{mn}S^{3}\right) \label{eq:HKT1c} ,\\
H_{b2} ={}& K_{b2}\sum_{mn\rho\tau}c^\dagger_{0m\rho} c_{0n\tau}\sigma^0_{\rho\tau}\left(\lambda^7_{mn}S^{10} - \lambda^5_{mn}S^{11} + \lambda^2_{mn}S^{12}\right) \label{eq:HKT1d} ,\\
H_{a3} ={}& K_{a3}\sum_{mn\rho\tau}c^\dagger_{0m\rho} c_{0n\tau}\left[(\sigma^y_{\rho\tau}\lambda^1_{mn} + \sigma^z_{\rho\tau}\lambda^4_{mn})S^1 + (\sigma^z_{\rho\tau}\lambda^6_{mn} + \sigma^x_{\rho\tau}\lambda^1_{mn})S^2 + (\sigma^x_{\rho\tau}\lambda^4_{mn} + \sigma^y_{\rho\tau}\lambda^6_{mn})S^3\right] \label{eq:HKT1e} ,\\
H_{b3} ={}& K_{b3}\sum_{mn\rho\tau}c^\dagger_{0m\rho} c_{0n\tau}\left[(\sigma^y_{\rho\tau}\lambda^1_{mn} + \sigma^z_{\rho\tau}\lambda^4_{mn})S^{10} + (\sigma^z_{\rho\tau}\lambda^6_{mn} + \sigma^x_{\rho\tau}\lambda^1_{mn})S^{11} + (\sigma^x_{\rho\tau}\lambda^4_{mn} + \sigma^y_{\rho\tau}\lambda^6_{mn})S^{12}\right] \label{eq:HKT1f} ,\\
H_{a4} ={}& K_{a4}\sum_{mn\rho\tau}c^\dagger_{0m\rho} c_{0n\tau}\left[\sigma^x_{\rho\tau}\left(\frac{\sqrt{3}}{2}\lambda_{mn}^3+\frac{1}{2}\lambda_{mn}^8\right)S^1 + \sigma^y_{\rho\tau}\left(-\frac{\sqrt{3}}{2}\lambda^3_{mn} + \frac{1}{2}\lambda^8_{mn}\right)S^2 - \sigma^z_{\rho\tau}\lambda^8_{mn}S^3\right] \label{eq:HKT1g} ,\\
H_{b4} ={}& K_{b4}\sum_{mn\rho\tau}c^\dagger_{0m\rho} c_{0n\tau}\left[\sigma^x_{\rho\tau}\left(\frac{\sqrt{3}}{2}\lambda_{mn}^3+\frac{1}{2}\lambda_{mn}^8\right)S^{10} + \sigma^y_{\rho\tau}\left(-\frac{\sqrt{3}}{2}\lambda^3_{mn} + \frac{1}{2}\lambda^8_{mn}\right)S^{11} - \sigma^z_{\rho\tau}\lambda^8_{mn}S^{12}\right] \label{eq:HKT1h} ,\\
H_{E1} ={}& K_{E1}\sum_{mn\rho\tau}c^\dagger_{0m\rho} c_{0n\tau}\sigma^0_{\rho\tau}\left(\lambda^3_{mn}S^4 - \lambda^8_{mn}S^5\right) \label{eq:HKEa} ,\\
H_{E2} ={}& K_{E2}\sum_{mn\rho\tau}c^\dagger_{0m\rho} c_{0n\tau}\left[\sigma^x_{\rho\tau}\lambda^7_{mn}\left(\frac{\sqrt{3}}{2}S^4 - \frac{1}{2}S^5\right) + \sigma^y_{\rho\tau}\lambda^5_{mn}\left(\frac{\sqrt{3}}{2}S^4 + \frac{1}{2}S^5\right) + \sigma^z_{\rho\tau}\lambda^2_{mn}S^5\right] \label{eq:HKEb} ,\\
H_{2\alpha\scriptscriptstyle{+}} ={}& K_{2\alpha\scriptscriptstyle{+}}\sum_{mn\rho\tau} c^\dagger_{0m\rho} c_{0n\tau}\sigma^0_{\rho\tau}\left(\lambda^6_{mn}S^6 + \lambda^4_{mn}S^7 + \lambda^1_{mn}S^8\right) \label{eq:HKT2pa} ,\\
H_{2\beta\scriptscriptstyle{+}} ={}& K_{2\beta\scriptscriptstyle{+}}\sum_{mn\rho\tau} c^\dagger_{0m\rho} c_{0n\tau}\left[(\sigma^y_{\rho\tau}\lambda^2_{mn} - \sigma^z_{\rho\tau}\lambda^5_{mn})S^6 + (\sigma^z_{\rho\tau}\lambda^7_{mn} + \sigma^x_{\rho\tau}\lambda^2_{mn})S^7 - (\sigma^x_{\rho\tau}\lambda^5_{mn} - \sigma^y_{\rho\tau}\lambda^7_{mn})S^8\right] \label{eq:HKT2pb} ,\\
H_{A} ={}& K_{A}\sum_{mn\rho\tau}c^\dagger_{0m\rho} c_{0n\tau}\left(\sigma^x_{\rho\tau}\lambda^6_{mn} + \sigma^y_{\rho\tau}\lambda^4 + \sigma^z_{\rho\tau}\lambda^1_{mn}\right)S^9 \label{eq:HKA2} ,\\
H_{2\alpha\scriptscriptstyle{-}} ={}& K_{2\alpha\scriptscriptstyle{-}}\sum_{mn\rho\tau}c^\dagger_{0m\rho} c_{0n\tau}\left[\sigma^x_{\rho\tau}\left(\frac{1}{2}\lambda^3_{mn} - \frac{\sqrt{3}}{2}\lambda^8\right)S^{13} +  \sigma^y_{\rho\tau}\left(\frac{1}{2}\lambda^3_{mn} +\frac{\sqrt{3}}{2}\lambda^8_{mn}\right)S^{14} - \sigma^z_{\rho\tau}\lambda^3_{mn}S^{15}\right] \label{eq:HKT2ma} ,\\
H_{2\beta\scriptscriptstyle{-}} ={}& K_{2\beta\scriptscriptstyle{-}}\sum_{mn\rho\tau}c^\dagger_{0m\rho} c_{0n\tau}\left[(\sigma^y_{\rho\tau}\lambda^1_{mn} - \sigma^z_{\rho\tau}\lambda^4_{mn})S^{13} + (\sigma^z_{\rho\tau}\lambda^6_{mn} - \sigma^x_{\rho\tau}\lambda^1_{mn})S^{14} + (\sigma^x_{\rho\tau}\lambda^4_{mn} - \sigma^y_{\rho\tau}\lambda^6_{mn})S^{15}\right] .\label{eq:HKT2mb}
\end{align}

\subsection{Bose-Kondo Coupling \label{app:bose_hamiltonians}}
\begin{figure}[t]
    \centering
    \includegraphics{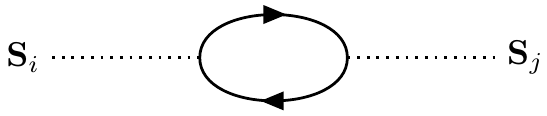}
    \caption{Effective Kondo Lattice RKKY Interaction; dotted lines refer to the spin operators and the solid lines are fermion propagators (not to be confused with dashed lines in other diagrams referring to pseudofermion propagators). The $\textbf{S}_i$ and $\textbf{S}_j$ are the (15 component) spin operators on different sites $i$ and $j$ in the parent Kondo lattice.}
    \label{fig:rkky}
\end{figure}

The dynamical bosonic bath field representing local magnetic fluctuations is derived to respect the local symmetry. To do this, we construct the effective interaction between spins in the corresponding parent Kondo lattice. Starting with the Fermi-Kondo Hamiltonians in Eqs.~\eqref{eq:HKT1a}-\eqref{eq:HKT2mb}, we can compute the effective interaction between two spins on different sites $i$ and $j$ by computing the diagram in Fig.~\ref{fig:rkky}.

By replacing one of the multipolar moment operators in the RKKY interaction by the dynamical bosonic field, we find the symmetry-allowed coupling between the local moment and bosonic bath. The allowed Bose-Kondo Hamiltonians are then Eqs.~\eqref{eq:HgT1a}-\eqref{eq:HgT2m}.

\begin{align}
H^B_{a} ={}& g_{a}(\phi^1_0,\phi^2_0,\phi^3_0)\cdot(S^1,S^2,S^3), \label{eq:HgT1a}\\
H^B_{E} ={}& g_{E}(\phi^4_0,\phi^5_0)\cdot(S^4,S^5), \\
H^B_{2+} ={}& g_{2\scriptscriptstyle{+}}(\phi^6_0,\phi^7_0,\phi^8_0)\cdot(S^6,S^7,S^8), \\
H^B_{A} ={}& g_{A}\phi^9_0S^9, \\
H^B_{b} ={}& g_{b}(\phi^{10}_0,\phi^{11}_0,\phi^{12}_0)\cdot(S^{10},S^{11},S^{12}), \\
H^B_{2-} ={}& g_{2\scriptscriptstyle{-}}(\phi^{13}_0,\phi^{14}_0,\phi^{15}_0)\cdot(S^{13},S^{14},S^{15}).\label{eq:HgT2m}
\end{align}

Here, the subscript 0 denotes the Bose bath at the impurity site, whose components can be expanded  as $\phi^i_0 = \sum_\mathbf{k}(\phi^i_\mathbf{k} + \phi^{i\dagger}_{-\mathbf{k}})$. We note that this is not the most general symmetry allowed Bose-Kondo Hamiltonian. It is in fact possible to have terms such as $(\phi^1_0,\phi^2_0,\phi^3_0)\cdot(S^{10},S^{11},S^{12})$. We ignore these mixing terms in our analysis \cite{Custers2012_SM, Shiina1997_SM}. 

\begin{figure}[b]
    \centering
    \includegraphics[width=0.55\linewidth]{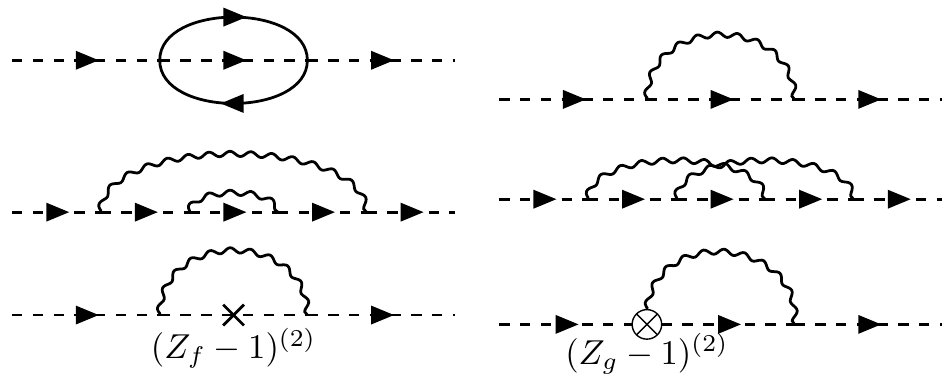}
    \caption{Pseudofermion self-energy, both direct and counterterm contribution.}
    \label{fig:pf_self_energy}
\end{figure}

\section{Details of the Renormalization Group Method}
\subsection{Dimensional regularization and \texorpdfstring{$\epsilon$}{epsilon} expansion}
The renormalization group analysis is carried out by using dimensional regularization with minimal subtraction \cite{Zhu2002_SM} in this work. In this scheme, the density of states of the conduction electrons and bosonic bath are parameterized by small factors $\epsilon'$ and $\epsilon$, respectively. This factor is already in place for the bosonic bath (see Eq.(1) of main text), and we now introduce the $\epsilon'$ factor for the fermionic density of states:
\begin{align}
    \sum_{\mathbf{k}}\delta(\omega-\xi_{\mathbf{k}})=N_{0}|\omega|^{-\epsilon'}.
\end{align}
These $\epsilon,\epsilon'$ factors are used in the minimal subtraction of poles, and $\epsilon'\to 0$ at the end of the calculation. This necessitates defining a renormalized field $f$ and dimensionless coupling constants $g_{i}$ and $K_{j}$,
\begin{align}
    f^{B}={}&Z_{f}^{1/2}f,\\
    g^{B}_{i}={}&g_{i} Z_{f}^{-1}Z_{g_{i}}\mu^{\epsilon/2}, \\
    K^{B}_{j}={}&K_{j} Z_{f}^{-1}Z_{K_{j}}\mu^{\epsilon'}, 
\end{align}
\noindent where $\mu$ is the renormalization energy scale, and $Z_{f}$, $Z_{g_i}$, and $Z_{K_j}$ are the renormalization constants for the pseudofermion $f$, bosonic couplings $g_{i}$ (here $i=a,b,2\pm,E,A$), and fermionic couplings $K_j$ (here $j=a(1,2,3,4),b(1,2,3,4),2(\alpha,\beta)\pm,E(1,2),A$). The superscript $B$ stands for the bare value which does not evolve under the RG flow. In addition, we absorb the density of states $N_{i}$ into the dimensionless couplings as $N_{0}K_{j}\rightarrow K_{j}$ and $N_1^{2}g_i\rightarrow g_i$, respectively. 
The details of the RG analysis and corresponding Feynman diagrams are enumerated in Supplementary Materials~\ref{app:rg}.

\subsection{Propagators and Counterterms}\label{app:rg}
\begin{figure}[t]
    \centering
    \includegraphics[width=0.45\linewidth]{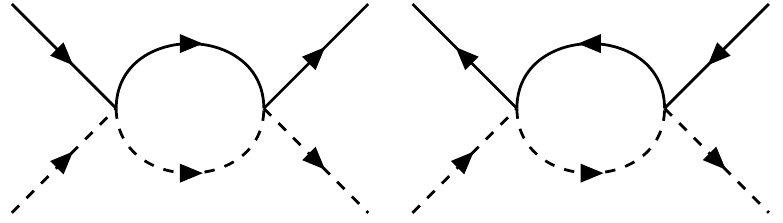}
    \caption{Order $K$ corrections to the Fermi-Kondo vertex. There are only direct contributions at this order.}
    \label{fig:J_1st_corr}
\end{figure}

\begin{figure}[t]
    \centering
    \includegraphics[width=0.45\linewidth]{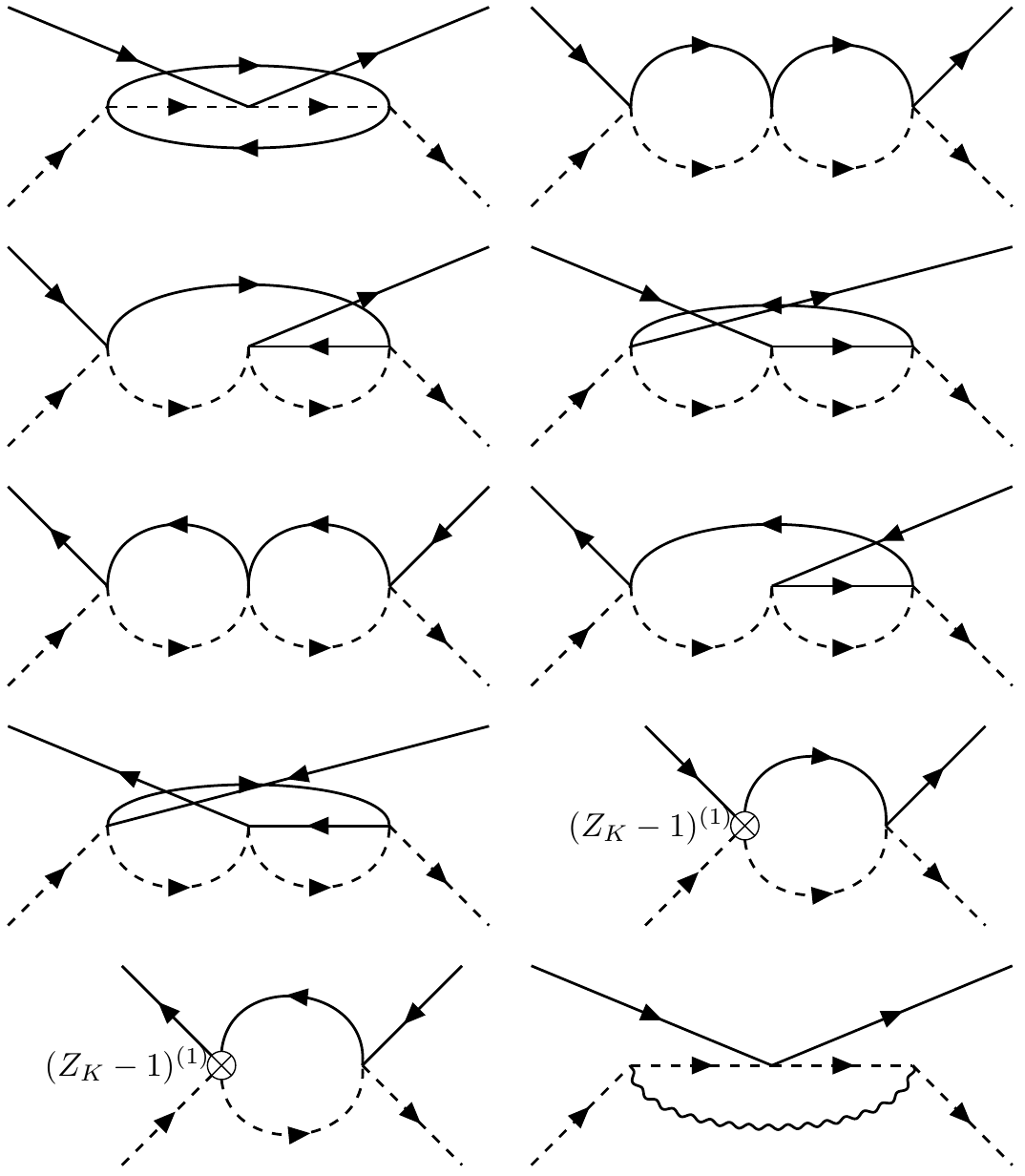}
    \caption{Order $K^2$ and $g^2$ corrections to the Fermi-Kondo vertex.}
    \label{fig:J_2nd_corr}
\end{figure}

\begin{figure}[t]
    \centering
    \includegraphics[width=0.45\linewidth]{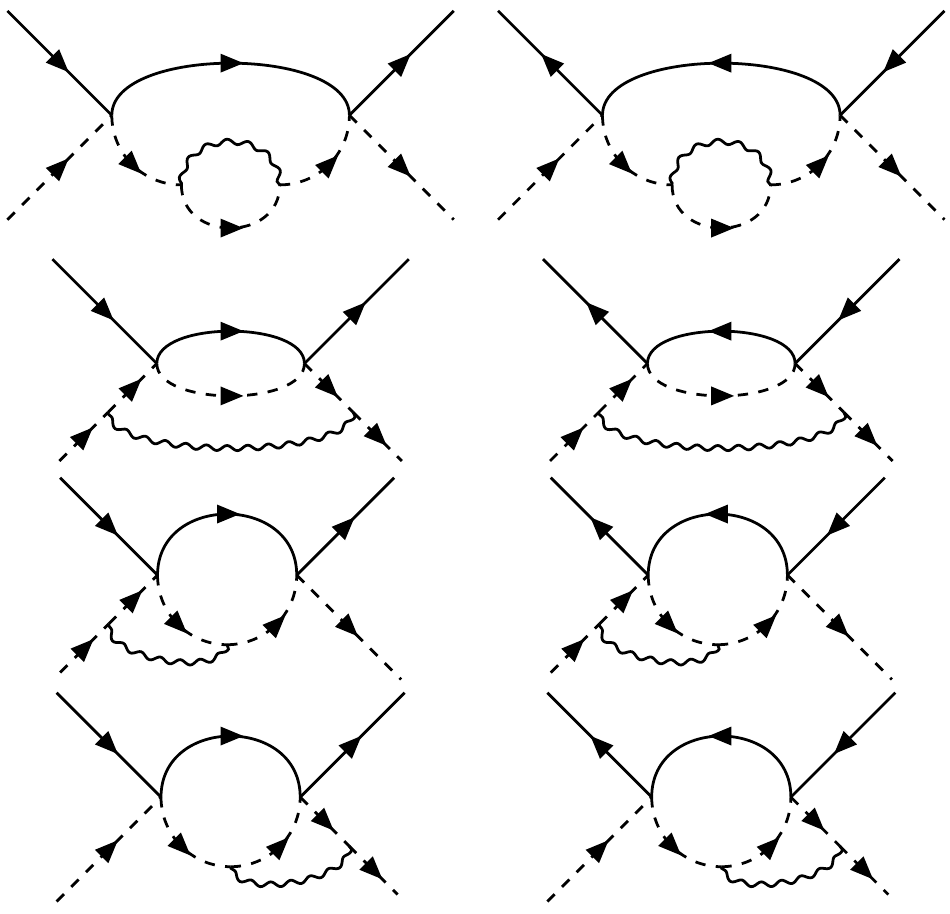}
    \caption{Order $Kg^2$ direct corrections to the Fermi-Kondo vertex.}
    \label{fig:J_3rd_corr_direct}
\end{figure}

\begin{figure}[t]
    \centering
    \includegraphics[width=0.45\linewidth]{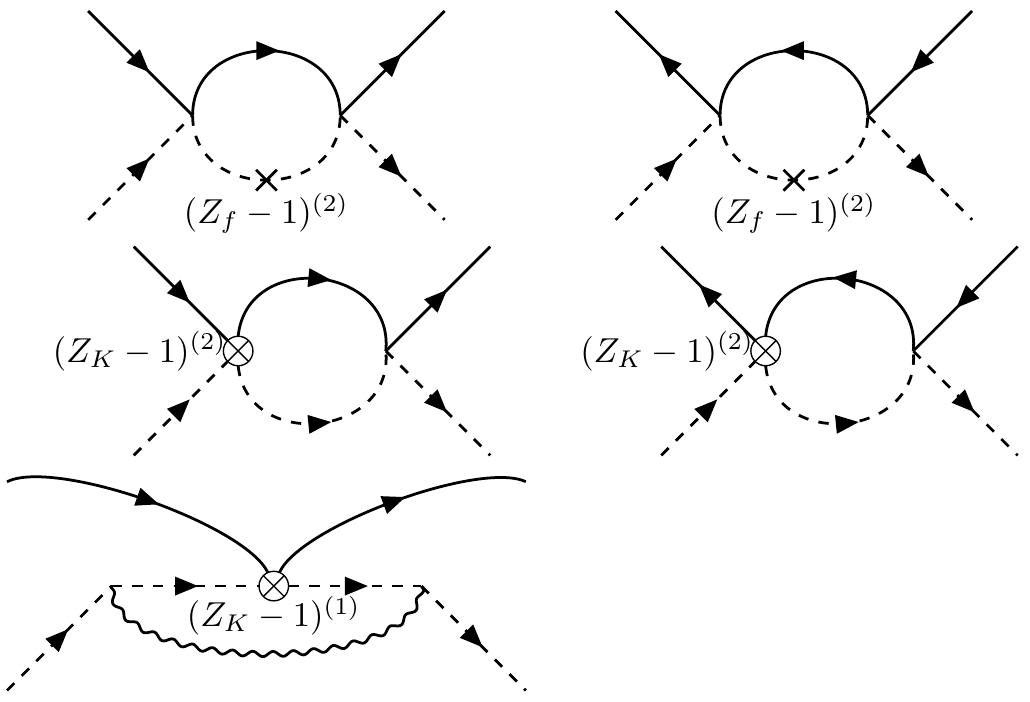}
    \caption{Order $Kg^2$ counterterm corrections to the Fermi-Kondo vertex.}
    \label{fig:J_3rd_corr_counter}
\end{figure}

\begin{figure}[t]
    \centering
    \includegraphics[width=0.45\linewidth]{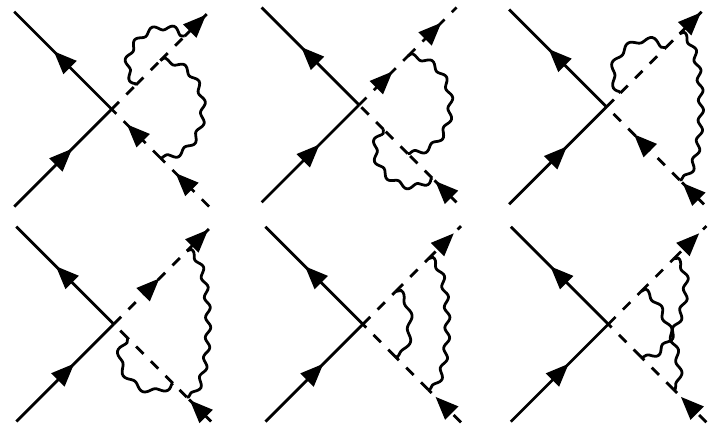}
    \caption{Order $g^4$ direct corrections to the Fermi-Kondo vertex.}
    \label{fig:J_4th_corr_direct}
\end{figure}

\begin{figure}[t]
    \centering
    \includegraphics[scale=0.8]{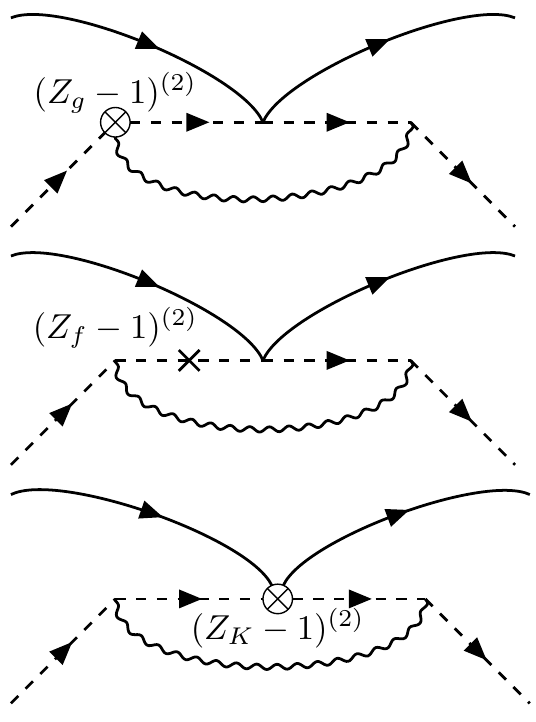}
    \caption{Order $g^4$ counterterm corrections to the Fermi-Kondo vertex.}
    \label{fig:J_4th_corr_counter}
\end{figure}

\begin{figure}[t]
    \centering
    \includegraphics[width=0.45\linewidth]{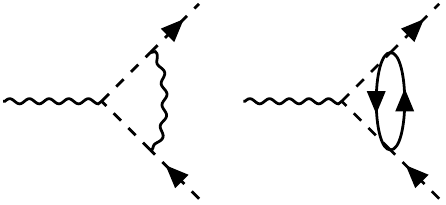}
    \caption{Order $g^2$ and $J^2$ corrections to the Bose-Kondo vertex.}
    \label{fig:g_2nd_corr}
\end{figure}

\begin{figure}[t]
    \centering
    \includegraphics[width=0.45\linewidth]{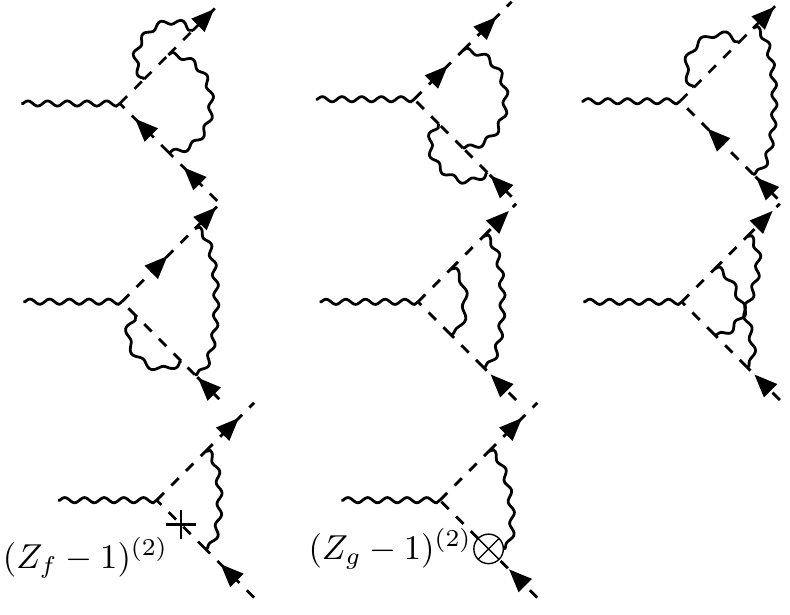}
    \caption{Order $g^4$ direct and counterterm corrections to the Bose-Kondo vertex.}
    \label{fig:g_4th_corr}
\end{figure}

From the bare kinetic Hamiltonian, Fermi-Kondo couplings, and bosonic bath couplings presented in the main text and appendix, we compute the vertex functions of the Fermi-Kondo and Bose-Kondo interactions and self energy of the pseudofermion to 2-loop order. Along the way, we introduce counterterms order by order to remove the divergences from the loop integrals. The counterterms at order $n$ for the pseudofermion, Fermi-Kondo couplings, and Bose-Kondo couplings are denoted $(Z_{f}-1)^{(n)}$, $(Z_{K_{i}}-1)^{(n)}$, and $(Z_{g_{i}}-1)^{(n)}$, respectively. The pseudofermion counterterm is denoted $\times$, and the Fermi-Kondo and Bose-Kondo counterterms are denoted $\otimes$; all are labelled in the appropriate Feynman diagrams. The corresponding Feynman diagrams for the pseudofermion self-energy are given in Fig.~\ref{fig:pf_self_energy}, the diagrams \cite{Ellis2017_SM} for the Fermi-Kondo vertex corrections are given in Figs.~\ref{fig:J_1st_corr}-\ref{fig:J_4th_corr_counter}, and the diagrams for the Bose-Kondo vertex corrections are given in Figs.~\ref{fig:g_2nd_corr}-\ref{fig:g_4th_corr}. The details of the calculation of renormalization constants from the Feynman diagrams are explained for a simpler case in a wonderful reference \cite{Zhu2002_SM}. In the Feynman diagrams of Figs.~\ref{fig:pf_self_energy}-\ref{fig:g_4th_corr}, solid, dashed, and squiggly lines refer to propagators of the conduction electrons (Eq.~\eqref{eq:fermion_gf}), pseudofermions (Eq.~\eqref{eq:pseudofermion_gf}), and bosonic bath (Eq.~\eqref{eq:boson_gf}), respectively, and their expressions are as follows:
\begin{align}
\mathcal{G}^c_0(i\omega,\mathbf{k}) ={}& \frac{1}{i\omega-\xi_{\mathbf{k}}}, \label{eq:fermion_gf} \\
\mathcal{G}^f_0(i\omega,\mathbf{k}) ={}& \frac{1}{i\omega-\lambda}, \label{eq:pseudofermion_gf} \\
\mathcal{G}^\phi_0(i\omega,\mathbf{k}) ={}& \frac{2\Omega_{\mathbf{k}}}{(i\omega-\Omega_{\mathbf{k}})(i\omega+\Omega_{\mathbf{k}})}. \label{eq:boson_gf}
\end{align}
For the explicit relation between the beta functions and renormalization constants, see Ref.~\cite{Han2022_SM}.


\section{Results for the Fermi-Kondo model}
\subsection{Beta Functions for the Fermi-Kondo Model \label{app:fermi_betas}}
The beta functions are very long so to save space we define a couple quantities in Eqs.\eqref{eq:WFKa}-\eqref{eq:WF2m} which appear repeatedly.

\begin{align}
W_{Ka} ={}& K_{a1}^2 + 4K_{b1}^2 + K_{a2}^2 + 4K_{b2}^2 + 2K_{a3}^2 + 8K_{b3}^2 + K_{a4}^2 + 4K_{b4}^2 + 2K_{E1}^2 + 3K_{E2}^2 + 2K_{2\alpha\scriptscriptstyle{+}}^2 \notag\\& + 4K_{2\beta\scriptscriptstyle{+}}^2 + 6K_{A}^2 +5K_{2\alpha\scriptscriptstyle{-}}^2 + 10K_{2\beta\scriptscriptstyle{-}}^2, \label{eq:WFKa} \\
W_{Kb} ={}& 4K_{a1}^2 + 4K_{b1}^2 + 4K_{a2}^2 + 4K_{b2}^2 + 8K_{a3}^2 + 8K_{b3}^2 + 4K_{a4}^2 + 4K_{b4}^2 \notag\\& + 4K_{2\alpha\scriptscriptstyle{+}}^2 + 8K_{2\beta\scriptscriptstyle{+}}^2 + 4K_{2\alpha\scriptscriptstyle{-}}^2 + 8K_{2\beta\scriptscriptstyle{-}}^2, \\
W_{KE} ={}& 3K_{a1}^2 + 3K_{a2}^2 + 6K_{a3}^2 + 3K_{a4}^2 + 2K_{E1}^2 + 3K_{E2}^2 + 6K_{2\alpha\scriptscriptstyle{+}}^2 + 12K_{2\beta\scriptscriptstyle{+}}^2 + 6K_{A}^2 +  3K_{2\alpha\scriptscriptstyle{-}}^2 + 6K_{2\beta\scriptscriptstyle{-}}^2, \\
W_{K2\scriptscriptstyle{+}} ={}& 2K_{a1}^2 + 4K_{b1}^2 + 2K_{a2}^2 + 4K_{b2}^2 + 4K_{a3}^2 + 8K_{b3}^2 + 2K_{a4}^2 + 4K_{b4}^2 + 4K_{E1}^2 + 6K_{E2}^2 \notag\\& + 4K_{2\alpha\scriptscriptstyle{+}}^2 + 8K_{2\beta\scriptscriptstyle{+}}^2 + 2K_{2\alpha\scriptscriptstyle{-}}^2 + 4K_{2\beta\scriptscriptstyle{-}}^2, \\
W_{A} ={}& 6K_{a1}^2 + 6K_{a2}^2 + 12K_{a3}^2 + 6K_{a4}^2 + 4K_{E1}^2 + 6K_{E2}^2 + 6K_{2\alpha\scriptscriptstyle{-}}^2 + 12K_{2\beta\scriptscriptstyle{-}}^2, \\
W_{2\scriptscriptstyle{-}} ={}& 5K_{a1}^2 + 4K_{b1}^2 + 5K_{a2}^2 + 4K_{b2}^2 + 10K_{a3}^2 + 8K_{b3}^2 + 5K_{a4}^2 + 4K_{b4}^2 + 2K_{E1}^2 \notag\\& + 3K_{E2}^2 + 2K_{2\alpha\scriptscriptstyle{+}}^2 + 4K_{2\beta\scriptscriptstyle{+}}^2 + 6K_{A}^2 + K_{2\alpha\scriptscriptstyle{-}}^2 + 2K_{2\beta\scriptscriptstyle{-}}^2. \label{eq:WF2m}
\end{align}

The beta functions for the Fermi-Kondo model are then given by Eqs.~\eqref{eq:beta_Ka1}-\eqref{eq:beta_K2bm}.
\begin{align}
\frac{d K_{a1}}{d\ln\mu} ={}&  K_{a1}(W_{Ka} - 4K_{b1}^2) - \frac{4 \sqrt{6} K_{A} K_{2\beta\scriptscriptstyle{-}}}{3} - 2 \sqrt{2} K_{E2} K_{2\beta\scriptscriptstyle{+}} - \frac{2 \sqrt{6} K_{a1} K_{b1}}{3} - 4 K_{b1} K_{a2} K_{b2} - 8 K_{b1} K_{a3} K_{b3}  \notag\\&- 4 K_{b1} K_{a4} K_{b4} + \frac{2 \sqrt{6} K_{a3} K_{b3}}{3} + 2 \sqrt{2} K_{b3} K_{2\beta\scriptscriptstyle{-}} + \frac{\sqrt{6} K_{a4} K_{b4}}{3} + \sqrt{6} K_{b4} K_{2\alpha\scriptscriptstyle{-}}, \label{eq:beta_Ka1} \\
\frac{d K_{b1}}{d\ln\mu} ={}& K_{b1}(W_{Kb} - 4K_{a1}^2) - \frac{\sqrt{6} K_{a1}^{2}}{3} - 4 K_{a1} K_{a2} K_{b2} - 8 K_{a1} K_{a3} K_{b3} - 4 K_{a1} K_{a4} K_{b4} + \frac{2 \sqrt{6} K_{b1}^{2}}{3} + \frac{\sqrt{6} K_{a3}^{2}}{3} \notag\\&- 2 \sqrt{2} K_{a3} K_{2\beta\scriptscriptstyle{-}} - \frac{2 \sqrt{6} K_{b3}^{2}}{3} + \frac{\sqrt{6} K_{a4}^{2}}{6} - \sqrt{6} K_{a4} K_{2\alpha\scriptscriptstyle{-}} - \frac{\sqrt{6} K_{b4}^{2}}{3} + \frac{\sqrt{6} K_{2\alpha\scriptscriptstyle{-}}^{2}}{6} - \frac{\sqrt{6} K_{2\beta\scriptscriptstyle{-}}^{2}}{3} - \frac{2 \sqrt{6} K_{2\beta\scriptscriptstyle{+}}^{2}}{3} ,\\
\frac{d K_{a2}}{d\ln \mu} ={}& K_{a2}(W_{Ka}-4K_{b2}^2) + 2 K_{A} K_{2\alpha\scriptscriptstyle{-}} - 2 K_{A} K_{2\beta\scriptscriptstyle{-}} - 2 K_{E1} K_{2\alpha\scriptscriptstyle{+}} - \sqrt{3} K_{E2} K_{2\beta\scriptscriptstyle{+}} - 4 K_{a1} K_{b1} K_{b2} - K_{a2} K_{b2} \notag\\
& - 8 K_{b2} K_{a3} K_{b3} - 4 K_{b2} K_{a4} K_{b4} - K_{a3} K_{b3} - \sqrt{3} K_{a3} K_{b4} - \sqrt{3} K_{b3} K_{a4} - \sqrt{3} K_{b3} K_{2\alpha\scriptscriptstyle{-}} + \sqrt{3} K_{b3} K_{2\beta\scriptscriptstyle{-}} \notag\\
&- 3 K_{b4} K_{2\beta\scriptscriptstyle{-}} ,\\
\frac{d K_{b2}}{d\ln\mu} ={}& K_{b2}(W_{Kb} - 4K_{a2}^2) - 4 K_{a1} K_{b1} K_{a2} - \frac{K_{a2}^{2}}{2} - 8 K_{a2} K_{a3} K_{b3} - 4 K_{a2} K_{a4} K_{b4} + K_{b2}^{2} - \frac{K_{a3}^{2}}{2} - \sqrt{3} K_{a3} K_{a4}  \notag\\
&+ \sqrt{3} K_{a3} K_{2\alpha\scriptscriptstyle{-}} - \sqrt{3} K_{a3} K_{2\beta\scriptscriptstyle{-}} + K_{b3}^{2} + 2 \sqrt{3} K_{b3} K_{b4} + 3 K_{a4} K_{2\beta\scriptscriptstyle{-}} + K_{2\alpha\scriptscriptstyle{-}} K_{2\beta\scriptscriptstyle{-}} + \frac{K_{2\beta\scriptscriptstyle{-}}^{2}}{2} - K_{2\alpha\scriptscriptstyle{+}}^{2} \notag\\&- K_{2\beta\scriptscriptstyle{+}}^{2} ,\\
\frac{d K_{a3}}{d\ln\mu} ={}& K_{a3}(W_{Ka} - 8K_{b3}^2) - K_{A} K_{2\alpha\scriptscriptstyle{-}} + K_{A} K_{2\beta\scriptscriptstyle{-}} - K_{E1} K_{2\beta\scriptscriptstyle{+}} - \frac{\sqrt{3} K_{E2} K_{2\alpha\scriptscriptstyle{+}}}{2} - 4 K_{a1} K_{b1} K_{b3} + \frac{\sqrt{6} K_{a1} K_{b3}}{3} + \notag\\
&\frac{\sqrt{6} K_{b1} K_{a3}}{3} - \sqrt{2} K_{b1} K_{2\beta\scriptscriptstyle{-}} - 4 K_{a2} K_{b2} K_{b3} - \frac{K_{a2} K_{b3}}{2} - \frac{\sqrt{3} K_{a2} K_{b4}}{2} - \frac{K_{b2} K_{a3}}{2} - \frac{\sqrt{3} K_{b2} K_{a4}}{2} \notag\\
&+ \frac{\sqrt{3} K_{b2} K_{2\alpha\scriptscriptstyle{-}}}{2} - \frac{\sqrt{3} K_{b2} K_{2\beta\scriptscriptstyle{-}}}{2}  - K_{a3} K_{b3} - \frac{\sqrt{3} K_{a3} K_{b4}}{3}  - 4 K_{b3} K_{a4} K_{b4} - \frac{\sqrt{3} K_{b3} K_{a4}}{3} + K_{b4} K_{2\beta\scriptscriptstyle{-}} ,\\
\frac{d K_{b3}}{d\ln\mu} ={}& K_{b3}(W_{Kb} - 8K_{a3}^2) - 4 K_{a1} K_{b1} K_{a3} + \frac{\sqrt{6} K_{a1} K_{a3}}{3} + \sqrt{2} K_{a1} K_{2\beta\scriptscriptstyle{-}} - \frac{2 \sqrt{6} K_{b1} K_{b3}}{3} - 4 K_{a2} K_{b2} K_{a3} - \frac{K_{a2} K_{a3}}{2} \notag\\
&- \frac{\sqrt{3} K_{a2} K_{a4}}{2} - \frac{\sqrt{3} K_{a2} K_{2\alpha\scriptscriptstyle{-}}}{2} + \frac{\sqrt{3} K_{a2} K_{2\beta\scriptscriptstyle{-}}}{2} + K_{b2} K_{b3} + \sqrt{3} K_{b2} K_{b4} - \frac{K_{a3}^{2}}{2} - 4 K_{a3} K_{a4} K_{b4} - \frac{\sqrt{3} K_{a3} K_{a4}}{3} \notag\\
&+ K_{b3}^{2} - K_{2\beta\scriptscriptstyle{+}}^{2}+ \frac{2 \sqrt{3} K_{b3} K_{b4}}{3} - K_{a4} K_{2\beta\scriptscriptstyle{-}} - \frac{K_{2\beta\scriptscriptstyle{-}}^{2}}{2} + K_{2\alpha\scriptscriptstyle{+}} K_{2\beta\scriptscriptstyle{+}}  ,\\
\frac{d K_{a4}}{d\ln\mu} ={}& K_{a4}(W_{Ka} - 4K_{b4}^2) - \frac{2 \sqrt{3} K_{A} K_{2\beta\scriptscriptstyle{-}}}{3} - K_{E2} K_{2\beta\scriptscriptstyle{+}} - 4 K_{a1} K_{b1} K_{b4} + \frac{\sqrt{6} K_{a1} K_{b4}}{3} + \frac{\sqrt{6} K_{b1} K_{a4}}{3} - \sqrt{6} K_{b1} K_{2\alpha\scriptscriptstyle{-}} \notag\\
&- 4 K_{a2} K_{b2} K_{b4} - \sqrt{3} K_{a2} K_{b3}  - \sqrt{3} K_{b2} K_{a3} + 3 K_{b2} K_{2\beta\scriptscriptstyle{-}} - 8 K_{a3} K_{b3} K_{b4} - \frac{2 \sqrt{3} K_{a3} K_{b3}}{3} - 2 K_{b3} K_{2\beta\scriptscriptstyle{-}} \notag\\
&- \frac{2 \sqrt{3} K_{a4} K_{b4}}{3}, \\
\frac{d K_{b4}}{d\ln\mu} ={}& K_{b4}(W_{Kb} - 4K_{a4}^2) - 4 K_{a1} K_{b1} K_{a4} + \frac{\sqrt{6} K_{a1} K_{a4}}{3} + \sqrt{6} K_{a1} K_{2\alpha\scriptscriptstyle{-}} - \frac{2 \sqrt{6} K_{b1} K_{b4}}{3} - 4 K_{a2} K_{b2} K_{a4} - \sqrt{3} K_{a2} K_{a3} \notag\\
&- 3 K_{a2} K_{2\beta\scriptscriptstyle{-}} + 2 \sqrt{3} K_{b2} K_{b3} - \frac{\sqrt{3} K_{a3}^{2}}{3} - 8 K_{a3} K_{b3} K_{a4} + 2 K_{a3} K_{2\beta\scriptscriptstyle{-}} + \frac{2 \sqrt{3} K_{b3}^{2}}{3} - \frac{\sqrt{3} K_{a4}^{2}}{3} + \frac{2 \sqrt{3} K_{b4}^{2}}{3} \notag\\
& + \frac{\sqrt{3} K_{2\alpha\scriptscriptstyle{-}}^{2}}{3} + \frac{\sqrt{3} K_{2\beta\scriptscriptstyle{-}}^{2}}{3} - 2 \sqrt{3} K_{2\alpha\scriptscriptstyle{+}} K_{2\beta\scriptscriptstyle{+}} + \frac{2 \sqrt{3} K_{2\beta\scriptscriptstyle{+}}^{2}}{3}, \\
\frac{d K_{E1}}{d\ln\mu} ={}& K_{E1}W_{KE} - 3 K_{A} K_{E2}  - 3 K_{a2} K_{2\alpha\scriptscriptstyle{+}} - 3 K_{a3} K_{2\beta\scriptscriptstyle{+}} - 3 \sqrt{3} K_{2\beta\scriptscriptstyle{-}} K_{2\beta\scriptscriptstyle{+}}, \\
\frac{d K_{E2}}{d\ln\mu} ={}& K_{E2}W_{KE} - 2 K_{A} K_{E1} + \sqrt{3} K_{A} K_{E2} - 2 \sqrt{2} K_{a1} K_{2\beta\scriptscriptstyle{+}} - \sqrt{3} K_{a2} K_{2\beta\scriptscriptstyle{+}} - \sqrt{3} K_{a3} K_{2\alpha\scriptscriptstyle{+}} - K_{a4} K_{2\beta\scriptscriptstyle{+}}  \notag\\
&+ 2 K_{2\alpha\scriptscriptstyle{-}} K_{2\alpha\scriptscriptstyle{+}}  - K_{2\alpha\scriptscriptstyle{-}} K_{2\beta\scriptscriptstyle{+}} - K_{2\beta\scriptscriptstyle{-}} K_{2\alpha\scriptscriptstyle{+}} + 2 K_{2\beta\scriptscriptstyle{-}} K_{2\beta\scriptscriptstyle{+}} ,\\
\frac{d K_{2\alpha\scriptscriptstyle{+}}}{d\ln\mu} ={}& K_{2\alpha\scriptscriptstyle{+}}W_{K2\scriptscriptstyle{+}} - 2 K_{E1} K_{a2} - \sqrt{3} K_{E2} K_{a3} + 2 K_{E2} K_{2\alpha\scriptscriptstyle{-}} - K_{E2} K_{2\beta\scriptscriptstyle{-}} - 2 K_{b2} K_{2\alpha\scriptscriptstyle{+}} + 2 K_{b3} K_{2\beta\scriptscriptstyle{+}}  \notag\\
&- 2 \sqrt{3} K_{b4} K_{2\beta\scriptscriptstyle{+}} ,\\
\frac{d K_{2\beta\scriptscriptstyle{+}}}{d\ln\mu} ={}& K_{2\beta\scriptscriptstyle{+}}W_{K2\scriptscriptstyle{+}} - K_{E1} K_{a3} - \sqrt{3} K_{E1} K_{2\beta\scriptscriptstyle{-}}  - \sqrt{2} K_{E2} K_{a1} - \frac{\sqrt{3} K_{E2} K_{a2}}{2} - \frac{K_{E2} K_{a4}}{2} - \frac{K_{E2} K_{2\alpha\scriptscriptstyle{-}}}{2} + K_{E2} K_{2\beta\scriptscriptstyle{-}}  \notag\\
&- \frac{2 \sqrt{6} K_{b1} K_{2\beta\scriptscriptstyle{+}}}{3} - K_{b2} K_{2\beta\scriptscriptstyle{+}} + K_{b3} K_{2\alpha\scriptscriptstyle{+}} - 2 K_{b3} K_{2\beta\scriptscriptstyle{+}} - \sqrt{3} K_{b4} K_{2\alpha\scriptscriptstyle{+}} + \frac{2 \sqrt{3} K_{b4} K_{2\beta\scriptscriptstyle{+}}}{3} ,\\
\frac{d K_{A}}{d\ln\mu} ={}&  K_{A}W_{A} - 2 K_{E1} K_{E2} + \frac{\sqrt{3} K_{E2}^{2}}{2} - \frac{4 \sqrt{6} K_{a1} K_{2\beta\scriptscriptstyle{-}}}{3} + 2 K_{a2} K_{2\alpha\scriptscriptstyle{-}} - 2 K_{a2} K_{2\beta\scriptscriptstyle{-}} - 2 K_{a3} K_{2\alpha\scriptscriptstyle{-}} + 2 K_{a3} K_{2\beta\scriptscriptstyle{-}} \notag\\
&- \frac{2 \sqrt{3} K_{a4} K_{2\beta\scriptscriptstyle{-}}}{3} ,\\
\frac{d K_{2\alpha\scriptscriptstyle{-}}}{d\ln\mu} ={}& K_{2\alpha\scriptscriptstyle{-}}W_{2\scriptscriptstyle{-}} + 2 K_{A} K_{a2} - 2 K_{A} K_{a3} + 2 K_{E2} K_{2\alpha\scriptscriptstyle{+}} - K_{E2} K_{2\beta\scriptscriptstyle{+}} + \sqrt{6} K_{a1} K_{b4} - \sqrt{6} K_{b1} K_{a4} + \frac{\sqrt{6} K_{b1} K_{2\alpha\scriptscriptstyle{-}}}{3}  \notag\\
&- \sqrt{3} K_{a2} K_{b3} + \sqrt{3} K_{b2} K_{a3} + K_{b2} K_{2\beta\scriptscriptstyle{-}} + \frac{2 \sqrt{3} K_{b4} K_{2\alpha\scriptscriptstyle{-}}}{3}, \\ 
\frac{d K_{2\beta\scriptscriptstyle{-}}}{d\ln\mu} ={}& K_{2\beta\scriptscriptstyle{-}}W_{2\scriptscriptstyle{-}} - \frac{2 \sqrt{6} K_{A} K_{a1}}{3} - K_{A} K_{a2} + K_{A} K_{a3} - \frac{\sqrt{3} K_{A} K_{a4}}{3} - \sqrt{3} K_{E1} K_{2\beta\scriptscriptstyle{+}} - \frac{K_{E2} K_{2\alpha\scriptscriptstyle{+}}}{2} + K_{E2} K_{2\beta\scriptscriptstyle{+}}  \notag\\
&+ \sqrt{2} K_{a1} K_{b3} - \sqrt{2} K_{b1} K_{a3} - \frac{\sqrt{6} K_{b1} K_{2\beta\scriptscriptstyle{-}}}{3}  + \frac{\sqrt{3} K_{a2} K_{b3}}{2} - \frac{3 K_{a2} K_{b4}}{2} - \frac{\sqrt{3} K_{b2} K_{a3}}{2} + \frac{3 K_{b2} K_{a4}}{2} \notag\\
&+ \frac{K_{b2} K_{2\alpha\scriptscriptstyle{-}}}{2} + \frac{K_{b2} K_{2\beta\scriptscriptstyle{-}}}{2} + K_{a3} K_{b4} - K_{b3} K_{a4} - K_{b3} K_{2\beta\scriptscriptstyle{-}} + \frac{\sqrt{3} K_{b4} K_{2\beta\scriptscriptstyle{-}}}{3}.\label{eq:beta_K2bm}
\end{align}
There are numerous fixed points of the Fermi-Kondo beta functions, but we only list the one which is useful for the two-stage Kondo destruction, $F$. Note that the fixed point $F$ should not be confused with the superscript $F$ which we will use below to denote the Fermi-Kondo beta functions. This fixed point is described in Supplementary Materials~\ref{app:fermi_beta_fxp}.
To simplify equations in the following appendices, we introduce the following for the beta functions in the Fermi-Kondo model:
\begin{align*}
    \beta^{F}(K_{i})\equiv \left[\frac{dK_{i}}{d\ln\mu}\right]_{F},
\end{align*}
where the superscript $F$ on the left hand side and the subscript $F$ on the right hand side stand for the Fermi-Kondo model only. In other words, the beta functions in Eqs.~\eqref{eq:beta_Ka1}-\eqref{eq:beta_K2bm} are given by $\beta^F(K_{i})$.

\subsection{Fermionic Kondo fixed points and their fixed point Hamiltonians}\label{app:fermi_beta_fxp}
When we consider the Fermi-Kondo Hamiltonians only, the beta functions are quite lengthy and are given in Supplementary Materials~\ref{app:fermi_betas}. From the beta functions, we find numerous stable fixed points, each of which corresponds to some low temperature phase of the high-energy Kondo models in Eqs.~\eqref{eq:HKT1a}-\eqref{eq:HKT2mb}. For the purposes of this work, we consider only a single fermionic Kondo fixed point $F$. It is characterized by $\Delta = 1$, where $1+\Delta$ is the scaling dimension of the leading irrelevant operator and $\Delta$ is the slope of the beta function at the fixed point. Note that, in the Kondo problem, it is possible for an apparently stable fixed point of the RG flow to actually be unstable. This occurs when the strong coupling limit of the fixed point Hamiltonian is stable, or, equivalently, has a unique ground state. This indicates that the fixed point at intermediate coupling is in fact unstable, and should actually flow to strong coupling \cite{Lavagna2003_SM, Bensimon2006a_SM}. For the case of $\text{SU}(N)$ $k$-channel Kondo problems, fixed points with unstable strong coupling limit correspond to non-Fermi liquid phases (for $k \geq N$), as shown through non-Abelian bosonization and conformal field theory approaches \cite{Affleck1991a_SM, Affleck1993b_SM, Ludwig1991_SM, Kimura2017b_SM}. In this work, we classify the fixed points into Fermi liquid phases and non-Fermi liquid phases depending only on whether they have unique or degenerate ground states in the strong coupling limit because none of the points can exactly be mapped to an $\text{SU}(N)$ $k$-channel Kondo model. A more rigorous classification may be provided in future works with nonperturbative techniques. 
From the Fermi-Kondo beta functions, we can obtain several fermionic Kondo fixed points. 
In this section, we will discuss properties of the Fermionic Kondo fixed points, in particular, we will focus on $F$ which is relevant to our discussion. 
The most important step is a change of basis for the conduction electrons. The original orbital basis (cubic harmonics) can be rewritten in terms of the angular momentum basis (spherical harmonics $\ket{\ell,m}$) \cite{Schultz2021b_SM, Patri2020e_SM},
\begin{align}
    \ket{x}={}&\frac{1}{\sqrt{2}}(-\ket{1,1}+\ket{1,-1}),\\
    \ket{y}={}&\frac{i}{\sqrt{2}}(\ket{1,1}+\ket{1,-1}),\\
    \ket{z}={}&\ket{1,0}.
\end{align}
By using the Clebsch-Gordan coefficients, the composite spin-orbit coupled angular momentum basis is given by
\begin{align}
    \ket{1,m}\Ket{\pm\tfrac{1}{2}}={}&\sqrt{\frac{2\pm m}{3}}\Ket{\tfrac{3}{2},m\pm\tfrac{1}{2}}\mp\sqrt{\frac{1\mp m}{3}}\Ket{\tfrac{1}{2},m\pm\tfrac{1}{2}}.
\end{align}
Let $U$ be the basis transformation from the cubic harmonic and spin basis to the total angular momentum basis. Then,
\begin{align}
    {}&(c_{x,\uparrow}\;c_{x,\downarrow}\;c_{y,\uparrow}\;c_{y,\downarrow}\;c_{z,\uparrow}\;c_{z,\downarrow}\;)^{\intercal}U^{\intercal}\notag\\
    {}&=(c_{\frac{3}{2},\frac{3}{2}}\;c_{\frac{3}{2},\frac{1}{2}}\;c_{\frac{3}{2},-\frac{1}{2}}\;c_{\frac{3}{2},-\frac{3}{2}}\;c_{\frac{1}{2},\frac{1}{2}}\;c_{\frac{1}{2},-\frac{1}{2}})^{\intercal},
\end{align}
where
\begin{align}
    U=\left(\begin{matrix}
 -\frac{1}{\sqrt{2}} & 0 & \frac{i}{\sqrt{2}} & 0 & 0 & 0 \\
 0 & -\frac{1}{\sqrt{6}} & 0 & \frac{i}{\sqrt{6}} & \sqrt{\frac{2}{3}} & 0 \\
 \frac{1}{\sqrt{6}} & 0 & \frac{i}{\sqrt{6}} & 0 & 0 & \sqrt{\frac{2}{3}} \\
 0 & \frac{1}{\sqrt{2}} & 0 & \frac{i}{\sqrt{2}} & 0 & 0 \\
 0 & -\frac{1}{\sqrt{3}} & 0 & \frac{i}{\sqrt{3}} & -\frac{1}{\sqrt{3}} & 0 \\
 -\frac{1}{\sqrt{3}} & 0 & -\frac{i}{\sqrt{3}} & 0 & 0 & \frac{1}{\sqrt{3}} \\
    \end{matrix}\right).
\end{align}

The fixed point values of $F$ are given by $(K_{a1}^{*},K_{b1}^{*},K_{a2}^{*},K_{b2}^{*},K_{a3}^{*},K_{b3}^{*},K_{a4}^{*},K_{b4}^{*},K_{E1}^{*},K_{E2}^{*},K_{2\alpha\scriptscriptstyle{+}}^{*},K_{2\beta\scriptscriptstyle{+}}^{*},K_{A}^{*},K_{2\alpha\scriptscriptstyle{-}}^{*},K_{2\beta\scriptscriptstyle{-}}^{*})=(0,\frac{1}{3\sqrt{6}},0,-\frac{1}{3},0,-\frac{1}{3},0,-\frac{2}{3\sqrt{3}},\pm\frac{1}{2\sqrt{3}},\mp\frac{1}{3},0,0,-\frac{1}{2\sqrt{3}},0,0)$. 
The fixed point Hamiltonian of $F$ corresponds to a 6 generator truncated SU(4) fixed point \cite{Schultz2021c_SM} as follows:
    \begin{align}
        H_{F}={}&\frac{1}{2}\sum_{\rho,\tau=1}^4\tilde{\psi}_{\rho}^{\dagger}[(\sigma^{0}\otimes\vec{\sigma})_{\rho\tau}\cdot(Q^{x},O^{y},Q^{z})]\tilde{\psi}_{\tau}\notag\\
        &+\sum_{\rho,\tau=1}^2 \psi_{\rho}^{\dagger}[\vec{\sigma}_{\rho\tau}\cdot(D^{x},D^{y},D^{z})]\psi_{\tau},
    \end{align}
    where $\vec{\sigma}=(\sigma^{x},\sigma^{y},\sigma^{z})$,  $\psi=c_{\tfrac{1}{2}}$ is a 2-component spinor, and $\tilde{\psi}$ is a 4-component spinor obtained by a further change of basis, for example, $\tilde{\psi}= U_{1}'c_{\frac{3}{2}}$,
    \begin{align}
        U_{1}'=\left(\begin{matrix}
 1 & 0 & 0 & 0 \\
 0 & 0 & 1 & 0 \\
 0 & 0 & 0 & 1 \\
 0 & 1 & 0 & 0 \\
        \end{matrix}\right),
    \end{align}
where the form of $U_{1}'$ may change slightly for a different chosen sign structure of the fixed point. Here, $Q^{x},O^{y},Q^{z}=S^{4,9,5}$ and $D^{x,y,z}=-S^{10,11,12}$ satisfy two mutually commuting SU(2) algebras.

\section{Results for the Bose-Kondo Model}

\subsection{Beta Functions for the Bose-Kondo Model \label{app:bose_betas}}
To neatly express the beta functions for the Bose-Kondo model, we define some commonly occurring quantities in Eqs.~\eqref{eq:Wga}-\eqref{eq:Wg2m}. 

\begin{align}
W_{ga} ={}& - \frac{g_{E}^{4}}{2} - \frac{g_{E}^{2} g_{A}^{2}}{2} - \frac{g_{E}^{2} g_{a}^{2}}{2} - \frac{3 g_{E}^{2} g_{2\scriptscriptstyle{+}}^{2}}{2} - \frac{g_{E}^{2} g_{2\scriptscriptstyle{-}}^{2}}{2} + \frac{g_{E}^{2}}{2} - \frac{5 g_{A}^{2} g_{a}^{2}}{4} - \frac{g_{A}^{2} g_{2\scriptscriptstyle{-}}^{2}}{4} + \frac{g_{A}^{2}}{2} - \frac{g_{a}^{4}}{4} - 2 g_{a}^{2} g_{b}^{2} - \frac{3 g_{a}^{2} g_{2\scriptscriptstyle{+}}^{2}}{4} \notag\\
& - 3 g_{a}^{2} g_{2\scriptscriptstyle{-}}^{2} + \frac{g_{a}^{2}}{4} - g_{b}^{4} - g_{b}^{2} g_{2\scriptscriptstyle{+}}^{2} - g_{b}^{2} g_{2\scriptscriptstyle{-}}^{2} + g_{b}^{2} - g_{2\scriptscriptstyle{+}}^{4} - \frac{3 g_{2\scriptscriptstyle{+}}^{2} g_{2\scriptscriptstyle{-}}^{2}}{4} + \frac{g_{2\scriptscriptstyle{+}}^{2}}{2} - \frac{g_{2\scriptscriptstyle{-}}^{4}}{4} + \frac{5 g_{2\scriptscriptstyle{-}}^{2}}{4}, \label{eq:Wga} \\
W_{gE} ={}& - \frac{g_{E}^{4}}{2} - \frac{g_{E}^{2} g_{A}^{2}}{2} - \frac{3 g_{E}^{2} g_{a}^{2}}{2} - \frac{3 g_{E}^{2} g_{2\scriptscriptstyle{+}}^{2}}{2} - \frac{3 g_{E}^{2} g_{2\scriptscriptstyle{-}}^{2}}{2} + \frac{g_{E}^{2}}{2} - \frac{3 g_{A}^{2} g_{a}^{2}}{4} - \frac{3 g_{A}^{2} g_{2\scriptscriptstyle{-}}^{2}}{4} + \frac{ g_{A}^{2}}{2} - \frac{3 g_{a}^{2} g_{b}^{2}}{2} - \frac{3 g_{a}^{2} g_{2\scriptscriptstyle{+}}^{2}}{4} \notag\\
&- \frac{3 g_{a}^{2} g_{2\scriptscriptstyle{-}}^{2}}{2} + \frac{3 g_{a}^{2}}{4} - 3 g_{b}^{2} g_{2\scriptscriptstyle{+}}^{2} - \frac{3 g_{b}^{2} g_{2\scriptscriptstyle{-}}^{2}}{2} - \frac{3 g_{2\scriptscriptstyle{+}}^{2} g_{2\scriptscriptstyle{-}}^{2}}{4} + \frac{3 g_{2\scriptscriptstyle{+}}^{2}}{2} + \frac{3 g_{2\scriptscriptstyle{-}}^{2}}{4}, \\
W_{g2\scriptscriptstyle{+}} ={}& - g_{E}^{2} g_{A}^{2} - g_{E}^{2} g_{a}^{2} - g_{E}^{2} g_{2\scriptscriptstyle{+}}^{2} - g_{E}^{2} g_{2\scriptscriptstyle{-}}^{2} + g_{E}^{2} - \frac{g_{A}^{2} g_{a}^{2} }{2} - \frac{g_{A}^{2} g_{2\scriptscriptstyle{-}}^{2}}{2} - \frac{g_{a}^{4} }{4} - g_{a}^{2} g_{b}^{2} - \frac{3 g_{a}^{2} g_{2\scriptscriptstyle{+}}^{2}}{2} - \frac{3 g_{a}^{2}  g_{2\scriptscriptstyle{-}}^{2}}{2} + \frac{g_{a}^{2} }{2} \notag\\
&- g_{b}^{4}  - 2 g_{b}^{2} g_{2\scriptscriptstyle{+}}^{2} - g_{b}^{2}  g_{2\scriptscriptstyle{-}}^{2} + g_{b}^{2} - g_{2\scriptscriptstyle{+}}^{4} - \frac{3 g_{2\scriptscriptstyle{+}}^{2} g_{2\scriptscriptstyle{-}}^{2}}{2} + g_{2\scriptscriptstyle{+}}^{2} - \frac{g_{2\scriptscriptstyle{-}}^{4}}{4} + \frac{g_{2\scriptscriptstyle{-}}^{2}}{2}, \\
W_{gA} ={}& - g_{E}^{2} g_{A}^{2} - 3 g_{E}^{2} g_{2\scriptscriptstyle{+}}^{2} + g_{E}^{2} - \frac{3 g_{A}^{2} g_{a}^{2}}{2} - \frac{3 g_{A}^{2} g_{2\scriptscriptstyle{-}}^{2}}{2} - 3 g_{a}^{2} g_{b}^{2} - \frac{3 g_{a}^{2} g_{2\scriptscriptstyle{+}}^{2}}{2} + \frac{3 g_{a}^{2}}{2} - 3 g_{b}^{2} g_{2\scriptscriptstyle{-}}^{2} - \frac{3 g_{2\scriptscriptstyle{+}}^{2} g_{2\scriptscriptstyle{-}}^{2}}{2} + \frac{3 g_{2\scriptscriptstyle{-}}^{2}}{2}, \\
W_{gb} ={}& - g_{E}^{2} g_{a}^{2} - 2 g_{E}^{2} g_{2\scriptscriptstyle{+}}^{2} - g_{E}^{2} g_{2\scriptscriptstyle{-}}^{2} - g_{A}^{2} g_{a}^{2} - g_{A}^{2} g_{2\scriptscriptstyle{-}}^{2} - \frac{g_{a}^{4}}{4} - 2 g_{a}^{2} g_{b}^{2} - \frac{3 g_{a}^{2} g_{2\scriptscriptstyle{-}}^{2}}{2} + g_{a}^{2} - g_{b}^{4} - 2 g_{b}^{2} g_{2\scriptscriptstyle{+}}^{2} - 2 g_{b}^{2} g_{2\scriptscriptstyle{-}}^{2} \notag\\
&+ g_{b}^{2} - g_{2\scriptscriptstyle{+}}^{4} + g_{2\scriptscriptstyle{+}}^{2} - \frac{g_{2\scriptscriptstyle{-}}^{4}}{4} + g_{2\scriptscriptstyle{-}}^{2}, \\
W_{g2\scriptscriptstyle{-}} ={}& - \frac{g_{E}^{4} }{2} - \frac{g_{E}^{2} g_{A}^{2} }{2} - \frac{g_{E}^{2} g_{a}^{2} }{2} - \frac{3 g_{E}^{2} g_{2\scriptscriptstyle{+}}^{2} }{2} - \frac{g_{E}^{2} g_{2\scriptscriptstyle{-}}^{2}}{2} + \frac{g_{E}^{2} }{2} - \frac{g_{A}^{2} g_{a}^{2} }{4} - \frac{5 g_{A}^{2} g_{2\scriptscriptstyle{-}}^{2}}{4} + \frac{g_{A}^{2} }{2} - \frac{g_{a}^{4} }{4} - g_{a}^{2} g_{b}^{2} - \frac{3 g_{a}^{2} g_{2\scriptscriptstyle{+}}^{2} }{4} \notag\\
&- 3 g_{a}^{2} g_{2\scriptscriptstyle{-}}^{2} + \frac{5 g_{a}^{2} }{4} - g_{b}^{4} - g_{b}^{2} g_{2\scriptscriptstyle{+}}^{2} - 2 g_{b}^{2} g_{2\scriptscriptstyle{-}}^{2} + g_{b}^{2} - g_{2\scriptscriptstyle{+}}^{4} - \frac{3 g_{2\scriptscriptstyle{+}}^{2} g_{2\scriptscriptstyle{-}}^{2}}{4} + \frac{g_{2\scriptscriptstyle{+}}^{2} g_{2\scriptscriptstyle{-}}}{2} - \frac{g_{2\scriptscriptstyle{-}}^{4}}{4} + \frac{g_{2\scriptscriptstyle{-}}^{2}}{4}. \label{eq:Wg2m}
\end{align}

\twocolumngrid
The beta functions for the Bose-Kondo model are then in Eqs.\eqref{eq:beta_ga}-\eqref{eq:beta_2gm}:

\begin{align}
\frac{d g_{a}}{d\ln\mu} ={}& -\frac{\epsilon}{2}g_{a} + g_{A}W_{ga}, \label{eq:beta_ga} \\
\frac{d g_E}{d\ln\mu} ={}& -\frac{\epsilon}{2}g_{T_E} + g_{E}W_{gE}, \\
\frac{d g_{2\scriptscriptstyle{+}}}{d\ln\mu} ={}& -\frac{\epsilon}{2}g_{2\scriptscriptstyle{+}} + g_{2\scriptscriptstyle{+}}W_{g2\scriptscriptstyle{+}}, \\
\frac{d g_{A}}{d\ln\mu} ={}& -\frac{\epsilon}{2}g_{A} + g_{A}W_{gA} , \\
\frac{d g_{b}}{d\ln\mu} ={}& -\frac{\epsilon}{2}g_{b} + g_{b}W_{gb}, \\
\frac{d g_{2\scriptscriptstyle{-}}}{d\ln \mu} ={}& -\frac{\epsilon}{2}g_{2\scriptscriptstyle{-}} + g_{2\scriptscriptstyle{-}}W_{g2\scriptscriptstyle{-}}. \label{eq:beta_2gm}
\end{align}
In addition, we introduce the notation for the beta functions in the Bose-Kondo model as follows
\begin{align*}
    \beta^{B}(g_{i})\equiv \left[\frac{dg_{i}}{d\ln\mu}\right]_{B},
\end{align*}
where the superscript $B$ on the left hand side and the subscript $B$ on the right hand side stand for the Bose-Kondo model. In other words, $\beta^{B}(g_{i})$ are the beta functions in Eqs.\eqref{eq:beta_ga}-\eqref{eq:beta_2gm}. 

\subsection{Analysis for the Bose-Kondo Hamiltonians \label{sec:bose_betas}}  
In this section, we will discuss the bosonic fixed point $B$ which is relevant to the two-stage Kondo destruction transition.
The fixed point value of $B$ is given by $(g^{*2}_{a},g^{*2}_{E},g^{*2}_{2\scriptscriptstyle{+}},g^{*2}_{A},g^{*2}_{b},g^{*2}_{2\scriptscriptstyle{-}})=(0,\epsilon+\epsilon^2+0,0,\frac{\epsilon}{2}+\frac{\epsilon^{2}}{4},0)$. This point has nonzero $g_{E}$ and $g_{b}$ bosonic bath couplings, so it corresponds to a mixed multipolar ordered fixed point which has quadrupolar ordering and dipolar ordering with $E$ and $T_{1}$ irreps respectively. $B$ is valid for $0<\epsilon<1/4$.

\section{Results for the Bose-Fermi Kondo model}
\subsection{Beta Functions for the Bose-Fermi Kondo Model \label{app:fermi_bose_betas}}
Because the full beta functions for the Bose-Fermi Kondo model extend the previous results, we can express them in terms of the Fermi-Kondo only or Bose-Kondo only results, plus some contributions due to the mixing between fermionic and bosonic fields. Furthermore, these beta functions also contain the commonly-occuring expressions defined in Appendices \ref{app:fermi_betas} and \ref{app:bose_betas}. Thus they are expressed succinctly as
\begin{align}
\frac{d K_{ai}}{d\ln\mu} ={}& \beta^F(K_{ai}) + K_{ai}W_{ga} ,\quad i=1,2,3,4,\\
\frac{d K_{bi}}{d\ln\mu} ={}& \beta^F(K_{bi}) + K_{bi}W_{gb} ,\quad i=1,2,3,4,\\
\frac{d K_{Ei}}{d\ln\mu} ={}& \beta^F(K_{Ei}) + K_{Ei}W_{gE} ,\quad i=1,2,\\
\frac{d K_{2i\scriptscriptstyle{+}}}{d\ln\mu} ={}& \beta^F(K_{2i\scriptscriptstyle{+}}) + K_{2i\scriptscriptstyle{+}}W_{g2\scriptscriptstyle{+}} ,\quad i=\alpha,\beta,\\
\frac{d K_{A}}{d\ln\mu} ={}& \beta^F(K_{A}) + K_{A}W_{gA} ,\\
\frac{d K_{2i\scriptscriptstyle{-}}}{d\ln\mu} ={}& \beta^F(K_{2i\scriptscriptstyle{-}}) + K_{2i\scriptscriptstyle{-}}W_{g2\scriptscriptstyle{-}} ,\quad i=\alpha,\beta,\\
\frac{d g_{a}}{d\ln\mu} ={}& \beta^{B}(g_{a})+g_{a}W_{Ka}-g_{b}W_{ab} ,\\
\frac{d g_{E}}{d\ln\mu} ={}& \beta^{B}(g_{E})+g_{E}W_{KE},\\
\frac{d g_{2\scriptscriptstyle{+}}}{d\ln\mu} ={}& \beta^{B}(g_{2\scriptscriptstyle{+}})+g_{2\scriptscriptstyle{+}}W_{K2\scriptscriptstyle{+}},\\
\frac{d g_{A}}{d\ln\mu} ={}& \beta^{B}(g_{A})+g_{E}W_{KA},\\
\frac{d g_{b}}{d\ln\mu} ={}& \beta^{B}(g_{b})+g_{b}W_{Kb}-g_{a}W_{ab} ,\\
\frac{d g_{2\scriptscriptstyle{-}}}{d\ln\mu} ={}& \beta^{B}(g_{2\scriptscriptstyle{-}})+g_{2\scriptscriptstyle{-}}W_{K2\scriptscriptstyle{-}},
\end{align}
where $W_{ab}\equiv 4(K_{a1}K_{b1}+K_{a2}K_{b2}+2K_{a3}K_{b3}+K_{a4}K_{b4})$.

\begin{figure*}[ht]
    \centering
    \subfigure[]{
    \includegraphics[height=0.235\linewidth]{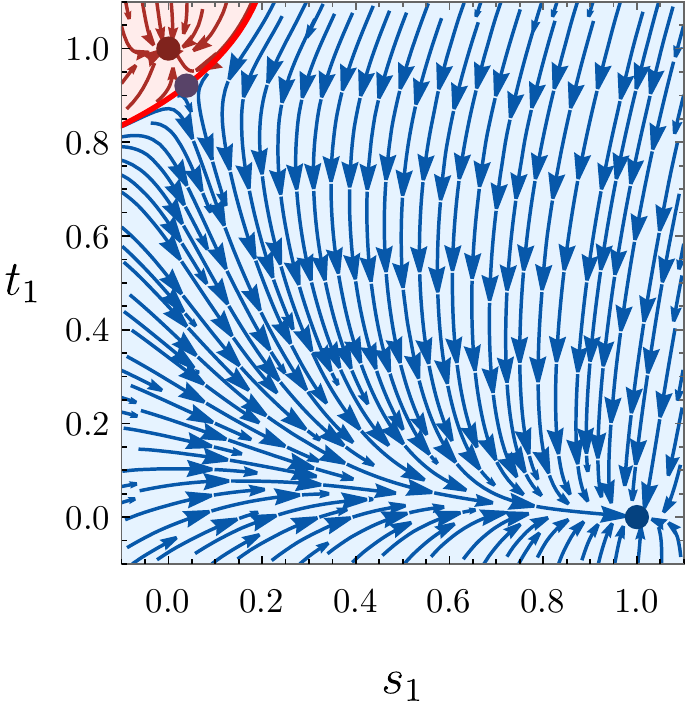}}
    \subfigure[]{
    \includegraphics[height=0.235\linewidth]{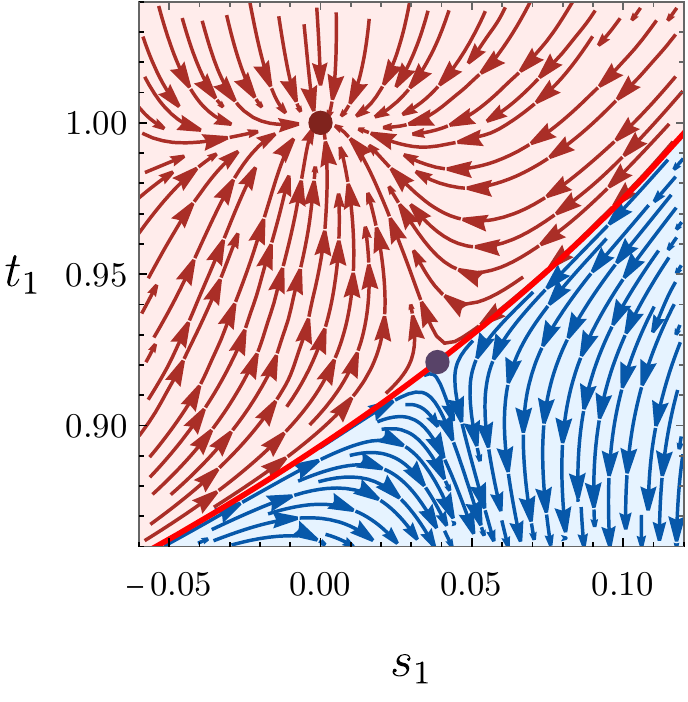}}
    \subfigure[]{
    \includegraphics[height=0.235\linewidth]{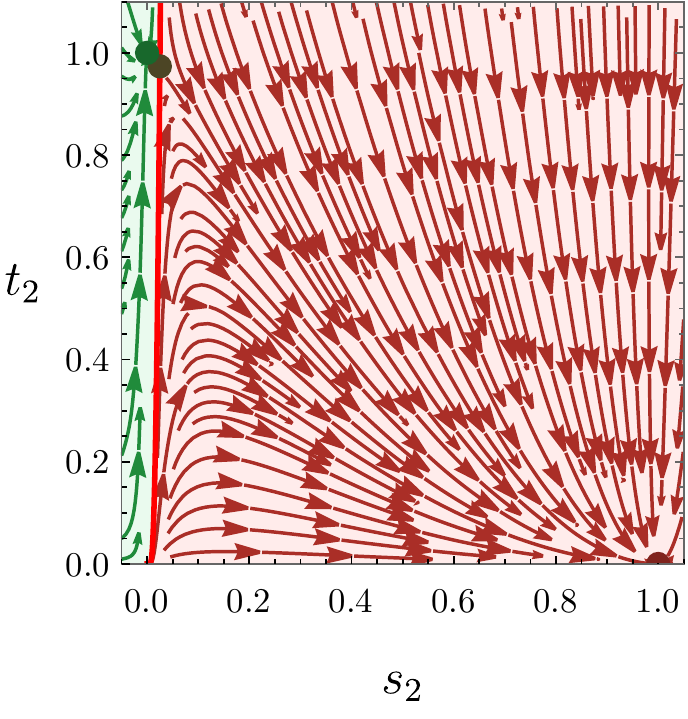}}
    \subfigure[]{
    \includegraphics[height=0.235\linewidth]{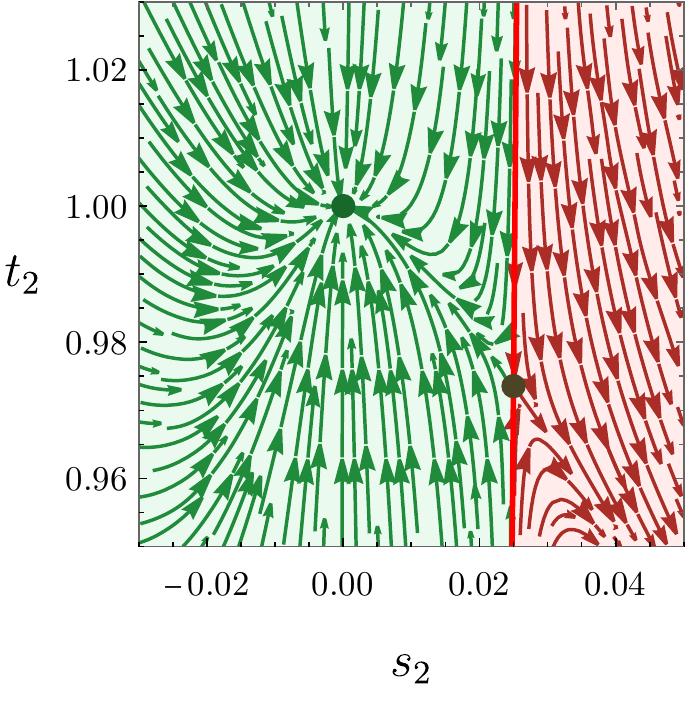}}
    \caption{The RG flow diagrams between $F$ and $P$, $B$ and $P$ when $\epsilon=0.1$. 
    (a-b) The RG flow diagrams between $F$ and $P$; (b) zooms in on the quantum critical point of (a). The blue and red points stand for $F$ and $P$. The purple point is the critical point $C_{FP}$ between $F$ and $P$.
    Here $s_{1}=-0.484\tilde{K}_{E}+1.484\tilde{K}_{A}$ and $t_{1}=1-1.236\tilde{K}_{E}+0.236\tilde{K}_{A}$
    with the constraint $\tilde{g}_{E}^{2}=1-1.675\tilde{K}_{E}+0.675\tilde{K}_{A}$ where $\tilde{K}_{E}=3K_{E1}=3K_{E2}$, $\tilde{K}_{A}=-2\sqrt{3}K_{A}$, and $\tilde{g}^{2}=8.873g^{2}$, and we set $(K_{b1},K_{b2},K_{b3},K_{b4})=(\frac{1}{3\sqrt{6}},-\frac{1}{3},-\frac{1}{3},-\frac{2}{3\sqrt{3}})$, and other coupling constants as zero.
    (c-d) The RG flow diagrams between $F$ and $P$; (d) zooms in on the critical point of (c). The green and red points stand for $B$ and $P$. The olive point is the critical point $C_{BP}$ between $B$ and $P$. Here, $s_{2}=3\sqrt{6}K_{b1}=-3K_{b2}=-3K_{b3}=-\frac{3\sqrt{3}}{2}K_{b4}$, $t_{2}=18.944g_{b}^{2}$, and we set $g_{E}^{2}=0.113$ and other coupling constants as zero.
    }
    \label{fig:rgflow_fmb}
\end{figure*}

From the above full beta functions, we can find numerous stable fixed points and critical points as presented in the main text. The nonzero fixed point values for the full beta functions up to $\epsilon^{2}$ order are presented in Tables~\ref{tab:two_stage_fixed_points}. 
The RG flow diagrams between $F$ and $P$, and between $P$ and $B$ are presented in Fig.~\ref{fig:rgflow_fmb}.
\onecolumngrid

\begin{table*}[!ht]
\centering
\begin{tabular}{|>{$}c<{$}|>{$}c<{$}|c|>{$}c<{$}|c|}
\hline
\text{Name}& \text{Nonzero Couplings} &Stability&\Delta& Between \\
\hline\hline
P & \left(K_{b1}^{*}, K_{b2}^{*}, K_{b3}^{*}, K_{b4}^{*},  g_{E}^{*2}\right)=\left(\frac{1}{3\sqrt{6}},-\frac{1}{3},-\frac{1}{3},-\frac{2}{3\sqrt{3}},\epsilon+\epsilon^2\right) & Stable &\frac{\epsilon}{2}  &-\\
\hline 
C_{FP} & \begin{tabular}{>{$}c<{$}}\left(K_{b1}^{*}, K_{b2}^{*}, K_{b3}^{*}, K_{b4}^{*}, K_{E1}^{*}, K_{E2}^{*}, K_{A}^{*}, g_{E}^{*2}\right) \\= \left(\frac{1}{3\sqrt{6}},-\frac{1}{3},-\frac{1}{3},-\frac{2}{3\sqrt{3}},\pm(\frac{\epsilon}{2\sqrt{6}}+\frac{\sqrt{3}\epsilon^{2}}{16\sqrt{2}}),\mp(\frac{\epsilon}{3\sqrt{2}}+\frac{\epsilon^{2}}{8\sqrt{2}}),-\frac{\epsilon}{4\sqrt{3}},\epsilon+\epsilon^2\right)\end{tabular} & Critical &\text{-}& $F$ and $P$\\
C_{BP} & \left(K_{b1}^{*}, K_{b2}^{*}, K_{b3}^{*}, K_{b4}^{*},  g_{E}^{*2},g_{b}^{*2}\right) = \left(\frac{\epsilon}{12\sqrt{6}},-\frac{\epsilon}{12},-\frac{\epsilon}{12},-\frac{\epsilon}{6\sqrt{3}},\epsilon+\epsilon^{2},\frac{\epsilon}{2}+\frac{\epsilon^{2}}{8}\right) & Critical &\text{-}& $B$ and $P$ \\
\hline
\end{tabular}
\caption{
The nonzero fixed point values of the partially Kondo destroyed fixed point $P$ and related critical points $C_{FP}$ and $C_{BP}$ in the full beta functions up to $\epsilon^{2}$ order. Here, $P$ is the partially Kondo destroyed fixed point, and $C_{FP}$ and $C_{BP}$ are the critical points between $F$ and $P$, and $P$ and $B$, respectively. 
}
\label{tab:two_stage_fixed_points}
\end{table*}

\subsection{Analysis for the Bose-Fermi Kondo Hamiltonians}
The beta functions, fixed point values, and flow diagrams for the Bose-Fermi Kondo Hamiltonians are enumerated in Supplementary Materials~\ref{app:fermi_bose_betas}. The fermionic and bosonic fixed points presented earlier are all still stable under the full renormalization group flow. In addition to these fixed points, we also find critical points between the fermionic and bosonic stable fixed points. We also find the stable fixed point $P$ and critical points $C_{FP}$ and $C_{BP}$ discussed in the main text.

\section{Elastic Free Energy and Multipolar Susceptibility}

\subsection{Multipolar susceptibility exponents}\label{app:susceptibility}

We can compute the multipolar susceptibility exponent by using the beta functions of the Bose-Kondo couplings.
Depending on the fixed point values of the Bose-Kondo couplings, the multipolar susceptibility exponent is given by \cite{Zhu2002_SM,Zarand2002_SM,Han2022_SM}:
\begin{equation}
\gamma_i = \begin{cases} \displaystyle{\epsilon +\left[\frac{2}{g_{i}}\frac{dg_{i}}{d\ln\mu}\right]_{\text{f.p.}}}=\epsilon, & g_{i}^{*} \neq 0, \\
 \displaystyle{\epsilon + 2\left[\frac{\partial}{\partial g_i}\frac{d g_i}{d\ln\mu}\right]_{\text{f.p.}}}, & g_{i}^{*} = 0, \\
\end{cases}
\end{equation}
where $\text{f.p.}$ stands for evaluation at the fixed point. Here, $\gamma_{i}$ becomes $\epsilon$ to all orders of $\epsilon$ by definition of the fixed point; $[({dg_{i}}/{d\ln\mu})/g_{i}]_{\text{f.p.}}=0$ when $g_{i}^{*}\neq0$ \cite{Zhu2002_SM,Zarand2002_SM}. In the other case where $g_{i}^{*}=0$, $\gamma_{i}$ is computed up to $\epsilon^{2}$ order. 

The table for the multipolar susceptibility exponent discussed in the main text is shown in Table~\ref{tab:susceptibility}.
\begin{table*}[t]
    \centering
    \begin{tabular}{|>{\footnotesize$}c<{$}|>{\footnotesize$}c<{$} >{\footnotesize$}c<{$} >{\footnotesize$}c<{$} >{\footnotesize$}c<{$} >{\footnotesize$}c<{$} >{\footnotesize$}c<{$}|}
    \hline
         \text{F.P.}& \gamma_{a} & \gamma_{E} & \gamma_{2\scriptscriptstyle{+}} & \gamma_{A} & \gamma_{b} & \gamma_{2\scriptscriptstyle{-}}\\
         \hline\hline
         B & 2\epsilon & \epsilon & 3\epsilon+2\epsilon^2 & 2\epsilon+2\epsilon^2 & \epsilon & 2\epsilon \\
         \hline
         P & 4+ \epsilon & \epsilon & 4+2 \epsilon+2 \epsilon^2 & 2 \epsilon+2 \epsilon^2 & 4 & 4+ \epsilon \\
         \hline
         C_{FP} & 4+ \epsilon & \epsilon & 4+2 \epsilon+\frac{3 \epsilon^2}{2} & 2 \epsilon+\frac{3 \epsilon^2}{2} & 4 & 4+ \epsilon \\
         C_{BP} & 2 \epsilon & \epsilon & 3 \epsilon+2 \epsilon^2 & 2 \epsilon+2 \epsilon^2 & \epsilon & 2 \epsilon \\
         \hline
    \end{tabular}
    \caption{Multipolar susceptibilities at different fixed points. `F.P.' stands for the fixed points. $\gamma_{i}$ stands for the multipolar susceptibility exponent for the multipolar moment in $i$-irrep. The exponents are calculated up to $\epsilon^{2}$ order. For the case when the bosonic bath coupling is nonzero at a particular fixed point, the exponent is exactly $\epsilon$ to all orders. We note that, when the leading contribution to the susceptibility exponent is not proportional to a power of $\epsilon$, then the perturbative estimation is not controlled by it. The leading nonperturbative contribution may therefore be different for these cases.}
    \label{tab:susceptibility}
\end{table*}

\subsection{Elastic free energy}
Although the dipolar susceptibilities can be detected by conventional techniques, the purely multipolar susceptibilities require elastic measurements. One way to achieve this is through ultrasound experiments. The symmetry-allowed free energy produces a linear coupling between strain and quadrupolar moments, which facilitates a relationship between elastic constants and quadrupolar susceptibilities. Furthermore, in the presence of an external magnetic field, a product of magnetic field and strain couples linearly to octupolar moments, adding octupolar susceptibility corrections to the elastic constants. The symmetry allowed elastic energy in the presence of a magnetic field $\mathbf{h} = (0,0,h_z)$ is then \cite{Patri2019d_SM,Sorensen2021_SM,Han2022_SM}
\begin{align}
\mathcal{F} ={}& \frac{C_{11}^{0} - C_{12}^{0}}{2}(\epsilon_{\mu}^2 + \epsilon_{\nu}^2) + \frac{C_{44}^{0}}{2}(\epsilon_{xy}^2 + \epsilon_{yz}^2 + \epsilon_{zx}^2) + \frac{C_{11}^{0}+2C_{12}^{0}}{2}\epsilon_{B}^{2}\notag\\
&-s_{E}(\epsilon_{\mu}\mathcal{O}_{22}+\epsilon_{\nu}\mathcal{O}_{20})-s_{+}(\epsilon_{xy}\mathcal{O}_{xy}+\epsilon_{yz}\mathcal{O}_{yz}+\epsilon_{zx}\mathcal{O}_{zx})\notag\notag\\
{}&-s_{A}\mathcal{T}_{xyz}(h_{x}\epsilon_{yz}+h_{y}\epsilon_{zx}+h_{z}\epsilon_{xy})-s_{-}(\mathcal{T}^{\beta}_{x}h_{x}(\epsilon_{yy}-\epsilon_{zz})+\mathcal{T}^{\beta}_{y}h_{y}(\epsilon_{zz}-\epsilon_{xx})+\mathcal{T}^{\beta}_{z}h_{z}(\epsilon_{xx}-\epsilon_{yy})),
\end{align}
where
$\epsilon_{\nu}\equiv(2\epsilon_{zz}-\epsilon_{xx}-\epsilon_{yy})/\sqrt{6}$, $\epsilon_{\mu}\equiv(\epsilon_{xx}-\epsilon_{yy})/\sqrt{2}$, and $\epsilon_{B}=(\epsilon_{xx}+\epsilon_{yy}+\epsilon_{zz})/\sqrt{3}$, $\epsilon_{ij}$ is the elastic tensor, and the different $s$ parameters are the couplings strengths between multipolar moments and the external fields. Under the assumption that we apply the magnetic field along the $z$-direction and that the field is sufficiently small such that the cubic symmetry is negligibly affected, by using the second-order perturbation theory, we can obtain the renormalized elastic constants in the main text as follows:
\begin{align}
    (C_{11}-C_{12})={}&(C_{11}^{0}-C_{12}^{0})-(s_{E}^{2})\chi'_{Q_E}-2(s_{-}^{2}h_{z}^{2})\chi_{O_{2}}',\\
    C_{44}={}&C_{44}^{0}-(s_{+}^{2})\chi'_{Q_{2}}-(s_{A}^{2}h_{z}^{2})\chi'_{O_A},
\end{align}
where $h_{z}$ is the magnetic field along $z$-direction, and $\chi_{Q_{E}}',\chi_{Q_{2}}',\chi_{O_{2}}',\chi_{O_{A}}'$ are the multipolar susceptibilities for quadrupolar moments in $E$ and $T_{2+}$ irreps., and octupolar moments in $T_{2-}$ and $A$ irreps., respectively. 


\begin{thebibliography}{72}%
\makeatletter
\providecommand \@ifxundefined [1]{%
 \@ifx{#1\undefined}
}%
\providecommand \@ifnum [1]{%
 \ifnum #1\expandafter \@firstoftwo
 \else \expandafter \@secondoftwo
 \fi
}%
\providecommand \@ifx [1]{%
 \ifx #1\expandafter \@firstoftwo
 \else \expandafter \@secondoftwo
 \fi
}%
\providecommand \natexlab [1]{#1}%
\providecommand \enquote  [1]{``#1''}%
\providecommand \bibnamefont  [1]{#1}%
\providecommand \bibfnamefont [1]{#1}%
\providecommand \citenamefont [1]{#1}%
\providecommand \href@noop [0]{\@secondoftwo}%
\providecommand \href [0]{\begingroup \@sanitize@url \@href}%
\providecommand \@href[1]{\@@startlink{#1}\@@href}%
\providecommand \@@href[1]{\endgroup#1\@@endlink}%
\providecommand \@sanitize@url [0]{\catcode `\\12\catcode `\$12\catcode
  `\&12\catcode `\#12\catcode `\^12\catcode `\_12\catcode `\%12\relax}%
\providecommand \@@startlink[1]{}%
\providecommand \@@endlink[0]{}%
\providecommand \url  [0]{\begingroup\@sanitize@url \@url }%
\providecommand \@url [1]{\endgroup\@href {#1}{\urlprefix }}%
\providecommand \urlprefix  [0]{URL }%
\providecommand \Eprint [0]{\href }%
\providecommand \doibase [0]{http://dx.doi.org/}%
\providecommand \selectlanguage [0]{\@gobble}%
\providecommand \bibinfo  [0]{\@secondoftwo}%
\providecommand \bibfield  [0]{\@secondoftwo}%
\providecommand \translation [1]{[#1]}%
\providecommand \BibitemOpen [0]{}%
\providecommand \bibitemStop [0]{}%
\providecommand \bibitemNoStop [0]{.\EOS\space}%
\providecommand \EOS [0]{\spacefactor3000\relax}%
\providecommand \BibitemShut  [1]{\csname bibitem#1\endcsname}%
\let\auto@bib@innerbib\@empty
\bibitem [{\citenamefont {Doniach}(1977)}]{Doniach1977a}%
  \BibitemOpen
  \bibfield  {author} {\bibinfo {author} {\bibfnamefont {S.}~\bibnamefont
  {Doniach}},\ }\href {\doibase 10.1016/0378-4363(77)90190-5} {\bibfield
  {journal} {\bibinfo  {journal} {Physica B+C}\ }\textbf {\bibinfo {volume}
  {91}},\ \bibinfo {pages} {231} (\bibinfo {year} {1977})}\BibitemShut
  {NoStop}%
\bibitem [{\citenamefont {Coleman}(1983)}]{Coleman1983}%
  \BibitemOpen
  \bibfield  {author} {\bibinfo {author} {\bibfnamefont {P.}~\bibnamefont
  {Coleman}},\ }\href {\doibase 10.1103/PhysRevB.28.5255} {\bibfield  {journal}
  {\bibinfo  {journal} {Physical Review B}\ }\textbf {\bibinfo {volume} {28}},\
  \bibinfo {pages} {5255} (\bibinfo {year} {1983})}\BibitemShut {NoStop}%
\bibitem [{\citenamefont {Schr{\"{o}}der}\ \emph {et~al.}(2000)\citenamefont
  {Schr{\"{o}}der}, \citenamefont {Aeppli}, \citenamefont {Coldea},
  \citenamefont {Adams}, \citenamefont {Stockert}, \citenamefont
  {L{\"{o}}hneysen}, \citenamefont {Bucher}, \citenamefont {Ramazashvili},\
  and\ \citenamefont {Coleman}}]{Schroder2000}%
  \BibitemOpen
  \bibfield  {author} {\bibinfo {author} {\bibfnamefont {A.}~\bibnamefont
  {Schr{\"{o}}der}}, \bibinfo {author} {\bibfnamefont {G.}~\bibnamefont
  {Aeppli}}, \bibinfo {author} {\bibfnamefont {R.}~\bibnamefont {Coldea}},
  \bibinfo {author} {\bibfnamefont {M.}~\bibnamefont {Adams}}, \bibinfo
  {author} {\bibfnamefont {O.}~\bibnamefont {Stockert}}, \bibinfo {author}
  {\bibfnamefont {H.}~\bibnamefont {L{\"{o}}hneysen}}, \bibinfo {author}
  {\bibfnamefont {E.}~\bibnamefont {Bucher}}, \bibinfo {author} {\bibfnamefont
  {R.}~\bibnamefont {Ramazashvili}}, \ and\ \bibinfo {author} {\bibfnamefont
  {P.}~\bibnamefont {Coleman}},\ }\href {\doibase 10.1038/35030039} {\bibfield
  {journal} {\bibinfo  {journal} {Nature}\ }\textbf {\bibinfo {volume} {407}},\
  \bibinfo {pages} {351} (\bibinfo {year} {2000})}\BibitemShut {NoStop}%
\bibitem [{\citenamefont {Aynajian}\ \emph {et~al.}(2012)\citenamefont
  {Aynajian}, \citenamefont {{Da Silva Neto}}, \citenamefont {Gyenis},
  \citenamefont {Baumbach}, \citenamefont {Thompson}, \citenamefont {Fisk},
  \citenamefont {Bauer},\ and\ \citenamefont {Yazdani}}]{Aynajian2012}%
  \BibitemOpen
  \bibfield  {author} {\bibinfo {author} {\bibfnamefont {P.}~\bibnamefont
  {Aynajian}}, \bibinfo {author} {\bibfnamefont {E.~H.}\ \bibnamefont {{Da
  Silva Neto}}}, \bibinfo {author} {\bibfnamefont {A.}~\bibnamefont {Gyenis}},
  \bibinfo {author} {\bibfnamefont {R.~E.}\ \bibnamefont {Baumbach}}, \bibinfo
  {author} {\bibfnamefont {J.~D.}\ \bibnamefont {Thompson}}, \bibinfo {author}
  {\bibfnamefont {Z.}~\bibnamefont {Fisk}}, \bibinfo {author} {\bibfnamefont
  {E.~D.}\ \bibnamefont {Bauer}}, \ and\ \bibinfo {author} {\bibfnamefont
  {A.}~\bibnamefont {Yazdani}},\ }\href {\doibase 10.1038/nature11204}
  {\bibfield  {journal} {\bibinfo  {journal} {Nature}\ }\textbf {\bibinfo
  {volume} {486}},\ \bibinfo {pages} {201} (\bibinfo {year}
  {2012})}\BibitemShut {NoStop}%
\bibitem [{\citenamefont {Yang}\ \emph {et~al.}(2017)\citenamefont {Yang},
  \citenamefont {Pines},\ and\ \citenamefont {Lonzarich}}]{Yang2017}%
  \BibitemOpen
  \bibfield  {author} {\bibinfo {author} {\bibfnamefont {Y.~F.}\ \bibnamefont
  {Yang}}, \bibinfo {author} {\bibfnamefont {D.}~\bibnamefont {Pines}}, \ and\
  \bibinfo {author} {\bibfnamefont {G.}~\bibnamefont {Lonzarich}},\ }\href
  {\doibase 10.1073/pnas.1703172114} {\bibfield  {journal} {\bibinfo  {journal}
  {Proceedings of the National Academy of Sciences of the United States of
  America}\ }\textbf {\bibinfo {volume} {114}},\ \bibinfo {pages} {6250}
  (\bibinfo {year} {2017})}\BibitemShut {NoStop}%
\bibitem [{\citenamefont {Kumar}\ \emph {et~al.}(2022)\citenamefont {Kumar},
  \citenamefont {Hu}, \citenamefont {MacDonald},\ and\ \citenamefont
  {Potter}}]{Kumar2021}%
  \BibitemOpen
  \bibfield  {author} {\bibinfo {author} {\bibfnamefont {A.}~\bibnamefont
  {Kumar}}, \bibinfo {author} {\bibfnamefont {N.~C.}\ \bibnamefont {Hu}},
  \bibinfo {author} {\bibfnamefont {A.~H.}\ \bibnamefont {MacDonald}}, \ and\
  \bibinfo {author} {\bibfnamefont {A.~C.}\ \bibnamefont {Potter}},\ }\href
  {\doibase 10.1103/PhysRevB.106.L041116} {\bibfield  {journal} {\bibinfo
  {journal} {Phys. Rev. B}\ }\textbf {\bibinfo {volume} {106}},\ \bibinfo
  {pages} {L041116} (\bibinfo {year} {2022})}\BibitemShut {NoStop}%
\bibitem [{\citenamefont {Si}\ and\ \citenamefont {Steglich}(2010)}]{Si2010a}%
  \BibitemOpen
  \bibfield  {author} {\bibinfo {author} {\bibfnamefont {Q.}~\bibnamefont
  {Si}}\ and\ \bibinfo {author} {\bibfnamefont {F.}~\bibnamefont {Steglich}},\
  }\href {\doibase 10.1126/science.1191195} {\bibfield  {journal} {\bibinfo
  {journal} {Science}\ }\textbf {\bibinfo {volume} {329}},\ \bibinfo {pages}
  {1161} (\bibinfo {year} {2010})}\BibitemShut {NoStop}%
\bibitem [{\citenamefont {Maksimovic}\ \emph {et~al.}(2022)\citenamefont
  {Maksimovic}, \citenamefont {Eilbott}, \citenamefont {Cookmeyer},
  \citenamefont {Wan}, \citenamefont {Rusz}, \citenamefont {Nagarajan},
  \citenamefont {Haley}, \citenamefont {Maniv}, \citenamefont {Gong},
  \citenamefont {Faubel}, \citenamefont {Hayes}, \citenamefont {Bangura},
  \citenamefont {Singleton}, \citenamefont {Palmstrom}, \citenamefont {Winter},
  \citenamefont {McDonald}, \citenamefont {Jang}, \citenamefont {Ai},
  \citenamefont {Lin}, \citenamefont {Ciocys}, \citenamefont {Gobbo},
  \citenamefont {Werman}, \citenamefont {Oppeneer}, \citenamefont {Altman},
  \citenamefont {Lanzara},\ and\ \citenamefont {Analytis}}]{Maksimovic2022}%
  \BibitemOpen
  \bibfield  {author} {\bibinfo {author} {\bibfnamefont {N.}~\bibnamefont
  {Maksimovic}}, \bibinfo {author} {\bibfnamefont {D.~H.}\ \bibnamefont
  {Eilbott}}, \bibinfo {author} {\bibfnamefont {T.}~\bibnamefont {Cookmeyer}},
  \bibinfo {author} {\bibfnamefont {F.}~\bibnamefont {Wan}}, \bibinfo {author}
  {\bibfnamefont {J.}~\bibnamefont {Rusz}}, \bibinfo {author} {\bibfnamefont
  {V.}~\bibnamefont {Nagarajan}}, \bibinfo {author} {\bibfnamefont {S.~C.}\
  \bibnamefont {Haley}}, \bibinfo {author} {\bibfnamefont {E.}~\bibnamefont
  {Maniv}}, \bibinfo {author} {\bibfnamefont {A.}~\bibnamefont {Gong}},
  \bibinfo {author} {\bibfnamefont {S.}~\bibnamefont {Faubel}}, \bibinfo
  {author} {\bibfnamefont {I.~M.}\ \bibnamefont {Hayes}}, \bibinfo {author}
  {\bibfnamefont {A.}~\bibnamefont {Bangura}}, \bibinfo {author} {\bibfnamefont
  {J.}~\bibnamefont {Singleton}}, \bibinfo {author} {\bibfnamefont {J.~C.}\
  \bibnamefont {Palmstrom}}, \bibinfo {author} {\bibfnamefont {L.}~\bibnamefont
  {Winter}}, \bibinfo {author} {\bibfnamefont {R.}~\bibnamefont {McDonald}},
  \bibinfo {author} {\bibfnamefont {S.}~\bibnamefont {Jang}}, \bibinfo {author}
  {\bibfnamefont {P.}~\bibnamefont {Ai}}, \bibinfo {author} {\bibfnamefont
  {Y.}~\bibnamefont {Lin}}, \bibinfo {author} {\bibfnamefont {S.}~\bibnamefont
  {Ciocys}}, \bibinfo {author} {\bibfnamefont {J.}~\bibnamefont {Gobbo}},
  \bibinfo {author} {\bibfnamefont {Y.}~\bibnamefont {Werman}}, \bibinfo
  {author} {\bibfnamefont {P.~M.}\ \bibnamefont {Oppeneer}}, \bibinfo {author}
  {\bibfnamefont {E.}~\bibnamefont {Altman}}, \bibinfo {author} {\bibfnamefont
  {A.}~\bibnamefont {Lanzara}}, \ and\ \bibinfo {author} {\bibfnamefont
  {J.~G.}\ \bibnamefont {Analytis}},\ }\href {\doibase 10.1126/science.aaz4566}
  {\bibfield  {journal} {\bibinfo  {journal} {Science}\ }\textbf {\bibinfo
  {volume} {375}},\ \bibinfo {pages} {76} (\bibinfo {year} {2022})}\BibitemShut
  {NoStop}%
\bibitem [{\citenamefont {Prokle\ifmmode~\check{s}\else \v{s}\fi{}ka}\ \emph
  {et~al.}(2015)\citenamefont {Prokle\ifmmode~\check{s}\else \v{s}\fi{}ka},
  \citenamefont {Kratochv\'{\i}lov\'a}, \citenamefont
  {Uhl\'{\i}\ifmmode~\check{r}\else \v{r}\fi{}ov\'a}, \citenamefont
  {Sechovsk\'y},\ and\ \citenamefont {Custers}}]{Prokleska2015}%
  \BibitemOpen
  \bibfield  {author} {\bibinfo {author} {\bibfnamefont {J.}~\bibnamefont
  {Prokle\ifmmode~\check{s}\else \v{s}\fi{}ka}}, \bibinfo {author}
  {\bibfnamefont {M.}~\bibnamefont {Kratochv\'{\i}lov\'a}}, \bibinfo {author}
  {\bibfnamefont {K.}~\bibnamefont {Uhl\'{\i}\ifmmode~\check{r}\else
  \v{r}\fi{}ov\'a}}, \bibinfo {author} {\bibfnamefont {V.}~\bibnamefont
  {Sechovsk\'y}}, \ and\ \bibinfo {author} {\bibfnamefont {J.}~\bibnamefont
  {Custers}},\ }\href {\doibase 10.1103/PhysRevB.92.161114} {\bibfield
  {journal} {\bibinfo  {journal} {Physical Review B}\ }\textbf {\bibinfo
  {volume} {92}},\ \bibinfo {pages} {161114(R)} (\bibinfo {year}
  {2015})}\BibitemShut {NoStop}%
\bibitem [{\citenamefont {Inui}\ and\ \citenamefont {Motome}(2020)}]{Inui2020}%
  \BibitemOpen
  \bibfield  {author} {\bibinfo {author} {\bibfnamefont {K.}~\bibnamefont
  {Inui}}\ and\ \bibinfo {author} {\bibfnamefont {Y.}~\bibnamefont {Motome}},\
  }\href {\doibase 10.1103/PhysRevB.102.155126} {\bibfield  {journal} {\bibinfo
   {journal} {Physical Review B}\ }\textbf {\bibinfo {volume} {102}},\ \bibinfo
  {pages} {155126} (\bibinfo {year} {2020})}\BibitemShut {NoStop}%
\bibitem [{\citenamefont {Knebel}\ \emph {et~al.}(2011)\citenamefont {Knebel},
  \citenamefont {Aoki},\ and\ \citenamefont {Flouquet}}]{Knebel2011}%
  \BibitemOpen
  \bibfield  {author} {\bibinfo {author} {\bibfnamefont {G.}~\bibnamefont
  {Knebel}}, \bibinfo {author} {\bibfnamefont {D.}~\bibnamefont {Aoki}}, \ and\
  \bibinfo {author} {\bibfnamefont {J.}~\bibnamefont {Flouquet}},\ }\href
  {\doibase 10.1016/j.crhy.2011.05.002} {\bibfield  {journal} {\bibinfo
  {journal} {Comptes Rendus Physique}\ }\textbf {\bibinfo {volume} {12}},\
  \bibinfo {pages} {542} (\bibinfo {year} {2011})}\BibitemShut {NoStop}%
\bibitem [{\citenamefont {Seiro}\ \emph {et~al.}(2018)\citenamefont {Seiro},
  \citenamefont {Jiao}, \citenamefont {Kirchner}, \citenamefont {Hartmann},
  \citenamefont {Friedemann}, \citenamefont {Krellner}, \citenamefont {Geibel},
  \citenamefont {Si}, \citenamefont {Steglich},\ and\ \citenamefont
  {Wirth}}]{Seiro2018}%
  \BibitemOpen
  \bibfield  {author} {\bibinfo {author} {\bibfnamefont {S.}~\bibnamefont
  {Seiro}}, \bibinfo {author} {\bibfnamefont {L.}~\bibnamefont {Jiao}},
  \bibinfo {author} {\bibfnamefont {S.}~\bibnamefont {Kirchner}}, \bibinfo
  {author} {\bibfnamefont {S.}~\bibnamefont {Hartmann}}, \bibinfo {author}
  {\bibfnamefont {S.}~\bibnamefont {Friedemann}}, \bibinfo {author}
  {\bibfnamefont {C.}~\bibnamefont {Krellner}}, \bibinfo {author}
  {\bibfnamefont {C.}~\bibnamefont {Geibel}}, \bibinfo {author} {\bibfnamefont
  {Q.}~\bibnamefont {Si}}, \bibinfo {author} {\bibfnamefont {F.}~\bibnamefont
  {Steglich}}, \ and\ \bibinfo {author} {\bibfnamefont {S.}~\bibnamefont
  {Wirth}},\ }\href {\doibase 10.1038/s41467-018-05801-5} {\bibfield  {journal}
  {\bibinfo  {journal} {Nature Communications}\ }\textbf {\bibinfo {volume}
  {9}},\ \bibinfo {pages} {3324} (\bibinfo {year} {2018})}\BibitemShut {NoStop}%
\bibitem [{\citenamefont {Oh}\ \emph {et~al.}(2019)\citenamefont {Oh},
  \citenamefont {Lee}, \citenamefont {Kim},\ and\ \citenamefont
  {Moon}}]{Oh2019}%
  \BibitemOpen
  \bibfield  {author} {\bibinfo {author} {\bibfnamefont {H.}~\bibnamefont
  {Oh}}, \bibinfo {author} {\bibfnamefont {S.}~\bibnamefont {Lee}}, \bibinfo
  {author} {\bibfnamefont {Y.~B.}\ \bibnamefont {Kim}}, \ and\ \bibinfo
  {author} {\bibfnamefont {E.~G.}\ \bibnamefont {Moon}},\ }\href {\doibase 10.1103/PhysRevLett.122.167201} {\bibfield  {journal} {\bibinfo  {journal}
  {Physical Review Letters}\ }\textbf {\bibinfo {volume} {122}},\ \bibinfo
  {pages} {167201} (\bibinfo {year} {2019})}\BibitemShut {NoStop}%
\bibitem [{\citenamefont {Fuhrman}\ \emph {et~al.}(2021)\citenamefont
  {Fuhrman}, \citenamefont {Sidorenko}, \citenamefont {H{\"{a}}nel},
  \citenamefont {Winkler}, \citenamefont {Prokofiev}, \citenamefont
  {Rodriguez-Rivera}, \citenamefont {Qiu}, \citenamefont {Blaha}, \citenamefont
  {Si}, \citenamefont {Broholm},\ and\ \citenamefont {Paschen}}]{Fuhrman2021}%
  \BibitemOpen
  \bibfield  {author} {\bibinfo {author} {\bibfnamefont {W.~T.}\ \bibnamefont
  {Fuhrman}}, \bibinfo {author} {\bibfnamefont {A.}~\bibnamefont {Sidorenko}},
  \bibinfo {author} {\bibfnamefont {J.}~\bibnamefont {H{\"{a}}nel}}, \bibinfo
  {author} {\bibfnamefont {H.}~\bibnamefont {Winkler}}, \bibinfo {author}
  {\bibfnamefont {A.}~\bibnamefont {Prokofiev}}, \bibinfo {author}
  {\bibfnamefont {J.~A.}\ \bibnamefont {Rodriguez-Rivera}}, \bibinfo {author}
  {\bibfnamefont {Y.}~\bibnamefont {Qiu}}, \bibinfo {author} {\bibfnamefont
  {P.}~\bibnamefont {Blaha}}, \bibinfo {author} {\bibfnamefont
  {Q.}~\bibnamefont {Si}}, \bibinfo {author} {\bibfnamefont {C.~L.}\
  \bibnamefont {Broholm}}, \ and\ \bibinfo {author} {\bibfnamefont
  {S.}~\bibnamefont {Paschen}},\ }\href {\doibase 10.1126/sciadv.abf9134}
  {\bibfield  {journal} {\bibinfo  {journal} {Science Advances}\ }\textbf
  {\bibinfo {volume} {7}},\ \bibinfo {pages} {eabf9134} (\bibinfo {year}
  {2021})}\BibitemShut {NoStop}%
\bibitem [{\citenamefont {Sorensen}\ and\ \citenamefont
  {Fisher}(2021)}]{Sorensen2021}%
  \BibitemOpen
  \bibfield  {author} {\bibinfo {author} {\bibfnamefont {M.~E.}\ \bibnamefont
  {Sorensen}}\ and\ \bibinfo {author} {\bibfnamefont {I.~R.}\ \bibnamefont
  {Fisher}},\ }\href {\doibase 10.1103/PhysRevB.103.155106} {\bibfield
  {journal} {\bibinfo  {journal} {Physical Review B}\ }\textbf {\bibinfo
  {volume} {103}},\ \bibinfo {pages} {155106} (\bibinfo {year}
  {2021})}\BibitemShut {NoStop}%
\bibitem [{\citenamefont {Thalmeier}\ \emph {et~al.}(2008)\citenamefont
  {Thalmeier}, \citenamefont {Takimoto}, \citenamefont {Chang},\ and\
  \citenamefont {Eremin}}]{Thalmeier2008}%
  \BibitemOpen
  \bibfield  {author} {\bibinfo {author} {\bibfnamefont {P.}~\bibnamefont
  {Thalmeier}}, \bibinfo {author} {\bibfnamefont {T.}~\bibnamefont {Takimoto}},
  \bibinfo {author} {\bibfnamefont {J.}~\bibnamefont {Chang}}, \ and\ \bibinfo
  {author} {\bibfnamefont {I.}~\bibnamefont {Eremin}},\ }\href {\doibase 10.1143/JPSJS.77SA.43} {\bibfield  {journal} {\bibinfo  {journal} {Journal of
  the Physical Society of Japan}\ }\textbf {\bibinfo {volume} {77}},\ \bibinfo
  {pages} {43} (\bibinfo {year} {2008})}\BibitemShut {NoStop}%
\bibitem [{\citenamefont {Koitzsch}\ \emph {et~al.}(2016)\citenamefont
  {Koitzsch}, \citenamefont {Heming}, \citenamefont {Knupfer}, \citenamefont
  {B{\"{u}}chner}, \citenamefont {Portnichenko}, \citenamefont {Dukhnenko},
  \citenamefont {Shitsevalova}, \citenamefont {Filipov}, \citenamefont {Lev},
  \citenamefont {Strocov}, \citenamefont {Ollivier},\ and\ \citenamefont
  {Inosov}}]{Koitzsch2016}%
  \BibitemOpen
  \bibfield  {author} {\bibinfo {author} {\bibfnamefont {A.}~\bibnamefont
  {Koitzsch}}, \bibinfo {author} {\bibfnamefont {N.}~\bibnamefont {Heming}},
  \bibinfo {author} {\bibfnamefont {M.}~\bibnamefont {Knupfer}}, \bibinfo
  {author} {\bibfnamefont {B.}~\bibnamefont {B{\"{u}}chner}}, \bibinfo {author}
  {\bibfnamefont {P.~Y.}\ \bibnamefont {Portnichenko}}, \bibinfo {author}
  {\bibfnamefont {A.~V.}\ \bibnamefont {Dukhnenko}}, \bibinfo {author}
  {\bibfnamefont {N.~Y.}\ \bibnamefont {Shitsevalova}}, \bibinfo {author}
  {\bibfnamefont {V.~B.}\ \bibnamefont {Filipov}}, \bibinfo {author}
  {\bibfnamefont {L.~L.}\ \bibnamefont {Lev}}, \bibinfo {author} {\bibfnamefont
  {V.~N.}\ \bibnamefont {Strocov}}, \bibinfo {author} {\bibfnamefont
  {J.}~\bibnamefont {Ollivier}}, \ and\ \bibinfo {author} {\bibfnamefont
  {D.~S.}\ \bibnamefont {Inosov}},\ }\href {\doibase 10.1038/ncomms10876}
  {\bibfield  {journal} {\bibinfo  {journal} {Nature Communications}\ }\textbf
  {\bibinfo {volume} {7}},\ \bibinfo {pages} {10876} (\bibinfo {year}
  {2016})}\BibitemShut {NoStop}%
\bibitem [{\citenamefont {Barman}\ \emph {et~al.}(2019)\citenamefont {Barman},
  \citenamefont {Singh}, \citenamefont {Johnson},\ and\ \citenamefont
  {Alam}}]{Barman2019a}%
  \BibitemOpen
  \bibfield  {author} {\bibinfo {author} {\bibfnamefont {C.~K.}\ \bibnamefont
  {Barman}}, \bibinfo {author} {\bibfnamefont {P.}~\bibnamefont {Singh}},
  \bibinfo {author} {\bibfnamefont {D.~D.}\ \bibnamefont {Johnson}}, \ and\
  \bibinfo {author} {\bibfnamefont {A.}~\bibnamefont {Alam}},\ }\href {\doibase 10.1103/PhysRevLett.122.076401} {\bibfield  {journal} {\bibinfo  {journal}
  {Physical Review Letters}\ }\textbf {\bibinfo {volume} {122}},\ \bibinfo
  {pages} {076401} (\bibinfo {year} {2019})}\BibitemShut {NoStop}%
\bibitem [{\citenamefont {Thalmeier}\ \emph {et~al.}(2014)\citenamefont
  {Thalmeier}, \citenamefont {Takimoto},\ and\ \citenamefont
  {Ikeda}}]{Thalmeier2014}%
  \BibitemOpen
  \bibfield  {author} {\bibinfo {author} {\bibfnamefont {P.}~\bibnamefont
  {Thalmeier}}, \bibinfo {author} {\bibfnamefont {T.}~\bibnamefont {Takimoto}},
  \ and\ \bibinfo {author} {\bibfnamefont {H.}~\bibnamefont {Ikeda}},\ }\href
  {\doibase 10.1080/14786435.2013.861615} {\bibfield  {journal} {\bibinfo
  {journal} {Philosophical Magazine}\ }\textbf {\bibinfo {volume} {94}},\
  \bibinfo {pages} {3863} (\bibinfo {year} {2014})}\BibitemShut {NoStop}%
\bibitem [{\citenamefont {Patri}\ \emph {et~al.}(2020)\citenamefont {Patri},
  \citenamefont {Khait},\ and\ \citenamefont {Kim}}]{Patri2020d}%
  \BibitemOpen
  \bibfield  {author} {\bibinfo {author} {\bibfnamefont {A.~S.}\ \bibnamefont
  {Patri}}, \bibinfo {author} {\bibfnamefont {I.}~\bibnamefont {Khait}}, \ and\
  \bibinfo {author} {\bibfnamefont {Y.~B.}\ \bibnamefont {Kim}},\ }\href
  {\doibase 10.1103/PhysRevResearch.2.013257} {\bibfield  {journal} {\bibinfo
  {journal} {Physical Review Research}\ }\textbf {\bibinfo {volume} {2}},\
  \bibinfo {pages} {013257} (\bibinfo {year} {2020})}\BibitemShut {NoStop}%
\bibitem [{\citenamefont {Patri}\ and\ \citenamefont {Kim}(2020)}]{Patri2020e}%
  \BibitemOpen
  \bibfield  {author} {\bibinfo {author} {\bibfnamefont {A.~S.}\ \bibnamefont
  {Patri}}\ and\ \bibinfo {author} {\bibfnamefont {Y.~B.}\ \bibnamefont
  {Kim}},\ }\href {\doibase 10.1103/PhysRevX.10.041021} {\bibfield  {journal}
  {\bibinfo  {journal} {Physical Review X}\ }\textbf {\bibinfo {volume} {10}},\
  \bibinfo {pages} {041021} (\bibinfo {year} {2020})}\BibitemShut {NoStop}%
\bibitem [{\citenamefont {Schultz}\ \emph
  {et~al.}(2021{\natexlab{a}})\citenamefont {Schultz}, \citenamefont {Patri},\
  and\ \citenamefont {Kim}}]{Schultz2021b}%
  \BibitemOpen
  \bibfield  {author} {\bibinfo {author} {\bibfnamefont {D.~J.}\ \bibnamefont
  {Schultz}}, \bibinfo {author} {\bibfnamefont {A.~S.}\ \bibnamefont {Patri}},
  \ and\ \bibinfo {author} {\bibfnamefont {Y.~B.}\ \bibnamefont {Kim}},\ }\href
  {\doibase 10.1103/PhysRevResearch.3.013189} {\bibfield  {journal} {\bibinfo
  {journal} {Physical Review Research}\ }\textbf {\bibinfo {volume} {3}},\
  \bibinfo {pages} {013189} (\bibinfo {year} {2021}{\natexlab{a}})}\BibitemShut
  {NoStop}%
\bibitem [{\citenamefont {Sakai}\ and\ \citenamefont
  {Nakatsuji}(2011)}]{Sakai2011b}%
  \BibitemOpen
  \bibfield  {author} {\bibinfo {author} {\bibfnamefont {A.}~\bibnamefont
  {Sakai}}\ and\ \bibinfo {author} {\bibfnamefont {S.}~\bibnamefont
  {Nakatsuji}},\ }\href {\doibase 10.1143/JPSJ.80.063701} {\bibfield  {journal}
  {\bibinfo  {journal} {Journal of the Physical Society of Japan}\ }\textbf
  {\bibinfo {volume} {80}},\ \bibinfo {pages} {063701} (\bibinfo {year}
  {2011})}\BibitemShut {NoStop}%
\bibitem [{\citenamefont {Thalmeier}\ \emph {et~al.}(2021)\citenamefont
  {Thalmeier}, \citenamefont {Akbari},\ and\ \citenamefont
  {Shiina}}]{Thalmeier2021}%
  \BibitemOpen
  \bibfield  {author} {\bibinfo {author} {\bibfnamefont {P.}~\bibnamefont
  {Thalmeier}}, \bibinfo {author} {\bibfnamefont {A.}~\bibnamefont {Akbari}}, \
  and\ \bibinfo {author} {\bibfnamefont {R.}~\bibnamefont {Shiina}},\ }\href
  {\doibase 10.1201/9781003146483-8} {\bibfield  {journal} {\bibinfo  {journal}
  {Rare-Earth Borides}\ ,\ \bibinfo {pages} {615}} (\bibinfo {year}
  {2021})}\BibitemShut {NoStop}%
\bibitem [{\citenamefont {Iwasa}\ \emph {et~al.}(2018)\citenamefont {Iwasa},
  \citenamefont {Onimaru}, \citenamefont {Takabatake}, \citenamefont
  {Higashinaka}, \citenamefont {Aoki}, \citenamefont {Ohira-Kawamura},\ and\
  \citenamefont {Nakajima}}]{Iwasa2018}%
  \BibitemOpen
  \bibfield  {author} {\bibinfo {author} {\bibfnamefont {K.}~\bibnamefont
  {Iwasa}}, \bibinfo {author} {\bibfnamefont {T.}~\bibnamefont {Onimaru}},
  \bibinfo {author} {\bibfnamefont {T.}~\bibnamefont {Takabatake}}, \bibinfo
  {author} {\bibfnamefont {R.}~\bibnamefont {Higashinaka}}, \bibinfo {author}
  {\bibfnamefont {Y.}~\bibnamefont {Aoki}}, \bibinfo {author} {\bibfnamefont
  {S.}~\bibnamefont {Ohira-Kawamura}}, \ and\ \bibinfo {author} {\bibfnamefont
  {K.}~\bibnamefont {Nakajima}},\ }\href {\doibase 10.1016/j.physb.2017.11.002}
  {\bibfield  {journal} {\bibinfo  {journal} {Physica B: Condensed Matter}\
  }\textbf {\bibinfo {volume} {551}},\ \bibinfo {pages} {37} (\bibinfo {year}
  {2018})}\BibitemShut {NoStop}%
\bibitem [{\citenamefont {Patri}\ and\ \citenamefont {Kim}(2022)}]{Patri2022}%
  \BibitemOpen
  \bibfield  {author} {\bibinfo {author} {\bibfnamefont {A.~S.}\ \bibnamefont
  {Patri}}\ and\ \bibinfo {author} {\bibfnamefont {Y.~B.}\ \bibnamefont
  {Kim}},\ }\href {\doibase 10.21468/SCIPOSTPHYS.12.2.057} {\bibfield
  {journal} {\bibinfo  {journal} {SciPost Physics}\ }\textbf {\bibinfo {volume}
  {12}},\ \bibinfo {pages} {057} (\bibinfo {year} {2022}),\ 10.21468/SCIPOSTPHYS.12.2.057}\BibitemShut
  {NoStop}%
\bibitem [{\citenamefont {Sato}\ \emph {et~al.}(2012)\citenamefont {Sato},
  \citenamefont {Ibuka}, \citenamefont {Nambu}, \citenamefont {Yamazaki},
  \citenamefont {Hong}, \citenamefont {Sakai},\ and\ \citenamefont
  {Nakatsuji}}]{Sato2012a}%
  \BibitemOpen
  \bibfield  {author} {\bibinfo {author} {\bibfnamefont {T.~J.}\ \bibnamefont
  {Sato}}, \bibinfo {author} {\bibfnamefont {S.}~\bibnamefont {Ibuka}},
  \bibinfo {author} {\bibfnamefont {Y.}~\bibnamefont {Nambu}}, \bibinfo
  {author} {\bibfnamefont {T.}~\bibnamefont {Yamazaki}}, \bibinfo {author}
  {\bibfnamefont {T.}~\bibnamefont {Hong}}, \bibinfo {author} {\bibfnamefont
  {A.}~\bibnamefont {Sakai}}, \ and\ \bibinfo {author} {\bibfnamefont
  {S.}~\bibnamefont {Nakatsuji}},\ }\href {\doibase 10.1103/PhysRevB.86.184419}
  {\bibfield  {journal} {\bibinfo  {journal} {Physical Review B - Condensed
  Matter and Materials Physics}\ }\textbf {\bibinfo {volume} {86}},\ \bibinfo
  {pages} {184419} (\bibinfo {year} {2012})}\BibitemShut {NoStop}%
\bibitem [{\citenamefont {Fu}\ \emph {et~al.}(2020)\citenamefont {Fu},
  \citenamefont {Sakai}, \citenamefont {Sogabe}, \citenamefont {Tsujimoto},
  \citenamefont {Matsumoto},\ and\ \citenamefont {Nakatsuji}}]{Fu2020a}%
  \BibitemOpen
  \bibfield  {author} {\bibinfo {author} {\bibfnamefont {M.}~\bibnamefont
  {Fu}}, \bibinfo {author} {\bibfnamefont {A.}~\bibnamefont {Sakai}}, \bibinfo
  {author} {\bibfnamefont {N.}~\bibnamefont {Sogabe}}, \bibinfo {author}
  {\bibfnamefont {M.}~\bibnamefont {Tsujimoto}}, \bibinfo {author}
  {\bibfnamefont {Y.}~\bibnamefont {Matsumoto}}, \ and\ \bibinfo {author}
  {\bibfnamefont {S.}~\bibnamefont {Nakatsuji}},\ }\href {\doibase 10.7566/JPSJ.89.013704} {\bibfield  {journal} {\bibinfo  {journal} {Journal
  of the Physical Society of Japan}\ }\textbf {\bibinfo {volume} {89}},\ \bibinfo {pages} {013704}
  (\bibinfo {year} {2020}),\ 10.7566/JPSJ.89.013704}\BibitemShut {NoStop}%
\bibitem [{\citenamefont {Shimura}\ \emph {et~al.}(2019)\citenamefont
  {Shimura}, \citenamefont {Zhang}, \citenamefont {Zeng}, \citenamefont
  {Rhodes}, \citenamefont {Sch{\"{o}}nemann}, \citenamefont {Tsujimoto},
  \citenamefont {Matsumoto}, \citenamefont {Sakai}, \citenamefont {Sakakibara},
  \citenamefont {Araki}, \citenamefont {Zheng}, \citenamefont {Zhou},
  \citenamefont {Balicas},\ and\ \citenamefont {Nakatsuji}}]{Shimura2019}%
  \BibitemOpen
  \bibfield  {author} {\bibinfo {author} {\bibfnamefont {Y.}~\bibnamefont
  {Shimura}}, \bibinfo {author} {\bibfnamefont {Q.}~\bibnamefont {Zhang}},
  \bibinfo {author} {\bibfnamefont {B.}~\bibnamefont {Zeng}}, \bibinfo {author}
  {\bibfnamefont {D.}~\bibnamefont {Rhodes}}, \bibinfo {author} {\bibfnamefont
  {R.}~\bibnamefont {Sch{\"{o}}nemann}}, \bibinfo {author} {\bibfnamefont
  {M.}~\bibnamefont {Tsujimoto}}, \bibinfo {author} {\bibfnamefont
  {Y.}~\bibnamefont {Matsumoto}}, \bibinfo {author} {\bibfnamefont
  {A.}~\bibnamefont {Sakai}}, \bibinfo {author} {\bibfnamefont
  {T.}~\bibnamefont {Sakakibara}}, \bibinfo {author} {\bibfnamefont
  {K.}~\bibnamefont {Araki}}, \bibinfo {author} {\bibfnamefont
  {W.}~\bibnamefont {Zheng}}, \bibinfo {author} {\bibfnamefont
  {Q.}~\bibnamefont {Zhou}}, \bibinfo {author} {\bibfnamefont {L.}~\bibnamefont
  {Balicas}}, \ and\ \bibinfo {author} {\bibfnamefont {S.}~\bibnamefont
  {Nakatsuji}},\ }\href {\doibase 10.1103/PhysRevLett.122.256601} {\bibfield
  {journal} {\bibinfo  {journal} {Physical Review Letters}\ }\textbf {\bibinfo
  {volume} {122}},\ \bibinfo {pages} {256601} (\bibinfo {year}
  {2019})}\BibitemShut {NoStop}%
\bibitem [{\citenamefont {Matsubayashi}\ \emph {et~al.}(2012)\citenamefont
  {Matsubayashi}, \citenamefont {Tanaka}, \citenamefont {Sakai}, \citenamefont
  {Nakatsuji}, \citenamefont {Kubo},\ and\ \citenamefont
  {Uwatoko}}]{Matsubayashi2012c}%
  \BibitemOpen
  \bibfield  {author} {\bibinfo {author} {\bibfnamefont {K.}~\bibnamefont
  {Matsubayashi}}, \bibinfo {author} {\bibfnamefont {T.}~\bibnamefont
  {Tanaka}}, \bibinfo {author} {\bibfnamefont {A.}~\bibnamefont {Sakai}},
  \bibinfo {author} {\bibfnamefont {S.}~\bibnamefont {Nakatsuji}}, \bibinfo
  {author} {\bibfnamefont {Y.}~\bibnamefont {Kubo}}, \ and\ \bibinfo {author}
  {\bibfnamefont {Y.}~\bibnamefont {Uwatoko}},\ }\href {\doibase 10.1103/PhysRevLett.109.187004} {\bibfield  {journal} {\bibinfo  {journal}
  {Physical Review Letters}\ }\textbf {\bibinfo {volume} {109}},\ \bibinfo
  {pages} {187004} (\bibinfo {year} {2012})}\BibitemShut {NoStop}%
\bibitem [{\citenamefont {Tsujimoto}\ \emph {et~al.}(2014)\citenamefont
  {Tsujimoto}, \citenamefont {Matsumoto}, \citenamefont {Tomita}, \citenamefont
  {Sakai},\ and\ \citenamefont {Nakatsuji}}]{Tsujimoto2014a}%
  \BibitemOpen
  \bibfield  {author} {\bibinfo {author} {\bibfnamefont {M.}~\bibnamefont
  {Tsujimoto}}, \bibinfo {author} {\bibfnamefont {Y.}~\bibnamefont
  {Matsumoto}}, \bibinfo {author} {\bibfnamefont {T.}~\bibnamefont {Tomita}},
  \bibinfo {author} {\bibfnamefont {A.}~\bibnamefont {Sakai}}, \ and\ \bibinfo
  {author} {\bibfnamefont {S.}~\bibnamefont {Nakatsuji}},\ }\href {\doibase 10.1103/PhysRevLett.113.267001} {\bibfield  {journal} {\bibinfo  {journal}
  {Physical Review Letters}\ }\textbf {\bibinfo {volume} {113}},\ \bibinfo
  {pages} {267001} (\bibinfo {year} {2014})}\BibitemShut {NoStop}%
\bibitem [{\citenamefont {Sakai}\ \emph {et~al.}(2012)\citenamefont {Sakai},
  \citenamefont {Kuga},\ and\ \citenamefont {Nakatsuji}}]{Sakai2012a}%
  \BibitemOpen
  \bibfield  {author} {\bibinfo {author} {\bibfnamefont {A.}~\bibnamefont
  {Sakai}}, \bibinfo {author} {\bibfnamefont {K.}~\bibnamefont {Kuga}}, \ and\
  \bibinfo {author} {\bibfnamefont {S.}~\bibnamefont {Nakatsuji}},\ }\href
  {\doibase 10.1143/JPSJ.81.083702} {\bibfield  {journal} {\bibinfo  {journal}
  {Journal of the Physical Society of Japan}\ }\textbf {\bibinfo {volume}
  {81}},\ \bibinfo {pages} {083702} (\bibinfo {year} {2012})}\BibitemShut {NoStop}%
\bibitem [{\citenamefont {Kitagawa}\ \emph {et~al.}(1997)\citenamefont
  {Kitagawa}, \citenamefont {Takeda}, \citenamefont {Ishikawa}, \citenamefont
  {Yoshida}, \citenamefont {Ishiguro},\ and\ \citenamefont
  {Komatsubara}}]{Kitagawa1997}%
  \BibitemOpen
  \bibfield  {author} {\bibinfo {author} {\bibfnamefont {J.}~\bibnamefont
  {Kitagawa}}, \bibinfo {author} {\bibfnamefont {N.}~\bibnamefont {Takeda}},
  \bibinfo {author} {\bibfnamefont {M.}~\bibnamefont {Ishikawa}}, \bibinfo
  {author} {\bibfnamefont {T.}~\bibnamefont {Yoshida}}, \bibinfo {author}
  {\bibfnamefont {A.}~\bibnamefont {Ishiguro}}, \ and\ \bibinfo {author}
  {\bibfnamefont {T.}~\bibnamefont {Komatsubara}},\ }\href {\doibase 10.1016/S0921-4526(96)00571-6} {\bibfield  {journal} {\bibinfo  {journal}
  {Physica B: Condensed Matter}\ }\textbf {\bibinfo {volume} {230-232}},\
  \bibinfo {pages} {139} (\bibinfo {year} {1997})}\BibitemShut {NoStop}%
\bibitem [{\citenamefont {Goto}\ \emph {et~al.}(2009)\citenamefont {Goto},
  \citenamefont {Watanabe}, \citenamefont {Tsuduku}, \citenamefont {Kobayashi},
  \citenamefont {Nemoto}, \citenamefont {Yanagisawa}, \citenamefont {Akatsu},
  \citenamefont {Ano}, \citenamefont {Suzuki}, \citenamefont {Takeda},
  \citenamefont {D{\"{o}}nni},\ and\ \citenamefont {Kitazawa}}]{Goto2009}%
  \BibitemOpen
  \bibfield  {author} {\bibinfo {author} {\bibfnamefont {T.}~\bibnamefont
  {Goto}}, \bibinfo {author} {\bibfnamefont {T.}~\bibnamefont {Watanabe}},
  \bibinfo {author} {\bibfnamefont {S.}~\bibnamefont {Tsuduku}}, \bibinfo
  {author} {\bibfnamefont {H.}~\bibnamefont {Kobayashi}}, \bibinfo {author}
  {\bibfnamefont {Y.}~\bibnamefont {Nemoto}}, \bibinfo {author} {\bibfnamefont
  {T.}~\bibnamefont {Yanagisawa}}, \bibinfo {author} {\bibfnamefont
  {M.}~\bibnamefont {Akatsu}}, \bibinfo {author} {\bibfnamefont
  {G.}~\bibnamefont {Ano}}, \bibinfo {author} {\bibfnamefont {O.}~\bibnamefont
  {Suzuki}}, \bibinfo {author} {\bibfnamefont {N.}~\bibnamefont {Takeda}},
  \bibinfo {author} {\bibfnamefont {A.}~\bibnamefont {D{\"{o}}nni}}, \ and\
  \bibinfo {author} {\bibfnamefont {H.}~\bibnamefont {Kitazawa}},\ }\href
  {\doibase 10.1143/JPSJ.78.024716} {\bibfield  {journal} {\bibinfo  {journal}
  {Journal of the Physical Society of Japan}\ }\textbf {\bibinfo {volume}
  {78}},\ \bibinfo {pages} {024716} (\bibinfo {year} {2009})}\BibitemShut {NoStop}%
\bibitem [{\citenamefont {Paschen}\ \emph {et~al.}(2008)\citenamefont
  {Paschen}, \citenamefont {Laumann}, \citenamefont {Prokofiev}, \citenamefont
  {Strydom}, \citenamefont {Deen}, \citenamefont {Stewart}, \citenamefont
  {Neumaier}, \citenamefont {Goukassov},\ and\ \citenamefont
  {Mignot}}]{Paschen2008}%
  \BibitemOpen
  \bibfield  {author} {\bibinfo {author} {\bibfnamefont {S.}~\bibnamefont
  {Paschen}}, \bibinfo {author} {\bibfnamefont {S.}~\bibnamefont {Laumann}},
  \bibinfo {author} {\bibfnamefont {A.}~\bibnamefont {Prokofiev}}, \bibinfo
  {author} {\bibfnamefont {A.~M.}\ \bibnamefont {Strydom}}, \bibinfo {author}
  {\bibfnamefont {P.~P.}\ \bibnamefont {Deen}}, \bibinfo {author}
  {\bibfnamefont {J.~R.}\ \bibnamefont {Stewart}}, \bibinfo {author}
  {\bibfnamefont {K.}~\bibnamefont {Neumaier}}, \bibinfo {author}
  {\bibfnamefont {A.}~\bibnamefont {Goukassov}}, \ and\ \bibinfo {author}
  {\bibfnamefont {J.~M.}\ \bibnamefont {Mignot}},\ }\href {\doibase 10.1016/j.physb.2007.10.358} {\bibfield  {journal} {\bibinfo  {journal}
  {Physica B: Condensed Matter}\ }\textbf {\bibinfo {volume} {403}},\ \bibinfo
  {pages} {1306} (\bibinfo {year} {2008})}\BibitemShut {NoStop}%
\bibitem [{\citenamefont {Shiina}\ \emph {et~al.}(1997)\citenamefont {Shiina},
  \citenamefont {Shiba},\ and\ \citenamefont {Thalmeier}}]{Shiina1997}%
  \BibitemOpen
  \bibfield  {author} {\bibinfo {author} {\bibfnamefont {R.}~\bibnamefont
  {Shiina}}, \bibinfo {author} {\bibfnamefont {H.}~\bibnamefont {Shiba}}, \
  and\ \bibinfo {author} {\bibfnamefont {P.}~\bibnamefont {Thalmeier}},\ }\href
  {\doibase 10.1143/JPSJ.66.1741} {\bibfield  {journal} {\bibinfo  {journal}
  {Journal of the Physical Society of Japan}\ }\textbf {\bibinfo {volume}
  {66}},\ \bibinfo {pages} {1741} (\bibinfo {year} {1997})}\BibitemShut
  {NoStop}%
\bibitem [{\citenamefont {Strydom}\ \emph {et~al.}(2006)\citenamefont
  {Strydom}, \citenamefont {Pikul}, \citenamefont {Steglich},\ and\
  \citenamefont {Paschen}}]{Strydom2006}%
  \BibitemOpen
  \bibfield  {author} {\bibinfo {author} {\bibfnamefont {A.~M.}\ \bibnamefont
  {Strydom}}, \bibinfo {author} {\bibfnamefont {A.}~\bibnamefont {Pikul}},
  \bibinfo {author} {\bibfnamefont {F.}~\bibnamefont {Steglich}}, \ and\
  \bibinfo {author} {\bibfnamefont {S.}~\bibnamefont {Paschen}},\ }\href
  {\doibase 10.1088/1742-6596/51/1/054} {\bibfield  {journal} {\bibinfo
  {journal} {Journal of Physics: Conference Series}\ }\textbf {\bibinfo
  {volume} {51}},\ \bibinfo {pages} {239} (\bibinfo {year} {2006})}\BibitemShut
  {NoStop}%
\bibitem [{\citenamefont {Mitamura}\ \emph {et~al.}(2010)\citenamefont
  {Mitamura}, \citenamefont {Sakuraba}, \citenamefont {Tayama}, \citenamefont
  {Sakakibara}, \citenamefont {Tsuduku}, \citenamefont {Ano}, \citenamefont
  {Ishii}, \citenamefont {Akatsu}, \citenamefont {Nemoto}, \citenamefont
  {Goto}, \citenamefont {Kikkawa},\ and\ \citenamefont
  {Kitazawa}}]{Mitamura2010}%
  \BibitemOpen
  \bibfield  {author} {\bibinfo {author} {\bibfnamefont {H.}~\bibnamefont
  {Mitamura}}, \bibinfo {author} {\bibfnamefont {T.}~\bibnamefont {Sakuraba}},
  \bibinfo {author} {\bibfnamefont {T.}~\bibnamefont {Tayama}}, \bibinfo
  {author} {\bibfnamefont {T.}~\bibnamefont {Sakakibara}}, \bibinfo {author}
  {\bibfnamefont {S.}~\bibnamefont {Tsuduku}}, \bibinfo {author} {\bibfnamefont
  {G.}~\bibnamefont {Ano}}, \bibinfo {author} {\bibfnamefont {I.}~\bibnamefont
  {Ishii}}, \bibinfo {author} {\bibfnamefont {M.}~\bibnamefont {Akatsu}},
  \bibinfo {author} {\bibfnamefont {Y.}~\bibnamefont {Nemoto}}, \bibinfo
  {author} {\bibfnamefont {T.}~\bibnamefont {Goto}}, \bibinfo {author}
  {\bibfnamefont {A.}~\bibnamefont {Kikkawa}}, \ and\ \bibinfo {author}
  {\bibfnamefont {H.}~\bibnamefont {Kitazawa}},\ }\href {\doibase 10.1088/1742-6596/200/1/012118} {\bibfield  {journal} {\bibinfo  {journal}
  {Journal of Physics: Conference Series}\ }\textbf {\bibinfo {volume} {200}},\ \bibinfo {pages} {012118}
  (\bibinfo {year} {2010}),\ 10.1088/1742-6596/200/1/012118}\BibitemShut
  {NoStop}%
\bibitem [{\citenamefont {Ono}\ \emph {et~al.}(2013)\citenamefont {Ono},
  \citenamefont {Nakano}, \citenamefont {Takeda}, \citenamefont {Ano},
  \citenamefont {Akatsu}, \citenamefont {Nemoto}, \citenamefont {Goto},
  \citenamefont {D{\"{o}}nni},\ and\ \citenamefont {Kitazawa}}]{Ono2013}%
  \BibitemOpen
  \bibfield  {author} {\bibinfo {author} {\bibfnamefont {H.}~\bibnamefont
  {Ono}}, \bibinfo {author} {\bibfnamefont {T.}~\bibnamefont {Nakano}},
  \bibinfo {author} {\bibfnamefont {N.}~\bibnamefont {Takeda}}, \bibinfo
  {author} {\bibfnamefont {G.}~\bibnamefont {Ano}}, \bibinfo {author}
  {\bibfnamefont {M.}~\bibnamefont {Akatsu}}, \bibinfo {author} {\bibfnamefont
  {Y.}~\bibnamefont {Nemoto}}, \bibinfo {author} {\bibfnamefont
  {T.}~\bibnamefont {Goto}}, \bibinfo {author} {\bibfnamefont {A.}~\bibnamefont
  {D{\"{o}}nni}}, \ and\ \bibinfo {author} {\bibfnamefont {H.}~\bibnamefont
  {Kitazawa}},\ }\href {\doibase 10.1088/0953-8984/25/12/126003} {\bibfield
  {journal} {\bibinfo  {journal} {Journal of Physics Condensed Matter}\
  }\textbf {\bibinfo {volume} {25}},\ \bibinfo {pages} {126003} (\bibinfo {year} {2013}),\
  10.1088/0953-8984/25/12/126003}\BibitemShut {NoStop}%
\bibitem [{\citenamefont {Martelli}\ \emph {et~al.}(2019)\citenamefont
  {Martelli}, \citenamefont {Cai}, \citenamefont {Nica}, \citenamefont
  {Taupin}, \citenamefont {Prokofiev}, \citenamefont {Liu}, \citenamefont
  {Lai}, \citenamefont {Yu}, \citenamefont {Ingersent}, \citenamefont
  {K{\"{u}}chler}, \citenamefont {Strydom}, \citenamefont {Geiger},
  \citenamefont {Haenel}, \citenamefont {Larrea}, \citenamefont {Si},\ and\
  \citenamefont {Paschen}}]{Martelli2019}%
  \BibitemOpen
  \bibfield  {author} {\bibinfo {author} {\bibfnamefont {V.}~\bibnamefont
  {Martelli}}, \bibinfo {author} {\bibfnamefont {A.}~\bibnamefont {Cai}},
  \bibinfo {author} {\bibfnamefont {E.~M.}\ \bibnamefont {Nica}}, \bibinfo
  {author} {\bibfnamefont {M.}~\bibnamefont {Taupin}}, \bibinfo {author}
  {\bibfnamefont {A.}~\bibnamefont {Prokofiev}}, \bibinfo {author}
  {\bibfnamefont {C.~C.}\ \bibnamefont {Liu}}, \bibinfo {author} {\bibfnamefont
  {H.~H.}\ \bibnamefont {Lai}}, \bibinfo {author} {\bibfnamefont
  {R.}~\bibnamefont {Yu}}, \bibinfo {author} {\bibfnamefont {K.}~\bibnamefont
  {Ingersent}}, \bibinfo {author} {\bibfnamefont {R.}~\bibnamefont
  {K{\"{u}}chler}}, \bibinfo {author} {\bibfnamefont {A.~M.}\ \bibnamefont
  {Strydom}}, \bibinfo {author} {\bibfnamefont {D.}~\bibnamefont {Geiger}},
  \bibinfo {author} {\bibfnamefont {J.}~\bibnamefont {Haenel}}, \bibinfo
  {author} {\bibfnamefont {J.}~\bibnamefont {Larrea}}, \bibinfo {author}
  {\bibfnamefont {Q.}~\bibnamefont {Si}}, \ and\ \bibinfo {author}
  {\bibfnamefont {S.}~\bibnamefont {Paschen}},\ }\href {\doibase 10.1073/pnas.1908101116} {\bibfield  {journal} {\bibinfo  {journal}
  {Proceedings of the National Academy of Sciences of the United States of
  America}\ }\textbf {\bibinfo {volume} {116}},\ \bibinfo {pages} {17701}
  (\bibinfo {year} {2019})}\BibitemShut {NoStop}%
\bibitem [{\citenamefont {Mazza}\ \emph {et~al.}(2022)\citenamefont {Mazza},
  \citenamefont {Portnichenko}, \citenamefont {Avdoshenko}, \citenamefont
  {Steffens}, \citenamefont {Boehm}, \citenamefont {Choi}, \citenamefont
  {Nikolo}, \citenamefont {Yan}, \citenamefont {Prokofiev}, \citenamefont
  {Paschen},\ and\ \citenamefont {Inosov}}]{Mazza2022}%
  \BibitemOpen
  \bibfield  {author} {\bibinfo {author} {\bibfnamefont {F.}~\bibnamefont
  {Mazza}}, \bibinfo {author} {\bibfnamefont {P.~Y.}\ \bibnamefont
  {Portnichenko}}, \bibinfo {author} {\bibfnamefont {S.}~\bibnamefont
  {Avdoshenko}}, \bibinfo {author} {\bibfnamefont {P.}~\bibnamefont
  {Steffens}}, \bibinfo {author} {\bibfnamefont {M.}~\bibnamefont {Boehm}},
  \bibinfo {author} {\bibfnamefont {E.~S.}\ \bibnamefont {Choi}}, \bibinfo
  {author} {\bibfnamefont {M.}~\bibnamefont {Nikolo}}, \bibinfo {author}
  {\bibfnamefont {X.}~\bibnamefont {Yan}}, \bibinfo {author} {\bibfnamefont
  {A.}~\bibnamefont {Prokofiev}}, \bibinfo {author} {\bibfnamefont
  {S.}~\bibnamefont {Paschen}}, \ and\ \bibinfo {author} {\bibfnamefont
  {D.~S.}\ \bibnamefont {Inosov}},\ }\href {\doibase 10.1103/physrevb.105.174429} {\bibfield  {journal} {\bibinfo  {journal}
  {Physical Review B}\ }\textbf {\bibinfo {volume} {105}},\ \bibinfo {pages}
  {174429} (\bibinfo {year} {2022})}\BibitemShut {NoStop}%
\bibitem [{\citenamefont {Coleman}\ \emph {et~al.}(2001)\citenamefont
  {Coleman}, \citenamefont {P{\'{e}}pin}, \citenamefont {Si},\ and\
  \citenamefont {Ramazashvili}}]{Coleman2001}%
  \BibitemOpen
  \bibfield  {author} {\bibinfo {author} {\bibfnamefont {P.}~\bibnamefont
  {Coleman}}, \bibinfo {author} {\bibfnamefont {C.}~\bibnamefont
  {P{\'{e}}pin}}, \bibinfo {author} {\bibfnamefont {Q.}~\bibnamefont {Si}}, \
  and\ \bibinfo {author} {\bibfnamefont {R.}~\bibnamefont {Ramazashvili}},\
  }\href {\doibase 10.1088/0953-8984/13/35/202} {\bibfield  {journal} {\bibinfo
   {journal} {Journal of Physics Condensed Matter}\ }\textbf {\bibinfo {volume}
  {13}},\ \bibinfo {pages} {R723} (\bibinfo {year} {2001})}\BibitemShut
  {NoStop}%
\bibitem [{\citenamefont {Si}\ \emph {et~al.}(2001)\citenamefont {Si},
  \citenamefont {Rabello}, \citenamefont {Ingersent},\ and\ \citenamefont
  {Smith}}]{Si2001}%
  \BibitemOpen
  \bibfield  {author} {\bibinfo {author} {\bibfnamefont {Q.}~\bibnamefont
  {Si}}, \bibinfo {author} {\bibfnamefont {S.}~\bibnamefont {Rabello}},
  \bibinfo {author} {\bibfnamefont {K.}~\bibnamefont {Ingersent}}, \ and\
  \bibinfo {author} {\bibfnamefont {J.~L.}\ \bibnamefont {Smith}},\ }\href
  {\doibase 10.1038/35101507} {\bibfield  {journal} {\bibinfo  {journal}
  {Nature}\ }\textbf {\bibinfo {volume} {413}},\ \bibinfo {pages} {804}
  (\bibinfo {year} {2001})}\BibitemShut {NoStop}%
\bibitem [{\citenamefont {Paschen}\ \emph {et~al.}(2004)\citenamefont
  {Paschen}, \citenamefont {L{\"{u}}hmann}, \citenamefont {Wirth},
  \citenamefont {Gegenwart}, \citenamefont {Trovarelli}, \citenamefont
  {Geibel}, \citenamefont {Steglich}, \citenamefont {Coleman},\ and\
  \citenamefont {Si}}]{Paschen2004}%
  \BibitemOpen
  \bibfield  {author} {\bibinfo {author} {\bibfnamefont {S.}~\bibnamefont
  {Paschen}}, \bibinfo {author} {\bibfnamefont {T.}~\bibnamefont
  {L{\"{u}}hmann}}, \bibinfo {author} {\bibfnamefont {S.}~\bibnamefont
  {Wirth}}, \bibinfo {author} {\bibfnamefont {P.}~\bibnamefont {Gegenwart}},
  \bibinfo {author} {\bibfnamefont {O.}~\bibnamefont {Trovarelli}}, \bibinfo
  {author} {\bibfnamefont {C.}~\bibnamefont {Geibel}}, \bibinfo {author}
  {\bibfnamefont {F.}~\bibnamefont {Steglich}}, \bibinfo {author}
  {\bibfnamefont {P.}~\bibnamefont {Coleman}}, \ and\ \bibinfo {author}
  {\bibfnamefont {Q.}~\bibnamefont {Si}},\ }\href {\doibase 10.1038/nature03129} {\bibfield  {journal} {\bibinfo  {journal} {Nature}\
  }\textbf {\bibinfo {volume} {432}},\ \bibinfo {pages} {881} (\bibinfo {year}
  {2004})}\BibitemShut {NoStop}%
\bibitem [{\citenamefont {Steglich}\ \emph {et~al.}(2014)\citenamefont
  {Steglich}, \citenamefont {Pfau}, \citenamefont {Lausberg}, \citenamefont
  {Hamann}, \citenamefont {Sun}, \citenamefont {Stockert}, \citenamefont
  {Brando}, \citenamefont {Friedemann}, \citenamefont {Krellner}, \citenamefont
  {Geibel}, \citenamefont {Wirth}, \citenamefont {Kirchner}, \citenamefont
  {Abrahams},\ and\ \citenamefont {Si}}]{Steglich2014}%
  \BibitemOpen
  \bibfield  {author} {\bibinfo {author} {\bibfnamefont {F.}~\bibnamefont
  {Steglich}}, \bibinfo {author} {\bibfnamefont {H.}~\bibnamefont {Pfau}},
  \bibinfo {author} {\bibfnamefont {S.}~\bibnamefont {Lausberg}}, \bibinfo
  {author} {\bibfnamefont {S.}~\bibnamefont {Hamann}}, \bibinfo {author}
  {\bibfnamefont {P.}~\bibnamefont {Sun}}, \bibinfo {author} {\bibfnamefont
  {U.}~\bibnamefont {Stockert}}, \bibinfo {author} {\bibfnamefont
  {M.}~\bibnamefont {Brando}}, \bibinfo {author} {\bibfnamefont
  {S.}~\bibnamefont {Friedemann}}, \bibinfo {author} {\bibfnamefont
  {C.}~\bibnamefont {Krellner}}, \bibinfo {author} {\bibfnamefont
  {C.}~\bibnamefont {Geibel}}, \bibinfo {author} {\bibfnamefont
  {S.}~\bibnamefont {Wirth}}, \bibinfo {author} {\bibfnamefont
  {S.}~\bibnamefont {Kirchner}}, \bibinfo {author} {\bibfnamefont
  {E.}~\bibnamefont {Abrahams}}, \ and\ \bibinfo {author} {\bibfnamefont
  {Q.}~\bibnamefont {Si}},\ }\href {\doibase 10.7566/JPSJ.83.061001} {\bibfield
   {journal} {\bibinfo  {journal} {Journal of the Physical Society of Japan}\
  }\textbf {\bibinfo {volume} {83}},\ \bibinfo {pages} {061001} (\bibinfo {year}
  {2014})}\BibitemShut {NoStop}%
\bibitem [{\citenamefont {Custers}\ \emph {et~al.}(2012)\citenamefont
  {Custers}, \citenamefont {Lorenzer}, \citenamefont {M{\"{u}}ller},
  \citenamefont {Prokofiev}, \citenamefont {Sidorenko}, \citenamefont
  {Winkler}, \citenamefont {Strydom}, \citenamefont {Shimura}, \citenamefont
  {Sakakibara}, \citenamefont {Yu}, \citenamefont {Si},\ and\ \citenamefont
  {Paschen}}]{Custers2012}%
  \BibitemOpen
  \bibfield  {author} {\bibinfo {author} {\bibfnamefont {J.}~\bibnamefont
  {Custers}}, \bibinfo {author} {\bibfnamefont {K.~A.}\ \bibnamefont
  {Lorenzer}}, \bibinfo {author} {\bibfnamefont {M.}~\bibnamefont
  {M{\"{u}}ller}}, \bibinfo {author} {\bibfnamefont {A.}~\bibnamefont
  {Prokofiev}}, \bibinfo {author} {\bibfnamefont {A.}~\bibnamefont
  {Sidorenko}}, \bibinfo {author} {\bibfnamefont {H.}~\bibnamefont {Winkler}},
  \bibinfo {author} {\bibfnamefont {A.~M.}\ \bibnamefont {Strydom}}, \bibinfo
  {author} {\bibfnamefont {Y.}~\bibnamefont {Shimura}}, \bibinfo {author}
  {\bibfnamefont {T.}~\bibnamefont {Sakakibara}}, \bibinfo {author}
  {\bibfnamefont {R.}~\bibnamefont {Yu}}, \bibinfo {author} {\bibfnamefont
  {Q.}~\bibnamefont {Si}}, \ and\ \bibinfo {author} {\bibfnamefont
  {S.}~\bibnamefont {Paschen}},\ }\href {\doibase 10.1038/nmat3214} {\bibfield
  {journal} {\bibinfo  {journal} {Nature Materials}\ }\textbf {\bibinfo
  {volume} {11}},\ \bibinfo {pages} {189} (\bibinfo {year} {2012})}\BibitemShut
  {NoStop}%
\bibitem [{\citenamefont {Friedemann}\ \emph
  {et~al.}(2010{\natexlab{a}})\citenamefont {Friedemann}, \citenamefont
  {Wirth}, \citenamefont {Oeschler}, \citenamefont {Krellner}, \citenamefont
  {Geibel}, \citenamefont {Steglich}, \citenamefont {Maquilon}, \citenamefont
  {Fisk}, \citenamefont {Paschen},\ and\ \citenamefont
  {Zwicknagl}}]{Friedemann2010}%
  \BibitemOpen
  \bibfield  {author} {\bibinfo {author} {\bibfnamefont {S.}~\bibnamefont
  {Friedemann}}, \bibinfo {author} {\bibfnamefont {S.}~\bibnamefont {Wirth}},
  \bibinfo {author} {\bibfnamefont {N.}~\bibnamefont {Oeschler}}, \bibinfo
  {author} {\bibfnamefont {C.}~\bibnamefont {Krellner}}, \bibinfo {author}
  {\bibfnamefont {C.}~\bibnamefont {Geibel}}, \bibinfo {author} {\bibfnamefont
  {F.}~\bibnamefont {Steglich}}, \bibinfo {author} {\bibfnamefont
  {S.}~\bibnamefont {Maquilon}}, \bibinfo {author} {\bibfnamefont
  {Z.}~\bibnamefont {Fisk}}, \bibinfo {author} {\bibfnamefont {S.}~\bibnamefont
  {Paschen}}, \ and\ \bibinfo {author} {\bibfnamefont {G.}~\bibnamefont
  {Zwicknagl}},\ }\href {\doibase 10.1103/PhysRevB.82.035103} {\bibfield
  {journal} {\bibinfo  {journal} {Physical Review B - Condensed Matter and
  Materials Physics}\ }\textbf {\bibinfo {volume} {82}},\ \bibinfo {pages}
  {035103} (\bibinfo {year} {2010}{\natexlab{a}})}\BibitemShut {NoStop}%
\bibitem [{\citenamefont {Friedemann}\ \emph
  {et~al.}(2010{\natexlab{b}})\citenamefont {Friedemann}, \citenamefont
  {Oeschler}, \citenamefont {Wirth}, \citenamefont {Krellner}, \citenamefont
  {Geibel}, \citenamefont {Steglich}, \citenamefont {Paschen}, \citenamefont
  {Kirchner},\ and\ \citenamefont {Si}}]{Friedemann2010a}%
  \BibitemOpen
  \bibfield  {author} {\bibinfo {author} {\bibfnamefont {S.}~\bibnamefont
  {Friedemann}}, \bibinfo {author} {\bibfnamefont {N.}~\bibnamefont
  {Oeschler}}, \bibinfo {author} {\bibfnamefont {S.}~\bibnamefont {Wirth}},
  \bibinfo {author} {\bibfnamefont {C.}~\bibnamefont {Krellner}}, \bibinfo
  {author} {\bibfnamefont {C.}~\bibnamefont {Geibel}}, \bibinfo {author}
  {\bibfnamefont {F.}~\bibnamefont {Steglich}}, \bibinfo {author}
  {\bibfnamefont {S.}~\bibnamefont {Paschen}}, \bibinfo {author} {\bibfnamefont
  {S.}~\bibnamefont {Kirchner}}, \ and\ \bibinfo {author} {\bibfnamefont
  {Q.}~\bibnamefont {Si}},\ }\href {\doibase 10.1073/pnas.1009202107}
  {\bibfield  {journal} {\bibinfo  {journal} {Proceedings of the National
  Academy of Sciences of the United States of America}\ }\textbf {\bibinfo
  {volume} {107}},\ \bibinfo {pages} {14547} (\bibinfo {year}
  {2010}{\natexlab{b}})}\BibitemShut {NoStop}%
\bibitem [{\citenamefont {Liu}\ \emph {et~al.}(2023)\citenamefont {Liu},
  \citenamefont {Paschen},\ and\ \citenamefont {Si}}]{Liu2021}%
  \BibitemOpen
  \bibfield  {author} {\bibinfo {author} {\bibfnamefont {C.-C.}\ \bibnamefont
  {Liu}}, \bibinfo {author} {\bibfnamefont {S.}~\bibnamefont {Paschen}}, \ and\
  \bibinfo {author} {\bibfnamefont {Q.}~\bibnamefont {Si}},\ }\href {\doibase 10.1073/pnas.2300903120} {\bibfield  {journal} {\bibinfo  {journal} {Proc.
  Natl. Acad. Sci. U.S.A.}\ }\textbf {\bibinfo {volume} {120}},\ \bibinfo
  {pages} {e2300903120} (\bibinfo {year} {2023})}\BibitemShut {NoStop}%
\bibitem [{\citenamefont {Si}\ and\ \citenamefont {Smith}(1996)}]{Si1996}%
  \BibitemOpen
  \bibfield  {author} {\bibinfo {author} {\bibfnamefont {Q.}~\bibnamefont
  {Si}}\ and\ \bibinfo {author} {\bibfnamefont {J.~L.}\ \bibnamefont {Smith}},\
  }\href {\doibase 10.1103/PhysRevLett.77.3391} {\bibfield  {journal} {\bibinfo
   {journal} {Physical Review Letters}\ }\textbf {\bibinfo {volume} {77}},\
  \bibinfo {pages} {3391} (\bibinfo {year} {1996})}\BibitemShut {NoStop}%
\bibitem [{\citenamefont {Smith}\ and\ \citenamefont {Si}(1999)}]{Smith1999}%
  \BibitemOpen
  \bibfield  {author} {\bibinfo {author} {\bibfnamefont {J.~L.}\ \bibnamefont
  {Smith}}\ and\ \bibinfo {author} {\bibfnamefont {Q.}~\bibnamefont {Si}},\
  }\href {\doibase 10.1209/epl/i1999-00151-4} {\bibfield  {journal} {\bibinfo
  {journal} {Europhysics Letters (EPL)}\ }\textbf {\bibinfo {volume} {45}},\
  \bibinfo {pages} {228} (\bibinfo {year} {1999})}\BibitemShut {NoStop}%
\bibitem [{\citenamefont {Sengupta}(2000)}]{Sengupta2000}%
  \BibitemOpen
  \bibfield  {author} {\bibinfo {author} {\bibfnamefont {A.~M.}\ \bibnamefont
  {Sengupta}},\ }\href {\doibase 10.1103/PhysRevB.61.4041} {\bibfield
  {journal} {\bibinfo  {journal} {Physical Review B}\ }\textbf {\bibinfo
  {volume} {61}},\ \bibinfo {pages} {4041} (\bibinfo {year}
  {2000})}\BibitemShut {NoStop}%
\bibitem [{\citenamefont {Zhu}\ and\ \citenamefont {Si}(2002)}]{Zhu2002}%
  \BibitemOpen
  \bibfield  {author} {\bibinfo {author} {\bibfnamefont {L.}~\bibnamefont
  {Zhu}}\ and\ \bibinfo {author} {\bibfnamefont {Q.}~\bibnamefont {Si}},\
  }\href {\doibase 10.1103/PhysRevB.66.024426} {\bibfield  {journal} {\bibinfo
  {journal} {Physical Review B - Condensed Matter and Materials Physics}\
  }\textbf {\bibinfo {volume} {66}},\ \bibinfo {pages} {024426} (\bibinfo
  {year} {2002})}\BibitemShut {NoStop}%
\bibitem [{\citenamefont {Zar{\'{a}}nd}\ and\ \citenamefont
  {Demler}(2002)}]{Zarand2002}%
  \BibitemOpen
  \bibfield  {author} {\bibinfo {author} {\bibfnamefont {G.}~\bibnamefont
  {Zar{\'{a}}nd}}\ and\ \bibinfo {author} {\bibfnamefont {E.}~\bibnamefont
  {Demler}},\ }\href {\doibase 10.1103/PhysRevB.66.024427} {\bibfield
  {journal} {\bibinfo  {journal} {Physical Review B - Condensed Matter and
  Materials Physics}\ }\textbf {\bibinfo {volume} {66}},\ \bibinfo {pages}
  {024427} (\bibinfo {year} {2002})}\BibitemShut {NoStop}%
\bibitem [{\citenamefont {Kirchner}\ \emph {et~al.}(2005)\citenamefont
  {Kirchner}, \citenamefont {Zhu},\ and\ \citenamefont {Si}}]{Kirchner2005}%
  \BibitemOpen
  \bibfield  {author} {\bibinfo {author} {\bibfnamefont {S.}~\bibnamefont
  {Kirchner}}, \bibinfo {author} {\bibfnamefont {L.}~\bibnamefont {Zhu}}, \
  and\ \bibinfo {author} {\bibfnamefont {Q.}~\bibnamefont {Si}},\ }\href
  {\doibase 10.1016/j.physb.2004.12.064} {\bibfield  {journal} {\bibinfo
  {journal} {Physica B: Condensed Matter}\ }\textbf {\bibinfo {volume}
  {359-361}},\ \bibinfo {pages} {83} (\bibinfo {year} {2005})}\BibitemShut
  {NoStop}%
\bibitem [{\citenamefont {D{\"{o}}nni}\ \emph {et~al.}(2000)\citenamefont
  {D{\"{o}}nni}, \citenamefont {Herrmannsd{\"{o}}rfer}, \citenamefont
  {Fischer}, \citenamefont {Keller}, \citenamefont {Fauth}, \citenamefont
  {McEwen}, \citenamefont {Goto},\ and\ \citenamefont
  {Komatsubara}}]{Donni2000}%
  \BibitemOpen
  \bibfield  {author} {\bibinfo {author} {\bibfnamefont {A.}~\bibnamefont
  {D{\"{o}}nni}}, \bibinfo {author} {\bibfnamefont {T.}~\bibnamefont
  {Herrmannsd{\"{o}}rfer}}, \bibinfo {author} {\bibfnamefont {P.}~\bibnamefont
  {Fischer}}, \bibinfo {author} {\bibfnamefont {L.}~\bibnamefont {Keller}},
  \bibinfo {author} {\bibfnamefont {F.}~\bibnamefont {Fauth}}, \bibinfo
  {author} {\bibfnamefont {K.~A.}\ \bibnamefont {McEwen}}, \bibinfo {author}
  {\bibfnamefont {T.}~\bibnamefont {Goto}}, \ and\ \bibinfo {author}
  {\bibfnamefont {T.}~\bibnamefont {Komatsubara}},\ }\href {\doibase 10.1088/0953-8984/12/45/307} {\bibfield  {journal} {\bibinfo  {journal}
  {Journal of Physics Condensed Matter}\ }\textbf {\bibinfo {volume} {12}},\
  \bibinfo {pages} {9441} (\bibinfo {year} {2000})}\BibitemShut {NoStop}%
\bibitem [{SM()}]{SM}%
  \BibitemOpen
  \href@noop {} {\ }\bibinfo {note} {See Supplemental Material at [URL will be
  inserted by publisher] for the details, which includes
  Refs.~\cite{Shiina1997,Zhu2002,Custers2012,Ellis2017,Han2022,Lavagna2003,Bensimon2006a,Affleck1991a,Affleck1993b,Ludwig1991,Kimura2017b,Schultz2021b,Patri2020e,Mora2009,Si2001,Si2010a,Steglich2014,Friedemann2010,Friedemann2010a,Paschen2004,Coleman2001,Zarand2002,Patri2019d,Sorensen2021,Schultz2021c}.}\BibitemShut
  {Stop}%
\bibitem [{\citenamefont {Ellis}(2017)}]{Ellis2017}%
  \BibitemOpen
  \bibfield  {author} {\bibinfo {author} {\bibfnamefont {J.~P.}\ \bibnamefont
  {Ellis}},\ }\href {\doibase 10.1016/j.cpc.2016.08.019} {\bibfield  {journal}
  {\bibinfo  {journal} {Computer Physics Communications}\ }\textbf {\bibinfo
  {volume} {210}},\ \bibinfo {pages} {103} (\bibinfo {year}
  {2017})}\BibitemShut {NoStop}%
\bibitem [{\citenamefont {Affleck}\ and\ \citenamefont
  {Ludwig}(1991)}]{Affleck1991a}%
  \BibitemOpen
  \bibfield  {author} {\bibinfo {author} {\bibfnamefont {I.}~\bibnamefont
  {Affleck}}\ and\ \bibinfo {author} {\bibfnamefont {A.~W.}\ \bibnamefont
  {Ludwig}},\ }\href {\doibase 10.1016/0550-3213(91)90419-X} {\bibfield
  {journal} {\bibinfo  {journal} {Nuclear Physics, Section B}\ }\textbf
  {\bibinfo {volume} {360}},\ \bibinfo {pages} {641} (\bibinfo {year}
  {1991})}\BibitemShut {NoStop}%
\bibitem [{\citenamefont {Affleck}\ and\ \citenamefont
  {Ludwig}(1993)}]{Affleck1993b}%
  \BibitemOpen
  \bibfield  {author} {\bibinfo {author} {\bibfnamefont {I.}~\bibnamefont
  {Affleck}}\ and\ \bibinfo {author} {\bibfnamefont {A.~W.~W.}\ \bibnamefont
  {Ludwig}},\ }\href {\doibase 10.1103/PhysRevB.48.7297} {\bibfield  {journal}
  {\bibinfo  {journal} {Physical Review B}\ }\textbf {\bibinfo {volume} {48}},\
  \bibinfo {pages} {7297} (\bibinfo {year} {1993})}\BibitemShut {NoStop}%
\bibitem [{\citenamefont {Ludwig}\ and\ \citenamefont
  {Affleck}(1991)}]{Ludwig1991}%
  \BibitemOpen
  \bibfield  {author} {\bibinfo {author} {\bibfnamefont {A.~W.~W.}\
  \bibnamefont {Ludwig}}\ and\ \bibinfo {author} {\bibfnamefont
  {I.}~\bibnamefont {Affleck}},\ }\href {\doibase 10.1103/PhysRevLett.67.3160}
  {\bibfield  {journal} {\bibinfo  {journal} {Physical Review Letters}\
  }\textbf {\bibinfo {volume} {67}},\ \bibinfo {pages} {3160} (\bibinfo {year}
  {1991})}\BibitemShut {NoStop}%
\bibitem [{\citenamefont {Mora}(2009)}]{Mora2009}%
  \BibitemOpen
  \bibfield  {author} {\bibinfo {author} {\bibfnamefont {C.}~\bibnamefont
  {Mora}},\ }\href {\doibase 10.1103/PhysRevB.80.125304} {\bibfield  {journal}
  {\bibinfo  {journal} {Physical Review B - Condensed Matter and Materials
  Physics}\ }\textbf {\bibinfo {volume} {80}},\ \bibinfo {pages} {125304}
  (\bibinfo {year} {2009})}\BibitemShut {NoStop}%
\bibitem [{\citenamefont {Kimura}\ and\ \citenamefont
  {Ozaki}(2017)}]{Kimura2017b}%
  \BibitemOpen
  \bibfield  {author} {\bibinfo {author} {\bibfnamefont {T.}~\bibnamefont
  {Kimura}}\ and\ \bibinfo {author} {\bibfnamefont {S.}~\bibnamefont {Ozaki}},\
  }\href {\doibase 10.7566/JPSJ.86.084703} {\bibfield  {journal} {\bibinfo
  {journal} {Journal of the Physical Society of Japan}\ }\textbf {\bibinfo
  {volume} {86}},\ \bibinfo {pages} {084703} (\bibinfo {year}
  {2017})}\BibitemShut {NoStop}%
\bibitem [{\citenamefont {Schultz}\ \emph
  {et~al.}(2021{\natexlab{b}})\citenamefont {Schultz}, \citenamefont {Patri},\
  and\ \citenamefont {Kim}}]{Schultz2021c}%
  \BibitemOpen
  \bibfield  {author} {\bibinfo {author} {\bibfnamefont {D.~J.}\ \bibnamefont
  {Schultz}}, \bibinfo {author} {\bibfnamefont {A.~S.}\ \bibnamefont {Patri}},
  \ and\ \bibinfo {author} {\bibfnamefont {Y.~B.}\ \bibnamefont {Kim}},\ }\href
  {\doibase 10.1103/PhysRevB.104.125144} {\bibfield  {journal} {\bibinfo
  {journal} {Physical Review B}\ }\textbf {\bibinfo {volume} {104}},\ \bibinfo
  {pages} {125144} (\bibinfo {year} {2021}{\natexlab{b}})}\BibitemShut
  {NoStop}%
\bibitem [{\citenamefont {Lavagna}\ \emph {et~al.}(2003)\citenamefont
  {Lavagna}, \citenamefont {Jerez},\ and\ \citenamefont
  {Bensimon}}]{Lavagna2003}%
  \BibitemOpen
  \bibfield  {author} {\bibinfo {author} {\bibfnamefont {M.}~\bibnamefont
  {Lavagna}}, \bibinfo {author} {\bibfnamefont {A.}~\bibnamefont {Jerez}}, \
  and\ \bibinfo {author} {\bibfnamefont {D.}~\bibnamefont {Bensimon}},\ }\href
  {\doibase 10.1007/978-94-010-0213-4_21} {\bibfield  {journal} {\bibinfo
  {journal} {Concepts in Electron Correlation}\ ,\ \bibinfo {pages} {219}}
  (\bibinfo {year} {2003})}\BibitemShut {NoStop}%
\bibitem [{\citenamefont {Bensimon}\ \emph {et~al.}(2006)\citenamefont
  {Bensimon}, \citenamefont {Jerez},\ and\ \citenamefont
  {Lavagna}}]{Bensimon2006a}%
  \BibitemOpen
  \bibfield  {author} {\bibinfo {author} {\bibfnamefont {D.}~\bibnamefont
  {Bensimon}}, \bibinfo {author} {\bibfnamefont {A.}~\bibnamefont {Jerez}}, \
  and\ \bibinfo {author} {\bibfnamefont {M.}~\bibnamefont {Lavagna}},\ }\href
  {\doibase 10.1103/PhysRevB.73.224445} {\bibfield  {journal} {\bibinfo
  {journal} {Physical Review B - Condensed Matter and Materials Physics}\
  }\textbf {\bibinfo {volume} {73}},\ \bibinfo {pages} {224445} (\bibinfo
  {year} {2006})}\BibitemShut {NoStop}%
\bibitem [{\citenamefont {Han}\ \emph {et~al.}(2022)\citenamefont {Han},
  \citenamefont {Schultz},\ and\ \citenamefont {Kim}}]{Han2022}%
  \BibitemOpen
  \bibfield  {author} {\bibinfo {author} {\bibfnamefont {S.E.}~\bibnamefont
  {Han}}, \bibinfo {author} {\bibfnamefont {D.~J.}\ \bibnamefont {Schultz}}, \
  and\ \bibinfo {author} {\bibfnamefont {Y.~B.}\ \bibnamefont {Kim}},\ }\href
  {\doibase 10.1103/PhysRevB.106.155155} {\bibfield  {journal} {\bibinfo
  {journal} {Phys. Rev. B}\ }\textbf {\bibinfo {volume} {106}},\ \bibinfo
  {pages} {155155} (\bibinfo {year} {2022})}\BibitemShut {NoStop}%
\bibitem [{\citenamefont {Patri}\ \emph {et~al.}(2019)\citenamefont {Patri},
  \citenamefont {Sakai}, \citenamefont {Lee}, \citenamefont {Paramekanti},
  \citenamefont {Nakatsuji},\ and\ \citenamefont {Kim}}]{Patri2019d}%
  \BibitemOpen
  \bibfield  {author} {\bibinfo {author} {\bibfnamefont {A.~S.}\ \bibnamefont
  {Patri}}, \bibinfo {author} {\bibfnamefont {A.}~\bibnamefont {Sakai}},
  \bibinfo {author} {\bibfnamefont {S.~B.}\ \bibnamefont {Lee}}, \bibinfo
  {author} {\bibfnamefont {A.}~\bibnamefont {Paramekanti}}, \bibinfo {author}
  {\bibfnamefont {S.}~\bibnamefont {Nakatsuji}}, \ and\ \bibinfo {author}
  {\bibfnamefont {Y.~B.}\ \bibnamefont {Kim}},\ }\href {\doibase 10.1038/s41467-019-11913-3} {\bibfield  {journal} {\bibinfo  {journal}
  {Nature Communications}\ }\textbf {\bibinfo {volume} {10}},\
  \bibinfo {pages} {4092} (\bibinfo {year}
  {2019}),\ 10.1038/s41467-019-11913-3}\BibitemShut {NoStop}%
\bibitem [{\citenamefont {Senthil}\ \emph {et~al.}(2003)\citenamefont
  {Senthil}, \citenamefont {Sachdev},\ and\ \citenamefont
  {Vojta}}]{Senthil2003}%
  \BibitemOpen
  \bibfield  {author} {\bibinfo {author} {\bibfnamefont {T.}~\bibnamefont
  {Senthil}}, \bibinfo {author} {\bibfnamefont {S.}~\bibnamefont {Sachdev}}, \
  and\ \bibinfo {author} {\bibfnamefont {M.}~\bibnamefont {Vojta}},\ }\href
  {\doibase 10.1103/PhysRevLett.90.216403} {\bibfield  {journal} {\bibinfo
  {journal} {Physical Review Letters}\ }\textbf {\bibinfo {volume} {90}},\
  \bibinfo {pages} {216403} (\bibinfo {year} {2003})}\BibitemShut {NoStop}%
\bibitem [{\citenamefont {Senthil}\ \emph {et~al.}(2004)\citenamefont
  {Senthil}, \citenamefont {Vojta},\ and\ \citenamefont
  {Sachdev}}]{Senthil2004}%
  \BibitemOpen
  \bibfield  {author} {\bibinfo {author} {\bibfnamefont {T.}~\bibnamefont
  {Senthil}}, \bibinfo {author} {\bibfnamefont {M.}~\bibnamefont {Vojta}}, \
  and\ \bibinfo {author} {\bibfnamefont {S.}~\bibnamefont {Sachdev}},\ }\href
  {\doibase 10.1103/PhysRevB.69.035111} {\bibfield  {journal} {\bibinfo
  {journal} {Physical Review B - Condensed Matter and Materials Physics}\
  }\textbf {\bibinfo {volume} {69}},\
  \bibinfo {pages} {035111} (\bibinfo {year} {2004}),\
  10.1103/PhysRevB.69.035111}\BibitemShut {NoStop}%
\bibitem [{\citenamefont {Rosenberg}\ \emph {et~al.}(2019)\citenamefont
  {Rosenberg}, \citenamefont {Chu}, \citenamefont {Ruff}, \citenamefont
  {Hristov},\ and\ \citenamefont {Fisher}}]{Rosenberg2019c}%
  \BibitemOpen
  \bibfield  {author} {\bibinfo {author} {\bibfnamefont {E.~W.}\ \bibnamefont
  {Rosenberg}}, \bibinfo {author} {\bibfnamefont {J.~H.}\ \bibnamefont {Chu}},
  \bibinfo {author} {\bibfnamefont {J.~P.}\ \bibnamefont {Ruff}}, \bibinfo
  {author} {\bibfnamefont {A.~T.}\ \bibnamefont {Hristov}}, \ and\ \bibinfo
  {author} {\bibfnamefont {I.~R.}\ \bibnamefont {Fisher}},\ }\href {\doibase 10.1073/pnas.1818910116} {\bibfield  {journal} {\bibinfo  {journal}
  {Proceedings of the National Academy of Sciences of the United States of
  America}\ }\textbf {\bibinfo {volume} {116}},\ \bibinfo {pages} {7232}
  (\bibinfo {year} {2019})}\BibitemShut {NoStop}%
\bibitem [{\citenamefont {Jeevan}\ \emph {et~al.}(2006)\citenamefont {Jeevan},
  \citenamefont {Geibel},\ and\ \citenamefont {Hossain}}]{Jeevan2006}%
  \BibitemOpen
  \bibfield  {author} {\bibinfo {author} {\bibfnamefont {H.~S.}\ \bibnamefont
  {Jeevan}}, \bibinfo {author} {\bibfnamefont {C.}~\bibnamefont {Geibel}}, \
  and\ \bibinfo {author} {\bibfnamefont {Z.}~\bibnamefont {Hossain}},\ }\href
  {\doibase 10.1103/PhysRevB.73.020407} {\bibfield  {journal} {\bibinfo
  {journal} {Physical Review B}\ }\textbf {\bibinfo {volume} {73}},\ \bibinfo
  {pages} {020407(R)} (\bibinfo {year} {2006})}\BibitemShut {NoStop}%
\end{thebibliography}

\begin{thebibliography}{25}%
\makeatletter
\providecommand \@ifxundefined [1]{%
 \@ifx{#1\undefined}
}%
\providecommand \@ifnum [1]{%
 \ifnum #1\expandafter \@firstoftwo
 \else \expandafter \@secondoftwo
 \fi
}%
\providecommand \@ifx [1]{%
 \ifx #1\expandafter \@firstoftwo
 \else \expandafter \@secondoftwo
 \fi
}%
\providecommand \natexlab [1]{#1}%
\providecommand \enquote  [1]{``#1''}%
\providecommand \bibnamefont  [1]{#1}%
\providecommand \bibfnamefont [1]{#1}%
\providecommand \citenamefont [1]{#1}%
\providecommand \href@noop [0]{\@secondoftwo}%
\providecommand \href [0]{\begingroup \@sanitize@url \@href}%
\providecommand \@href[1]{\@@startlink{#1}\@@href}%
\providecommand \@@href[1]{\endgroup#1\@@endlink}%
\providecommand \@sanitize@url [0]{\catcode `\\12\catcode `\$12\catcode
  `\&12\catcode `\#12\catcode `\^12\catcode `\_12\catcode `\%12\relax}%
\providecommand \@@startlink[1]{}%
\providecommand \@@endlink[0]{}%
\providecommand \url  [0]{\begingroup\@sanitize@url \@url }%
\providecommand \@url [1]{\endgroup\@href {#1}{\urlprefix }}%
\providecommand \urlprefix  [0]{URL }%
\providecommand \Eprint [0]{\href }%
\providecommand \doibase [0]{http://dx.doi.org/}%
\providecommand \selectlanguage [0]{\@gobble}%
\providecommand \bibinfo  [0]{\@secondoftwo}%
\providecommand \bibfield  [0]{\@secondoftwo}%
\providecommand \translation [1]{[#1]}%
\providecommand \BibitemOpen [0]{}%
\providecommand \bibitemStop [0]{}%
\providecommand \bibitemNoStop [0]{.\EOS\space}%
\providecommand \EOS [0]{\spacefactor3000\relax}%
\providecommand \BibitemShut  [1]{\csname bibitem#1\endcsname}%
\let\auto@bib@innerbib\@empty
\bibitem [{\citenamefont {Shiina}\ \emph {et~al.}(1997)\citenamefont {Shiina},
  \citenamefont {Shiba},\ and\ \citenamefont {Thalmeier}}]{Shiina1997_SM}%
  \BibitemOpen
  \bibfield  {author} {\bibinfo {author} {\bibfnamefont {R.}~\bibnamefont
  {Shiina}}, \bibinfo {author} {\bibfnamefont {H.}~\bibnamefont {Shiba}}, \
  and\ \bibinfo {author} {\bibfnamefont {P.}~\bibnamefont {Thalmeier}},\ }\href
  {\doibase 10.1143/JPSJ.66.1741} {\bibfield  {journal} {\bibinfo  {journal}
  {Journal of the Physical Society of Japan}\ }\textbf {\bibinfo {volume}
  {66}},\ \bibinfo {pages} {1741} (\bibinfo {year} {1997})}\BibitemShut
  {NoStop}%
\bibitem [{\citenamefont {Custers}\ \emph {et~al.}(2012)\citenamefont
  {Custers}, \citenamefont {Lorenzer}, \citenamefont {M{\"{u}}ller},
  \citenamefont {Prokofiev}, \citenamefont {Sidorenko}, \citenamefont
  {Winkler}, \citenamefont {Strydom}, \citenamefont {Shimura}, \citenamefont
  {Sakakibara}, \citenamefont {Yu}, \citenamefont {Si},\ and\ \citenamefont
  {Paschen}}]{Custers2012_SM}%
  \BibitemOpen
  \bibfield  {author} {\bibinfo {author} {\bibfnamefont {J.}~\bibnamefont
  {Custers}}, \bibinfo {author} {\bibfnamefont {K.~A.}\ \bibnamefont
  {Lorenzer}}, \bibinfo {author} {\bibfnamefont {M.}~\bibnamefont
  {M{\"{u}}ller}}, \bibinfo {author} {\bibfnamefont {A.}~\bibnamefont
  {Prokofiev}}, \bibinfo {author} {\bibfnamefont {A.}~\bibnamefont
  {Sidorenko}}, \bibinfo {author} {\bibfnamefont {H.}~\bibnamefont {Winkler}},
  \bibinfo {author} {\bibfnamefont {A.~M.}\ \bibnamefont {Strydom}}, \bibinfo
  {author} {\bibfnamefont {Y.}~\bibnamefont {Shimura}}, \bibinfo {author}
  {\bibfnamefont {T.}~\bibnamefont {Sakakibara}}, \bibinfo {author}
  {\bibfnamefont {R.}~\bibnamefont {Yu}}, \bibinfo {author} {\bibfnamefont
  {Q.}~\bibnamefont {Si}}, \ and\ \bibinfo {author} {\bibfnamefont
  {S.}~\bibnamefont {Paschen}},\ }\href {\doibase 10.1038/nmat3214} {\bibfield
  {journal} {\bibinfo  {journal} {Nature Materials}\ }\textbf {\bibinfo
  {volume} {11}},\ \bibinfo {pages} {189} (\bibinfo {year} {2012})}\BibitemShut
  {NoStop}%
\bibitem [{\citenamefont {Zhu}\ and\ \citenamefont {Si}(2002)}]{Zhu2002_SM}%
  \BibitemOpen
  \bibfield  {author} {\bibinfo {author} {\bibfnamefont {L.}~\bibnamefont
  {Zhu}}\ and\ \bibinfo {author} {\bibfnamefont {Q.}~\bibnamefont {Si}},\
  }\href {\doibase 10.1103/PhysRevB.66.024426} {\bibfield  {journal} {\bibinfo
  {journal} {Physical Review B - Condensed Matter and Materials Physics}\
  }\textbf {\bibinfo {volume} {66}},\ \bibinfo {pages} {024426} (\bibinfo
  {year} {2002})}\BibitemShut {NoStop}%
\bibitem [{\citenamefont {Ellis}(2017)}]{Ellis2017_SM}%
  \BibitemOpen
  \bibfield  {author} {\bibinfo {author} {\bibfnamefont {J.~P.}\ \bibnamefont
  {Ellis}},\ }\href {\doibase 10.1016/j.cpc.2016.08.019} {\bibfield  {journal}
  {\bibinfo  {journal} {Computer Physics Communications}\ }\textbf {\bibinfo
  {volume} {210}},\ \bibinfo {pages} {103} (\bibinfo {year}
  {2017})}\BibitemShut {NoStop}%
\bibitem [{\citenamefont {Han}\ \emph {et~al.}(2022)\citenamefont {Han},
  \citenamefont {Schultz},\ and\ \citenamefont {Kim}}]{Han2022_SM}%
  \BibitemOpen
  \bibfield  {author} {\bibinfo {author} {\bibfnamefont {S.}~\bibnamefont
  {Han}}, \bibinfo {author} {\bibfnamefont {D.~J.}\ \bibnamefont {Schultz}}, \
  and\ \bibinfo {author} {\bibfnamefont {Y.~B.}\ \bibnamefont {Kim}},\ }\href
  {\doibase 10.1103/PhysRevB.106.155155} {\bibfield  {journal} {\bibinfo
  {journal} {Phys. Rev. B}\ }\textbf {\bibinfo {volume} {106}},\ \bibinfo
  {pages} {155155} (\bibinfo {year} {2022})}\BibitemShut {NoStop}%
\bibitem [{\citenamefont {Lavagna}\ \emph {et~al.}(2003)\citenamefont
  {Lavagna}, \citenamefont {Jerez},\ and\ \citenamefont
  {Bensimon}}]{Lavagna2003_SM}%
  \BibitemOpen
  \bibfield  {author} {\bibinfo {author} {\bibfnamefont {M.}~\bibnamefont
  {Lavagna}}, \bibinfo {author} {\bibfnamefont {A.}~\bibnamefont {Jerez}}, \
  and\ \bibinfo {author} {\bibfnamefont {D.}~\bibnamefont {Bensimon}},\ }\href
  {\doibase 10.1007/978-94-010-0213-4_21} {\bibfield  {journal} {\bibinfo
  {journal} {Concepts in Electron Correlation}\ ,\ \bibinfo {pages} {219}}
  (\bibinfo {year} {2003})}\BibitemShut {NoStop}%
\bibitem [{\citenamefont {Bensimon}\ \emph {et~al.}(2006)\citenamefont
  {Bensimon}, \citenamefont {Jerez},\ and\ \citenamefont
  {Lavagna}}]{Bensimon2006a_SM}%
  \BibitemOpen
  \bibfield  {author} {\bibinfo {author} {\bibfnamefont {D.}~\bibnamefont
  {Bensimon}}, \bibinfo {author} {\bibfnamefont {A.}~\bibnamefont {Jerez}}, \
  and\ \bibinfo {author} {\bibfnamefont {M.}~\bibnamefont {Lavagna}},\ }\href
  {\doibase 10.1103/PhysRevB.73.224445} {\bibfield  {journal} {\bibinfo
  {journal} {Physical Review B - Condensed Matter and Materials Physics}\
  }\textbf {\bibinfo {volume} {73}},\ \bibinfo {pages} {224445} (\bibinfo
  {year} {2006})}\BibitemShut {NoStop}%
\bibitem [{\citenamefont {Affleck}\ and\ \citenamefont
  {Ludwig}(1991)}]{Affleck1991a_SM}%
  \BibitemOpen
  \bibfield  {author} {\bibinfo {author} {\bibfnamefont {I.}~\bibnamefont
  {Affleck}}\ and\ \bibinfo {author} {\bibfnamefont {A.~W.}\ \bibnamefont
  {Ludwig}},\ }\href {\doibase 10.1016/0550-3213(91)90419-X} {\bibfield
  {journal} {\bibinfo  {journal} {Nuclear Physics, Section B}\ }\textbf
  {\bibinfo {volume} {360}},\ \bibinfo {pages} {641} (\bibinfo {year}
  {1991})}\BibitemShut {NoStop}%
\bibitem [{\citenamefont {Affleck}\ and\ \citenamefont
  {Ludwig}(1993)}]{Affleck1993b_SM}%
  \BibitemOpen
  \bibfield  {author} {\bibinfo {author} {\bibfnamefont {I.}~\bibnamefont
  {Affleck}}\ and\ \bibinfo {author} {\bibfnamefont {A.~W.~W.}\ \bibnamefont
  {Ludwig}},\ }\href {\doibase 10.1103/PhysRevB.48.7297} {\bibfield  {journal}
  {\bibinfo  {journal} {Physical Review B}\ }\textbf {\bibinfo {volume} {48}},\
  \bibinfo {pages} {7297} (\bibinfo {year} {1993})}\BibitemShut {NoStop}%
\bibitem [{\citenamefont {Ludwig}\ and\ \citenamefont
  {Affleck}(1991)}]{Ludwig1991_SM}%
  \BibitemOpen
  \bibfield  {author} {\bibinfo {author} {\bibfnamefont {A.~W.~W.}\
  \bibnamefont {Ludwig}}\ and\ \bibinfo {author} {\bibfnamefont
  {I.}~\bibnamefont {Affleck}},\ }\href {\doibase 10.1103/PhysRevLett.67.3160}
  {\bibfield  {journal} {\bibinfo  {journal} {Physical Review Letters}\
  }\textbf {\bibinfo {volume} {67}},\ \bibinfo {pages} {3160} (\bibinfo {year}
  {1991})}\BibitemShut {NoStop}%
\bibitem [{\citenamefont {Kimura}\ and\ \citenamefont
  {Ozaki}(2017)}]{Kimura2017b_SM}%
  \BibitemOpen
  \bibfield  {author} {\bibinfo {author} {\bibfnamefont {T.}~\bibnamefont
  {Kimura}}\ and\ \bibinfo {author} {\bibfnamefont {S.}~\bibnamefont {Ozaki}},\
  }\href {\doibase 10.7566/JPSJ.86.084703} {\bibfield  {journal} {\bibinfo
  {journal} {Journal of the Physical Society of Japan}\ }\textbf {\bibinfo
  {volume} {86}},\ \bibinfo {pages} {084703} (\bibinfo {year}
  {2017})}\BibitemShut {NoStop}%
\bibitem [{\citenamefont {Schultz}\ \emph
  {et~al.}(2021{\natexlab{a}})\citenamefont {Schultz}, \citenamefont {Patri},\
  and\ \citenamefont {Kim}}]{Schultz2021b_SM}%
  \BibitemOpen
  \bibfield  {author} {\bibinfo {author} {\bibfnamefont {D.~J.}\ \bibnamefont
  {Schultz}}, \bibinfo {author} {\bibfnamefont {A.~S.}\ \bibnamefont {Patri}},
  \ and\ \bibinfo {author} {\bibfnamefont {Y.~B.}\ \bibnamefont {Kim}},\ }\href
  {\doibase 10.1103/PhysRevResearch.3.013189} {\bibfield  {journal} {\bibinfo
  {journal} {Physical Review Research}\ }\textbf {\bibinfo {volume} {3}},\
  \bibinfo {pages} {013189} (\bibinfo {year} {2021}{\natexlab{a}})}\BibitemShut
  {NoStop}%
\bibitem [{\citenamefont {Patri}\ and\ \citenamefont
  {Kim}(2020)}]{Patri2020e_SM}%
  \BibitemOpen
  \bibfield  {author} {\bibinfo {author} {\bibfnamefont {A.~S.}\ \bibnamefont
  {Patri}}\ and\ \bibinfo {author} {\bibfnamefont {Y.~B.}\ \bibnamefont
  {Kim}},\ }\href {\doibase 10.1103/PhysRevX.10.041021} {\bibfield  {journal}
  {\bibinfo  {journal} {Physical Review X}\ }\textbf {\bibinfo {volume} {10}},\
  \bibinfo {pages} {041021} (\bibinfo {year} {2020})}\BibitemShut {NoStop}%
\bibitem [{\citenamefont {Schultz}\ \emph
  {et~al.}(2021{\natexlab{b}})\citenamefont {Schultz}, \citenamefont {Patri},\
  and\ \citenamefont {Kim}}]{Schultz2021c_SM}%
  \BibitemOpen
  \bibfield  {author} {\bibinfo {author} {\bibfnamefont {D.~J.}\ \bibnamefont
  {Schultz}}, \bibinfo {author} {\bibfnamefont {A.~S.}\ \bibnamefont {Patri}},
  \ and\ \bibinfo {author} {\bibfnamefont {Y.~B.}\ \bibnamefont {Kim}},\ }\href
  {\doibase 10.1103/PhysRevB.104.125144} {\bibfield  {journal} {\bibinfo
  {journal} {Physical Review B}\ }\textbf {\bibinfo {volume} {104}},\ \bibinfo
  {pages} {125144} (\bibinfo {year} {2021}{\natexlab{b}})}\BibitemShut
  {NoStop}%
\bibitem [{\citenamefont {Mora}(2009)}]{Mora2009_SM}%
  \BibitemOpen
  \bibfield  {author} {\bibinfo {author} {\bibfnamefont {C.}~\bibnamefont
  {Mora}},\ }\href {\doibase 10.1103/PhysRevB.80.125304} {\bibfield  {journal}
  {\bibinfo  {journal} {Physical Review B - Condensed Matter and Materials
  Physics}\ }\textbf {\bibinfo {volume} {80}},\ \bibinfo {pages} {125304}
  (\bibinfo {year} {2009})}\BibitemShut {NoStop}%
\bibitem [{\citenamefont {Si}\ \emph {et~al.}(2001)\citenamefont {Si},
  \citenamefont {Rabello}, \citenamefont {Ingersent},\ and\ \citenamefont
  {Smith}}]{Si2001_SM}%
  \BibitemOpen
  \bibfield  {author} {\bibinfo {author} {\bibfnamefont {Q.}~\bibnamefont
  {Si}}, \bibinfo {author} {\bibfnamefont {S.}~\bibnamefont {Rabello}},
  \bibinfo {author} {\bibfnamefont {K.}~\bibnamefont {Ingersent}}, \ and\
  \bibinfo {author} {\bibfnamefont {J.~L.}\ \bibnamefont {Smith}},\ }\href
  {\doibase 10.1038/35101507} {\bibfield  {journal} {\bibinfo  {journal}
  {Nature}\ }\textbf {\bibinfo {volume} {413}},\ \bibinfo {pages} {804}
  (\bibinfo {year} {2001})}\BibitemShut {NoStop}%
\bibitem [{\citenamefont {Si}\ and\ \citenamefont
  {Steglich}(2010)}]{Si2010a_SM}%
  \BibitemOpen
  \bibfield  {author} {\bibinfo {author} {\bibfnamefont {Q.}~\bibnamefont
  {Si}}\ and\ \bibinfo {author} {\bibfnamefont {F.}~\bibnamefont {Steglich}},\
  }\href {\doibase 10.1126/science.1191195} {\bibfield  {journal} {\bibinfo
  {journal} {Science}\ }\textbf {\bibinfo {volume} {329}},\ \bibinfo {pages}
  {1161} (\bibinfo {year} {2010})}\BibitemShut {NoStop}%
\bibitem [{\citenamefont {Steglich}\ \emph {et~al.}(2014)\citenamefont
  {Steglich}, \citenamefont {Pfau}, \citenamefont {Lausberg}, \citenamefont
  {Hamann}, \citenamefont {Sun}, \citenamefont {Stockert}, \citenamefont
  {Brando}, \citenamefont {Friedemann}, \citenamefont {Krellner}, \citenamefont
  {Geibel}, \citenamefont {Wirth}, \citenamefont {Kirchner}, \citenamefont
  {Abrahams},\ and\ \citenamefont {Si}}]{Steglich2014_SM}%
  \BibitemOpen
  \bibfield  {author} {\bibinfo {author} {\bibfnamefont {F.}~\bibnamefont
  {Steglich}}, \bibinfo {author} {\bibfnamefont {H.}~\bibnamefont {Pfau}},
  \bibinfo {author} {\bibfnamefont {S.}~\bibnamefont {Lausberg}}, \bibinfo
  {author} {\bibfnamefont {S.}~\bibnamefont {Hamann}}, \bibinfo {author}
  {\bibfnamefont {P.}~\bibnamefont {Sun}}, \bibinfo {author} {\bibfnamefont
  {U.}~\bibnamefont {Stockert}}, \bibinfo {author} {\bibfnamefont
  {M.}~\bibnamefont {Brando}}, \bibinfo {author} {\bibfnamefont
  {S.}~\bibnamefont {Friedemann}}, \bibinfo {author} {\bibfnamefont
  {C.}~\bibnamefont {Krellner}}, \bibinfo {author} {\bibfnamefont
  {C.}~\bibnamefont {Geibel}}, \bibinfo {author} {\bibfnamefont
  {S.}~\bibnamefont {Wirth}}, \bibinfo {author} {\bibfnamefont
  {S.}~\bibnamefont {Kirchner}}, \bibinfo {author} {\bibfnamefont
  {E.}~\bibnamefont {Abrahams}}, \ and\ \bibinfo {author} {\bibfnamefont
  {Q.}~\bibnamefont {Si}},\ }\href {\doibase 10.7566/JPSJ.83.061001} {\bibfield
   {journal} {\bibinfo  {journal} {Journal of the Physical Society of Japan}\
  }\textbf {\bibinfo {volume} {83}},\ \bibinfo {pages} {061001} (\bibinfo {year}
  {2014})}\BibitemShut {NoStop}%
\bibitem [{\citenamefont {Friedemann}\ \emph
  {et~al.}(2010{\natexlab{a}})\citenamefont {Friedemann}, \citenamefont
  {Wirth}, \citenamefont {Oeschler}, \citenamefont {Krellner}, \citenamefont
  {Geibel}, \citenamefont {Steglich}, \citenamefont {Maquilon}, \citenamefont
  {Fisk}, \citenamefont {Paschen},\ and\ \citenamefont
  {Zwicknagl}}]{Friedemann2010_SM}%
  \BibitemOpen
  \bibfield  {author} {\bibinfo {author} {\bibfnamefont {S.}~\bibnamefont
  {Friedemann}}, \bibinfo {author} {\bibfnamefont {S.}~\bibnamefont {Wirth}},
  \bibinfo {author} {\bibfnamefont {N.}~\bibnamefont {Oeschler}}, \bibinfo
  {author} {\bibfnamefont {C.}~\bibnamefont {Krellner}}, \bibinfo {author}
  {\bibfnamefont {C.}~\bibnamefont {Geibel}}, \bibinfo {author} {\bibfnamefont
  {F.}~\bibnamefont {Steglich}}, \bibinfo {author} {\bibfnamefont
  {S.}~\bibnamefont {Maquilon}}, \bibinfo {author} {\bibfnamefont
  {Z.}~\bibnamefont {Fisk}}, \bibinfo {author} {\bibfnamefont {S.}~\bibnamefont
  {Paschen}}, \ and\ \bibinfo {author} {\bibfnamefont {G.}~\bibnamefont
  {Zwicknagl}},\ }\href {\doibase 10.1103/PhysRevB.82.035103} {\bibfield
  {journal} {\bibinfo  {journal} {Physical Review B - Condensed Matter and
  Materials Physics}\ }\textbf {\bibinfo {volume} {82}},\ \bibinfo {pages}
  {035103} (\bibinfo {year} {2010}{\natexlab{a}})}\BibitemShut {NoStop}%
\bibitem [{\citenamefont {Friedemann}\ \emph
  {et~al.}(2010{\natexlab{b}})\citenamefont {Friedemann}, \citenamefont
  {Oeschler}, \citenamefont {Wirth}, \citenamefont {Krellner}, \citenamefont
  {Geibel}, \citenamefont {Steglich}, \citenamefont {Paschen}, \citenamefont
  {Kirchner},\ and\ \citenamefont {Si}}]{Friedemann2010a_SM}%
  \BibitemOpen
  \bibfield  {author} {\bibinfo {author} {\bibfnamefont {S.}~\bibnamefont
  {Friedemann}}, \bibinfo {author} {\bibfnamefont {N.}~\bibnamefont
  {Oeschler}}, \bibinfo {author} {\bibfnamefont {S.}~\bibnamefont {Wirth}},
  \bibinfo {author} {\bibfnamefont {C.}~\bibnamefont {Krellner}}, \bibinfo
  {author} {\bibfnamefont {C.}~\bibnamefont {Geibel}}, \bibinfo {author}
  {\bibfnamefont {F.}~\bibnamefont {Steglich}}, \bibinfo {author}
  {\bibfnamefont {S.}~\bibnamefont {Paschen}}, \bibinfo {author} {\bibfnamefont
  {S.}~\bibnamefont {Kirchner}}, \ and\ \bibinfo {author} {\bibfnamefont
  {Q.}~\bibnamefont {Si}},\ }\href {\doibase 10.1073/pnas.1009202107}
  {\bibfield  {journal} {\bibinfo  {journal} {Proceedings of the National
  Academy of Sciences of the United States of America}\ }\textbf {\bibinfo
  {volume} {107}},\ \bibinfo {pages} {14547} (\bibinfo {year}
  {2010}{\natexlab{b}})}\BibitemShut {NoStop}%
\bibitem [{\citenamefont {Paschen}\ \emph {et~al.}(2004)\citenamefont
  {Paschen}, \citenamefont {L{\"{u}}hmann}, \citenamefont {Wirth},
  \citenamefont {Gegenwart}, \citenamefont {Trovarelli}, \citenamefont
  {Geibel}, \citenamefont {Steglich}, \citenamefont {Coleman},\ and\
  \citenamefont {Si}}]{Paschen2004_SM}%
  \BibitemOpen
  \bibfield  {author} {\bibinfo {author} {\bibfnamefont {S.}~\bibnamefont
  {Paschen}}, \bibinfo {author} {\bibfnamefont {T.}~\bibnamefont
  {L{\"{u}}hmann}}, \bibinfo {author} {\bibfnamefont {S.}~\bibnamefont
  {Wirth}}, \bibinfo {author} {\bibfnamefont {P.}~\bibnamefont {Gegenwart}},
  \bibinfo {author} {\bibfnamefont {O.}~\bibnamefont {Trovarelli}}, \bibinfo
  {author} {\bibfnamefont {C.}~\bibnamefont {Geibel}}, \bibinfo {author}
  {\bibfnamefont {F.}~\bibnamefont {Steglich}}, \bibinfo {author}
  {\bibfnamefont {P.}~\bibnamefont {Coleman}}, \ and\ \bibinfo {author}
  {\bibfnamefont {Q.}~\bibnamefont {Si}},\ }\href {\doibase 10.1038/nature03129} {\bibfield  {journal} {\bibinfo  {journal} {Nature}\
  }\textbf {\bibinfo {volume} {432}},\ \bibinfo {pages} {881} (\bibinfo {year}
  {2004})}\BibitemShut {NoStop}%
\bibitem [{\citenamefont {Coleman}\ \emph {et~al.}(2001)\citenamefont
  {Coleman}, \citenamefont {P{\'{e}}pin}, \citenamefont {Si},\ and\
  \citenamefont {Ramazashvili}}]{Coleman2001_SM}%
  \BibitemOpen
  \bibfield  {author} {\bibinfo {author} {\bibfnamefont {P.}~\bibnamefont
  {Coleman}}, \bibinfo {author} {\bibfnamefont {C.}~\bibnamefont
  {P{\'{e}}pin}}, \bibinfo {author} {\bibfnamefont {Q.}~\bibnamefont {Si}}, \
  and\ \bibinfo {author} {\bibfnamefont {R.}~\bibnamefont {Ramazashvili}},\
  }\href {\doibase 10.1088/0953-8984/13/35/202} {\bibfield  {journal} {\bibinfo
   {journal} {Journal of Physics Condensed Matter}\ }\textbf {\bibinfo {volume}
  {13}},\ \bibinfo {pages} {723} (\bibinfo {year} {2001})}\BibitemShut
  {NoStop}%
\bibitem [{\citenamefont {Zar{\'{a}}nd}\ and\ \citenamefont
  {Demler}(2002)}]{Zarand2002_SM}%
  \BibitemOpen
  \bibfield  {author} {\bibinfo {author} {\bibfnamefont {G.}~\bibnamefont
  {Zar{\'{a}}nd}}\ and\ \bibinfo {author} {\bibfnamefont {E.}~\bibnamefont
  {Demler}},\ }\href {\doibase 10.1103/PhysRevB.66.024427} {\bibfield
  {journal} {\bibinfo  {journal} {Physical Review B - Condensed Matter and
  Materials Physics}\ }\textbf {\bibinfo {volume} {66}},\ \bibinfo {pages}
  {024427} (\bibinfo {year} {2002})}\BibitemShut {NoStop}%
\bibitem [{\citenamefont {Patri}\ \emph {et~al.}(2019)\citenamefont {Patri},
  \citenamefont {Sakai}, \citenamefont {Lee}, \citenamefont {Paramekanti},
  \citenamefont {Nakatsuji},\ and\ \citenamefont {Kim}}]{Patri2019d_SM}%
  \BibitemOpen
  \bibfield  {author} {\bibinfo {author} {\bibfnamefont {A.~S.}\ \bibnamefont
  {Patri}}, \bibinfo {author} {\bibfnamefont {A.}~\bibnamefont {Sakai}},
  \bibinfo {author} {\bibfnamefont {S.~B.}\ \bibnamefont {Lee}}, \bibinfo
  {author} {\bibfnamefont {A.}~\bibnamefont {Paramekanti}}, \bibinfo {author}
  {\bibfnamefont {S.}~\bibnamefont {Nakatsuji}}, \ and\ \bibinfo {author}
  {\bibfnamefont {Y.~B.}\ \bibnamefont {Kim}},\ }\href {\doibase 10.1038/s41467-019-11913-3} {\bibfield  {journal} {\bibinfo  {journal}
  {Nature Communications}\ }\textbf {\bibinfo {volume} {10}} (\bibinfo {year}
  {2019}),\ 10.1038/s41467-019-11913-3}\BibitemShut {NoStop}%
\bibitem [{\citenamefont {Sorensen}\ and\ \citenamefont
  {Fisher}(2021)}]{Sorensen2021_SM}%
  \BibitemOpen
  \bibfield  {author} {\bibinfo {author} {\bibfnamefont {M.~E.}\ \bibnamefont
  {Sorensen}}\ and\ \bibinfo {author} {\bibfnamefont {I.~R.}\ \bibnamefont
  {Fisher}},\ }\href {\doibase 10.1103/PhysRevB.103.155106} {\bibfield
  {journal} {\bibinfo  {journal} {Physical Review B}\ }\textbf {\bibinfo
  {volume} {103}},\ \bibinfo {pages} {155106} (\bibinfo {year}
  {2021})}\BibitemShut {NoStop}%
\end{thebibliography}

%

\end{document}